\documentclass[a4paper,fleqn,usenatbib]{mnras}

\usepackage{newtxtext}
\usepackage{newtxmath}

\usepackage[T1]{fontenc}
\usepackage{ae,aecompl}

\usepackage[normalem]{ulem}  %

\usepackage{graphicx}	%

\ifdefined\Bbbk{}
    \let\Bbbk\relax
\else
\fi
\ifdefined\bigtimes{}
    \let\bigtimes\relax
\else
\fi
\usepackage[separate-uncertainty=true,multi-part-units=single,alsoload=astro]{siunitx}
\usepackage{pgf}
\usepackage{xprintlen}
\usepackage{booktabs}
\usepackage{mathabx}
\usepackage{wasysym}

\usepackage{orcidlink}

\hypersetup{pdfauthor={D. Evensberget et al.}}  %

\newcommand{\batsrus}{{\textsc{bats-r-us}}}
\newcommand{\swmf}{{\textsc{swmf}}}
\newcommand{\awsom}{{\textsc{awsom}}}
\renewcommand{\pluto}{{\textsc{pluto}}} %
\newcommand{\toupies}{{\textsc{toupies}}}

\newcommand{\Wind}{\textsc{w}} %
\newcommand{\Alfven}{\textsc{a}} %
\newcommand{\Star}{\text{\bigstar}} %

\renewcommand{\Earth}{{\mathchoice{}{}{\scriptscriptstyle}{}\oplus}} %
\renewcommand{\Sun}{{\mathchoice{}{}{\scriptscriptstyle}{}\odot}} %
\renewcommand{\Star}{{\mathchoice{}{}{\scriptscriptstyle}{}\bigstar}} %

\defcitealias{2018MNRAS.474.4956F}{F18}

\newcommand{\ZDI}{{\mathchoice{}{}{\scriptscriptstyle}{}\mathrm{ZDI}}}

\DeclareSIUnit\year{yr}
\DeclareSIUnit\astronomicalunit{au}
\DeclareSIUnit\parsec{pc}
\DeclareSIUnit\erg{erg}
\DeclareSIUnit\gauss{G}
\DeclareSIUnit\mSun{\mbox{\(M_\Sun\)}}
\DeclareSIUnit\rSun{\mbox{\(R_\Sun\)}}
\DeclareSIUnit\mEarth{\mbox{\(M_\Earth\)}}
\DeclareSIUnit\rEarth{\mbox{\(R_\Earth\)}}

\renewcommand{\vec}[1]{\boldsymbol{\mathbf{#1}}} %
\newcommand{\uvec}[1]{\boldsymbol{\mathbf{\hat{#1}}}} %

\newcommand\thefont{\expandafter\string\the\font} %

\renewcommand{\propto}{\mathrel{\mathchar"939}} %
\newcommand{\mysim}{{\sim}} %
\title[Winds of the Hyades]{The winds of young Solar-type stars in the Hyades}
\author[D. Evensberget et al.]{D. Evensberget \orcidlink{0000-0001-7810-8028}\(^{1}\)\thanks{E-mail: dag.evensberget@usq.edu.au (USQ)},
    B. D. Carter \orcidlink{0000-0003-0035-8769}\(^{1}\),
    S. C. Marsden \orcidlink{0000-0001-5522-8887}\(^{1}\),
    L. Brookshaw \orcidlink{0000-0001-8503-0062}\(^{1}\), and 
    \newauthor
    C. P. Folsom \orcidlink{0000-0002-9023-7890}\(^{2,3,4}\)
    \\
    \(^{1}\)Centre for Astrophysics, University of Southern Queensland, Toowoomba, Queensland 4350, Australia\\
    \(^{2}\)Department of Physics \& Space Science, Royal Military College of Canada, PO Box 17000 Station Forces, Kingston, ON, Canada K7K 0C6\\
    \(^{3}\)Institut de Recherche en Astrophysique et Plan\'{e}tologie, Universit\'{e} de Toulouse, CNRS, CNES, 14 avenue Edouard Belin, 31400 Toulouse, France\\
    \(^{4}\)Tartu Observatory, University of Tartu, Observatooriumi 1, T\~{o}ravere, 61602 Tartumaa, Estonia
}

\date{Accepted XXX.\@ Received YYY;\@ in original form ZZZ}
\pubyear{2020}
\begin{document}
\label{firstpage}
\pagerange{\pageref{firstpage}--\pageref{lastpage}}
\maketitle
\begin{abstract}

Stellar winds govern the spin-down of Solar-type stars as they age, and play an important role in determining planetary habitability, as powerful winds can lead to atmospheric erosion. 
We calculate three-dimensional stellar wind models for five young Solar-type stars in the Hyades cluster, using TOUPIES survey stellar magnetograms and state-of-the-art Alfvén wave driven wind modelling. The stars have the same 0.6~Gyr age  and similar fundamental parameters, and we account for the uncertainty in and underestimation of absolute field strength inherent in Zeeman-Doppler imaging by adopting both unscaled and scaled (by a factor of five) field strengths. For the unscaled fields, the resulting stellar wind mass loss is 2–4 times greater and the angular momentum loss 2–10 times greater than for the Sun today, with the scaled results correspondingly greater. We compare our results with a range published of wind models and for the Alfvén wave driven modelling see evidence of mass loss saturation at \(\num{\sim 10} \dot M_\Sun\). 
\end{abstract}


\section{Introduction}

As they age on the main sequence, Solar-type stars undergo significant changes. Main sequence Solar-type stars tend to develop magnetic fields that extend from the inner radiative layers to the stellar surface \citep{2003ApJ...586..464B}. This permits the entire star to shed angular momentum via the stellar wind. This mechanism is believed to explain the observationally-derived \citet[][]{1972ApJ...171..565S} law \(\Omega \propto t^{-1/2}\) relating angular velocity \(\Omega\) and age \(t\).
The stellar surface magnetic field strength \(B\), which may be inferred from spectropolarimetric observations~\cite[e.g.\@][]{2014MNRAS.444.3517M} is also correlated with age and rotation rate \citep{2014MNRAS.441.2361V}, as would be expected from dynamo models. Spectropolarimetric observations also permit the reconstruction of the large-scale stellar surface magnetic field through Zeeman-Doppler imaging \citep[ZDI,\@][]{1989A&A...225..456S}.

Angular momentum shedding is enhanced by magnetic forces which cause the stellar wind to co-rotate with the star. The co-rotation extends out to the radius \(R_\Alfven \propto B\) where the wind speed exceeds the propagation speed of Alfvén waves \citep{1942Natur.150..405A} and thus becomes decoupled from the stellar magnetic field. This co-rotation radius functions as a lever arm that greatly increases the amount of angular momentum lost via stellar winds~\citep{1962AnAp...25...18S}. 
The one-dimensional \citet{1967ApJ...148..217W} model of angular momentum loss \(\dot J = \frac{2}{3} \dot M \Omega R_\Alfven^2 \) can be used to derive a theoretical explanation of the Skumanich law. 

The wind mass loss \(\dot M\) for Solar-type stars is not well constrained. The most stringent upper limits on stellar wind mass loss come from observing the Lyman-\(\alpha\) line \citep{2014ApJ...781L..33W} in the outer astrosphere. Going backwards in time from the present-day Sun, the Lyman-\(\alpha\) observations predicts mass loss values rising up to \( \mysim  10^2 \dot M_\Sun\) before the relation breaks down at the `wind dividing line' age of \( \mysim \SI{0.7}{\giga\year}\). Stars beyond the wind dividing line have been found to have lower values of wind mass loss.

The stellar wind pressure in the habitable zone is dominated by the ram pressure term, which is proportional to
the product of the wind speed and \(\dot M\).
The wind pressure plays an important role in planetary habitability, as powerful winds can lead to atmospheric erosion \citep{2008SSRv..139..399L}. A clear understanding of the history of stellar winds in our Solar system can inform habitability studies for both exoplanets and Solar system planets. 

The similarity of the Solar-type Hyades stars make them well suited for studying the direct effect of surface magnetic field strength and complexity on the stellar wind, and to isolate and clarify the natural wind variability resulting from magnetic changes.
In addition to having the same age of \SI{0.6}{\giga\year}, the Hyades stars have the same provenance and therefore the same composition. 
In contrast to younger clusters,
the Hyades consists of slow rotators and exhibits little variation in terms of stellar rotation rates~\citep{2011MNRAS.413.2218D}. 
The small variation is consistent with theories of angular momentum evolution \citep[e.g.\@][]{1997A&A...326.1023B} and the Skumanich law, as rapid rotators would have had sufficient time to shed their initial angular momentum. 
These similarities between the stars mean that wind variations due to differences in stellar mass, age, composition, and period of rotation are reduced to a minimum.

By letting the observation-based ZDI surface magnetic field maps of Solar-type Hyades stars published by ~\citet{2018MNRAS.474.4956F}, hereafter~\citetalias{2018MNRAS.474.4956F}, drive
a state of the art numerical space weather code \citep{1999JCoPh.154..284P,2012JCoPh.231..870T} incorporating Alfvén wave heating and two-temperature effects \citep{2013ApJ...764...23S, 2014ApJ...782...81V} we produce fully three-dimensional wind maps extending from the transition region, past the corona and out to distances of several \si{\astronomicalunit{}}. The wind maps comprise the regular magnetohydrodynamic quantities, separate electron and ion energies, and Alfvén wave energy. 

The resulting observation-based wind maps permit us 
to calculate aggregate wind quantities such as wind mass loss \(\dot M\) and angular momentum loss \(\dot J\), and study spatial wind variations in wind pressure encountered by an Earth-like planet as it orbits the star.
The age of the Hyades make the models and wind maps particularly interesting as the stellar ages are close to the age of the Sun when life arose on the Earth \citep{1996Natur.384...55M},
 and close to the wind dividing line of \citet{2014ApJ...781L..33W}.

This paper is outlined as follows: 
In Section~\ref{sec:Observations} we describe the surface magnetic maps and how they are obtained using Zeeman-Doppler imaging; 
In Section~\ref{sec:Simulations} we describe the model equations and numerical model;
In Section~\ref{sec:results} we give an overview of our model results including aggregate quantities calculated from the wind models such as mass loss; 
In Section~\ref{sec:Discussion} we examine trends in aggregate quantities within our own dataset, and compare these trends with Solar values and values from similar studies in the literature; we also consider the implications of our results for the young Sun and Earth.
In Section~\ref{sec:Conclusions} we conclude and summarise our findings.

\section{Observations}\label{sec:Observations}
The stellar wind maps we present are based on observations carried out as part of the `TOwards Understanding the sPIn Evolution of Stars' (\toupies{}) project\footnote{\url{http://ipag.osug.fr/Anr_Toupies/}}, which used the ESPaDOnS instrument~\citep{2003ASPC..307...41D,2012MNRAS.426.1003S} at the Canada-Hawaii-France Telescope, and the Narval instrument~\citep{2003EAS.....9..105A} at the Télescope Bernard Lyot. ESPaDOnS and Narval are spectropolarimeters with a resolution \(R  \sim  65000\) that cover a wavelength range \SIrange{370}{1050}{\nano\meter}. In the \toupies{} study the instruments were configured to simultaneously record the circularly polarised Stokes \(V\) spectrum and the total intensity Stokes \(I\) spectrum. The particular observations of the Hyades stars which are used in this study were carried out as part of the \emph{History of the Magnetic Sun} Large Program at CFHT and are thus all made with ESPaDOnS.

\begin{figure}
        \centering
        \includegraphics{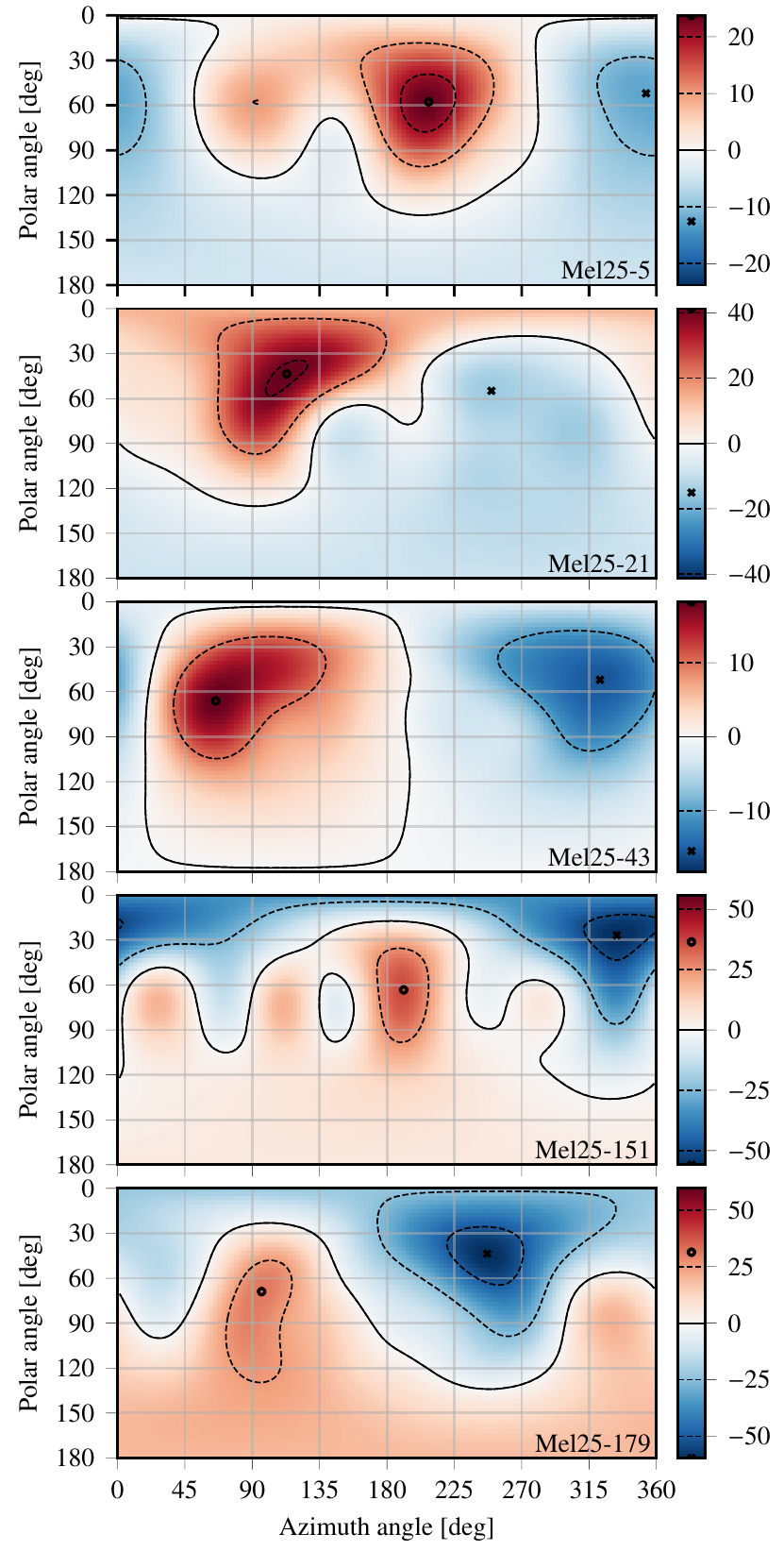}
        \caption{Radial magnetic field strength in Gauss for the Hyades stars modelled in this work based on the radial magnetic field coefficients derived in \citetalias{2018MNRAS.474.4956F}. 
        The azimuth angle is measured around the stellar equator. 
        The polar angle is measured southwards from the rotational north pole. 
        The circle and cross show the position and value of the maximum and minimum. The fully drawn contour line is where the radial magnetic field strength is zero, and the dashed contour lines represent increments as shown in the corresponding colour bar on the right of each plot. Note that the colour scale is not fixed between the plots.
        }\label{fig:magnetograms-zdi}
    \end{figure}

The stellar targets were observed over a two week period to minimise intrinsic magnetic field variations during the 
observations, while observing the star at different phases in order to map the entire visible stellar surface, and to simultaneously collect sufficient data to obtain an acceptable signal-to-noise ratio.
The radial component of the magnetograms which were derived in~\citetalias{2018MNRAS.474.4956F} are shown in Fig.~\ref{fig:magnetograms-zdi}; the derivation process is briefly described in Section~\ref{sec:mag-map-zdi}, we refer the reader to~\citetalias{2018MNRAS.474.4956F} and references therein for more information on the ZDI observations.

\begin{figure}
        \centering
        \includegraphics{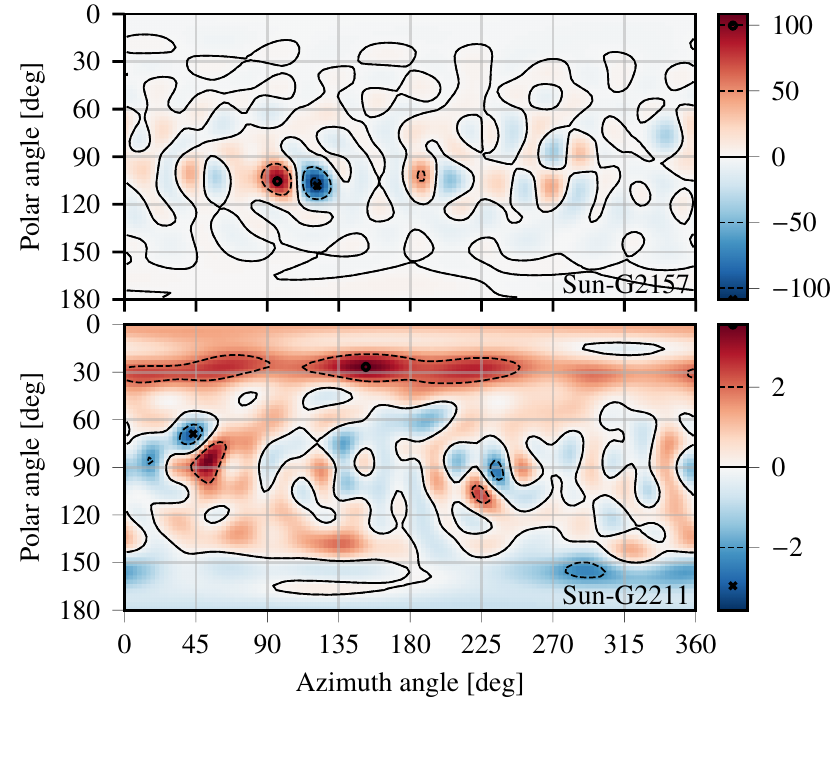}
        \caption{
            Radial magnetic field strength in Gauss for Solar cases (top: Solar maximum, bottom: Solar minimum) modelled in this work, based on magnetograms from the National Solar Observatory Global Oscillation Network Group 
            \citep[GONG, e.g.\@][]{1996Sci...272.1284H} 
            {\it CR Spherical Harmonic Transform Coefficients\/}
            product, filtered to the same formal resolution and using the same plotting scheme in Fig.~\ref{fig:magnetograms-zdi}. 
            The minimum angular length scale of represented features is \(\mysim\SI{12}{\degree}\). The difference in effective resolution between the Hyades magnetograms and the filtered Solar magnetograms is discussed in Section~\ref{sec:ZDI-missing-field}.
            Note the different colour scales for the two plots.
    }\label{fig:magnetograms-sun}
    \end{figure}

For comparison we also create wind models based on two filtered Solar magnetograms representing conditions at Solar maximum and Solar minimum, see Fig.~\ref{fig:magnetograms-sun}. The Solar magnetograms coefficients are obtained from the National Solar Observatory Global Oscillation Network Group \citep[GONG, e.g.\@][]{1996Sci...272.1284H} {\it CR Spherical Harmonic Transform Coefficients\/}\footnote{Available from \url{https://gong.nso.edu/}} product. The choice of Solar magnetogram and the filtering applied is described in Section~\ref{sec:Solar-magnetograms}.

\subsection{Magnetic mapping with Zeeman-Doppler imaging}\label{sec:mag-map-zdi}
In this section we briefly describe the methodology for generating magnetic maps from stellar spectropolarimetric data.

\begin{table}
    \centering
        \caption{
        Relevant stellar quantities from~\citetalias{2018MNRAS.474.4956F} and references therein. The stellar mass \(M\), radius \(R\), and period of rotation \(P_\text{rot}\) are input to the numerical simulations in Section~\ref{sec:numerical_model}. In the remainder of this work the stars will be referred by abbreviated names, e.g.\@ Mel25-5 refers to the star Cl~Melotte~25~5.
        }
    
        \addtolength{\tabcolsep}{-1pt}    
        \begin{tabular}{lcccr}
            \toprule
            Name 
            & Type
            & \(M\)            \(\left(\si{\mSun}\right)\)
            & \(R\)            \(\left(\si{\rSun}\right)\)
            & \(P_\text{rot}\) \(\left(\si{\day}\right)\)
            \\
            \midrule
            \href{http://simbad.u-strasbg.fr/simbad/sim-id?Ident=Cl+Melotte+25+5  }{Cl Melotte 25 5  } & K0 & \(0.85\pm{0.05}\) & \(0.91\pm{0.04}\) & \(10.6\pm{0.1}\) \\ %
            \href{http://simbad.u-strasbg.fr/simbad/sim-id?Ident=Cl+Melotte+25+21 }{Cl Melotte 25 21 } & G5 & \(0.90\pm{0.05}\) & \(0.91\pm{0.04}\) & \( 9.7\pm{0.2}\) \\ %
            \href{http://simbad.u-strasbg.fr/simbad/sim-id?Ident=Cl+Melotte+25+43 }{Cl Melotte 25 43 } & K0 & \(0.90\pm{0.05}\) & \(0.79\pm{0.08}\) & \( 9.9\pm{0.1}\) \\ %
            \href{http://simbad.u-strasbg.fr/simbad/sim-id?Ident=Cl+Melotte+25+151}{Cl Melotte 25 151} & K2 & \(0.85\pm{0.05}\) & \(0.82\pm{0.13}\) & \(10.4\pm{0.1}\) \\ %
            \href{http://simbad.u-strasbg.fr/simbad/sim-id?Ident=Cl+Melotte+25+179}{Cl Melotte 25 179} & K0 & \(0.85\pm{0.04}\) & \(0.84\pm{0.03}\) & \( 9.7\pm{0.1}\) \\ %
            \bottomrule 
        \end{tabular}\label{tab:observed_quantities}
        \addtolength{\tabcolsep}{1pt}
    \end{table}

    
Stellar magnetic fields can be reconstructed from spectropolarimetric observations. \citet{1989A&A...225..456S} pioneered the application of maximum entropy image reconstruction~\citep{1984MNRAS.211..111S} to this problem. The resulting technique, called Zeeman-Doppler imaging (ZDI) has been used to produce synoptic maps of the magnetic fields of numerous cool stars; the review by~\citet{2009ARA&A..47..333D} gives an overview of the application of ZDI to various types of stars including Solar-type stars.

As individual spectral lines from stars generally lack sufficient signal-to-noise in the Stokes \(V\) parameter for the Zeeman effect to be distinguished from noise, least square deconvolution~\citep[LSD,][]{1997MNRAS.291..658D,2010A&A...524A...5K} is used to combine spectral lines into a single weighted average LSD profile with a much higher signal-to-noise ratio.

Modern ZDI describes the surface magnetic field in terms of spherical harmonic coefficients~\citep{1999MNRAS.305L..35J,2006MNRAS.370..629D}.
The coefficients are found by applying the maximum entropy image reconstruction method to the LSD profile. 
This method selects a set of coefficients that simultaneously maximises `entropy' and satisfies a bound on the \(\chi^2\) value of the fit.
We direct the reader to~\citet{2016MNRAS.457..580F} and~\citetalias{2018MNRAS.474.4956F} for details concerning the derivation of the surface magnetic maps. In their methodology, all three vector components of the surface
magnetic field are recovered. In this work we only use the radial field to drive our simulations, as is common in MHD wind modelling (see Section~\ref{sec:numerical_model}). 
Here, we restate the formula for the radial component: The stellar surface radial magnetic field is represented as the real part of an orthogonal sum of the form
\begin{equation}\label{eq:zdi_br}
    B_r(\theta, \varphi) =
    \sum_{\ell=1}^{\ell_\text{max}} \,
    \sum_{m=0}^\ell \alpha_{\ell m} 
    \sqrt{\frac{2\ell+1}{4\pi} \frac{(\ell-m)!}{(\ell+m)!}} 
    P_{\ell m}(\cos \theta) e^{i m\varphi} 
\end{equation}
where \(\alpha_{\ell m} \) are the complex-valued spherical harmonics coefficients of the radial field, \(P_{\ell m}(\cos \theta)\) is the associated Legendre polynomial order \(m\) and degree \(\ell\),
and \(\theta, \varphi\) are the polar and azimuthal angles in spherical coordinates. 
As only the real part of equation~\eqref{eq:zdi_br} is used, it is sufficient to let \(m\) range from \(0\) to \(\ell\), i.e.\ negative \(m\) values are omitted from the sum. The amount of detail in the representation is controlled by the \(\ell_\text{max}\) parameter; the smallest features that can be reproduced are 
\( \mysim \SI{180}{\degree}/\ell_\text{max}\) in angular diameter.

\subsubsection{Estimates of ZDI uncertainty}\label{sec:ZDI-uncertainties}
The ZDI method does not propagate the uncertainty estimates which are implied by the signal-to-noise ratio of the input Stokes \(I\) and \(V\) profile data. Consequently the method does not produce error estimates along with the magnetic maps. 
It is, however, generally assumed that ZDI reproduces the large-scale magnetic field, although the absolute field strength is subject to considerable uncertainty and may be underreported.
The ability of ZDI to resolve the large-scale field is supported by observations of polarity reversals, presumably as part of a stellar magnetic cycle. Observations of the same star at different epochs have shown periodic polarity reversals in the stars
\(\tau\)~Bo\"{o}tis~\citep{2008MNRAS.385.1179D,2009MNRAS.398.1383F,2013MNRAS.435.1451F,2016MNRAS.459.4325M} and HD75332~\citep{2021MNRAS.501.3981B}; 
\citet{2011AN....332..866M} found evidence of field reversals in HD~78366 and HD~190771, and a more complex cycle in \(\xi\)~Bo\"{o}tis~A. 
We consider the effect of magnetic cycles on this work in Section~\ref{sec:magnetic_cycles}.

A study of two separate ZDI implementations by \citet{2000MNRAS.318..961H} confirmed that the polarity structure of the large-scale field is reconstructed accurately and robustly with respect to implementation details. 
The ZDI method has also been criticised: the review by~\citet{2016LNP...914..177K} notes significant differences between two ZDI field reconstructions by~\citet{2010MNRAS.403..159S} and by~\citet{2012A&A...548A..95C} based on observations made in overlapping time periods, and advocates the inclusion of radiative transfer modelling and inclusion of the linear polarisation signal when available. The linear polarisation signal is, however, \( \mysim  10\) times weaker than the circular polarisation signal, and hence will remain unavailable except in special cases such as the work of \citet{2015ApJ...805..169R}. For further discussions about the inherent uncertainty in ZDI we refer the reader to the papers of \citet{1997A&A...326.1135D} and \citet{2010MNRAS.407.2269M}. 

We have noted, as was observed by e.g.~\citet{2016MNRAS.459.4325M} that the ZDI reconstructed average absolute field strength is sensitive to the choice of target \(\chi^2\) value. As the \(\chi^2\) value is reduced, the risk of overfitting is increased. Overfitting may result in a non-physical increase in reconstructed field strength. 
For ZDI methods that do not use the maximum entropy image reconstruction of~\citet{1984MNRAS.211..111S}, a regularisation parameter must similarly be chosen in order not to under-fit or over-fit the observations.
\citet{2015A&A...582A..38A} proposed selecting a \(\chi^2\) value by maximising the second derivative of entropy as a function of \(\chi^2\). We refer the reader to \citetalias{2018MNRAS.474.4956F} for details about the choice of target \(\chi^2\) value used in this work.

\subsubsection{Missing field of Zeeman-Doppler imaging}\label{sec:ZDI-missing-field}
While ZDI is able to reproduce the large-scale magnetic field, the effective resolution of ZDI is limited in comparison to the Solar magnetograms. 
The ZDI magnetograms in Fig.~\ref{fig:magnetograms-zdi} and the Solar magnetograms in Fig.~\ref{fig:magnetograms-sun} both have \(\ell_\text{max}=15\). Consequently both sets of magnetograms have a formal minimum angular length scale of \(\SI{12}{\degree}\). A visual comparison of Fig.~\ref{fig:magnetograms-zdi} and Fig.~\ref{fig:magnetograms-sun}, however, indicates that the formal minimum angular length scale of the ZDI magnetograms is not attained for the Hyades stars: the smallest features are closer to \(\SI{45}{\degree}\) in angular diameter, giving an effective \(\ell_\text{max}\sim 4\). The effective resolution of ZDI is dependent on the stellar inclination and rate of rotation \citep{2010MNRAS.407.2269M} and the \(\ell_\text{max}=15\) value in \citetalias{2018MNRAS.474.4956F} was chosen in order to match the methodology of the earlier work on younger, more rapidly rotating stars in \citet{2016MNRAS.457..580F}. In the future we intend to extend our work to these stars as well; this justifies the choice of \(\ell_\text{max}=15\) for the magnetograms in this work.

An indication of the magnitude of the missing ZDI magnetic field can be found by comparing to a complementary technique known as Zeeman broadening.
The Zeeman broadening is sensitive to the absolute field strength of small-scale and large-scale fields, while ZDI is prone to cancellation effects which leads to an underestimation of the field strength. 
\citet{2015ApJ...813L..31Y} found that only about \SI{20}{\percent} of the magnetic flux found with Zeeman broadening is recovered using ZDI.\@
Similarly,~\citet{2018MNRAS.480..477V}  found that ZDI reproduces about \SIrange{10}{20}{\percent} of the magnetic field.
A similar result was found by~\citet{2019ApJ...876..118S} who found values ranging from a few percent to \SI{20}{\percent}.
Most recently,~\citet{2020A&A...635A.142K} found that ZDI reconstructs between \SI{10}{\percent} and \SI{1}{\percent} of the magnetic field energy (which is proportional to \(B^2\)) depending on the field strength.  
 
In an approach based on simulated photospheric fields,~\citet{2019MNRAS.483.5246L} found that the polarity structure of the radial magnetic field could be recovered, while the magnitude of the field could not in general be recovered. Rather, the individual energies of the spherical harmonics coefficients of the ZDI reconstructed field could vary from \SIrange{10}{110}{\percent} of the original value. 

Both the missing small-scale field and the potential underreporting of the large-scale field are issues that should be accounted for when comparing Solar and stellar magnetograms and the wind models derived from them. We discuss the issues of using stellar magnetograms in place of Solar magnetograms in Section~\ref{sect:effect-of-zdi-limitations}.

\subsection{The Solar magnetograms}\label{sec:Solar-magnetograms}
To verify that the wind model 
produces
reasonable results we compare our modelling results to Solar cases corresponding to Solar minimum and Solar maximum.
We chose the GONG magnetograms from Solar cycle 24 (December 2008--December 2019) that exhibited the highest and lowest average surface magnetic field strength: these were the magnetograms from Carrington Rotation 2157 (Solar maximum, November 2014) and 2211 (Solar minimum, December 2018). 

Unlike the ZDI magnetograms which are reconstructed from spectropolarimetric data as the star rotates, the Solar magnetograms are reconstructed from  observations of the extended Solar disc. 
The GONG Solar magnetograms have degree \(\ell_\text{max}=60\), i.e.\ the smallest length scale is \SI{3}{\degree}; in comparison to the ZDI magnetograms they are highly detailed. For a better comparison with the \citetalias{2018MNRAS.474.4956F} magnetograms we truncate the Solar magnetograms at the same order as the stellar magnetograms, \(\ell_\text{max}=15\). The two resulting magnetograms are shown in Fig.~\ref{fig:magnetograms-sun}.

While the Solar magnetograms are not affected by the ZDI error sources, comparisons of Solar magnetograms from different observatories have shown systematic variations that warrant the application of scaling factors, e.g.\@
a factor of 2--3 when comparing MDI magnetograms and GONG magnetograms~\citep{2014SoPh..289..769R}. We return to some of these issues in Section~\ref{sect:effect-of-zdi-limitations}.
\section{Simulations}\label{sec:Simulations}
In this section we provide an overview of the numerical simulations carried out as part of this work. Section~\ref{sec:model-equations} describes the differential equations governing our models,
and
Section~\ref{sec:numerical_model} describes the numerical model and boundary conditions. 

\subsection{Model equations}\label{sec:model-equations}

In this work we use the Space Weather Modelling Framework~\cite[\swmf{}, ][]{2005JGRA..11012226T,2012JCoPh.231..870T}, in particular the Alfvén Wave Solar Model~\cite[\awsom{}, ][]{2013ApJ...764...23S, 2014ApJ...782...81V} to simulate stellar winds driven by the \toupies{} magnetic maps described in Section~\ref{sec:mag-map-zdi}. The \awsom{} model is an extension of the \batsrus{} model \citep{1999JCoPh.154..284P,2012JCoPh.231..870T}. The \swmf{} permits us to simulate both the stellar coronae and the resulting stellar winds out to planetary distances; the outer boundary of our model is set to \(450 R_\Star\). For readers unfamiliar with \awsom{} we recommend the review by~\citet{2018LRSP...15....4G}.

Alfvén waves (travelling oscillations of the magnetic field) emanating from inside the Sun are one mechanism of coronal heating~\citep{1968ApJ...154..751B} and has been thought to have sufficient energy~\citep{1968ApJ...153..371C} to power Solar winds. 
Other proposed methods of heating include magnetic reconnection events 
\citep{1972ApJ...174..499P}, and 
type II spicules \citep{2011Sci...331...55D}. 
Indeed, a full explanation may involve multiple phenomena \citep{2019ARA&A..57..157C} in different regions.

In the \awsom{} model, the corona is heated by Alfvén waves in a physics-based, self-consistent manner. The model includes a physical model of the transition region in which the wind is heated to coronal temperatures by Alfvén wave energy emanating from deeper stellar layers, resulting in a Poynting flux \(\Pi_\Alfven\) proportional to the local \(|\vec{B}|\) value at the inner model boundary.  \citet{2013ApJ...764...23S} note that the \(\Pi_\Alfven \propto |\vec{B}|\) assumption is compatible with the models of \citet{2001JGR...10615849F}, \citet{2003ApJ...598.1387P}, \citet{2006ApJ...640L..75S} and \citet{2010ApJ...710..676C}. In the \awsom{} model, the inclusion of Alfvén wave energy is accomplished by complementing the MHD equations by a phenomenological model of Alfvén wave energy propagation, reflection, and dissipation.

Other heat exchange and cooling terms are also included; we will briefly describe them as we go through the model equations in the rest of this section. Heating and cooling terms are necessary~\citep{2003ApJ...595L..57R} to reproduce the bimodality of the slow and fast Solar wind.

\begin{subequations}
The set of equations we are solving comprise the 
two-temperature MHD equations and
two further equations that describe Alfvén wave energy travelling along magnetic field lines in parallel and antiparallel directions.
The mass conservation equation is
\begin{equation}
	\label{eq:mass_conservation}
	\frac{\partial \rho}{\partial t} + \nabla \left(\rho \vec{u}\right) = 0, 
\end{equation}
where \(\rho\) is the mass density and \(\vec{u}\) is the flow velocity;
the induction equation is
\begin{equation}
	\label{eq:induction}
	\frac{\partial \vec{B}}{\partial t} + \nabla \left(\vec{u}\vec{B} - \vec{B}\vec{u}\right) = 0,
\end{equation}
where \(\vec{B}\) is the magnetic field.

The energy densities of Alfvén waves traveling along the magnetic field in parallel and antiparallel directions, denoted by \(w^+\) and \(w^-\) respectively, are governed by

\begin{equation}
	\label{eq:alfven_waves}
	\frac{\partial w^\pm}{\partial t} + \nabla \left((\vec{u} \pm \vec{v}_\Alfven ) w^\pm \right) + \frac{w^\pm}{2}\left(\nabla \cdot \vec{u}\right) 
	= \mp R \sqrt{w^{-} w^+} - Q^\pm_{\text{w}}
\end{equation}
where \(\vec{v}_\Alfven= \vec{B}/\sqrt{\mu_0 \rho}\) is the Alfvén velocity, \(\mp R \sqrt{w^{-} w^+}\) are reflection rates transferring energy between \(w^+\) and \(w^-\), and the dissipation is given by \(Q^\pm_{\text{w}}\). 

The momentum conservation equation is
\begin{multline}
    \label{eq:momentum_conservation}
	\frac{\partial \left(\rho \vec{u}\right)}{\partial t} 
	+ \nabla \left( \rho \vec{u}\vec{u} - \frac{\vec{B}\vec{B}}{\mu_0}
    + p_\text{i} + p_\text{e} 
    + \frac{B^2}{2\mu_0} + p_\Alfven \right)
	= -\rho \frac{GM \vec{r}}{r^3}
\end{multline}
where \(p=p_\text{i} + p_\text{e}\) and \(p_\Alfven = \left(w^+ + w^-\right)/2\) are the sum of the ion and electron thermal pressure and the Alfvén wave pressures respectively. The constants \(G\) and \(\mu_0\) are the gravitational constant and the vacuum permeability, \(M\) is the stellar mass, and \(\vec{r}\) is the position relative to the stellar centre.

Finally there are separate energy equations for ions and electrons (here expressed in terms of thermal pressures). The ion energy equation is 
\begin{equation}
    \label{eq:pressure_ion}
    \frac{\dfrac{\partial p_\text{i}}{\partial t} + \nabla \left(p_\text{i}\vec{u}\right) }
    {\left(\gamma -1\right) }
    + p_\text{i}\nabla\vec{u} =
     \frac{p_\text{e}-p_\text{i}}{\tau_\text{eq}} 
    + f_\text{i} Q_\text{w},
    - \rho \frac{GM \vec{r} \cdot \vec{u} }{r^3},
\end{equation}
and the electron pressure equation is
\begin{equation}
    \label{eq:pressure_electron}
    \frac{\dfrac{\partial p_\text{e}}{\partial t} + \nabla \left(p_\text{e}\vec{u}\right)}{\gamma-1} 
    + p_\text{e}\nabla\vec{u}
    = 
    - \frac{p_\text{e}-p_\text{i}}{\tau_\text{eq}} 
    + (1-f_\text{i})Q_\text{w}
    -Q_\text{rad}
    -\nabla\vec{q}_\text{e},
\end{equation}
where \(\gamma=5/3\) is the ratio of specific heats. 
The equation of state is \(p = N  k_\textsc{b} T\)  where \(N = N_\text{i} + N_\text{e}\) is the sum of the ion and electron number densities (quasi-neutrality gives \(N_\text{i} = N_\text{e}\)) and 
\(T_\text{i}\) and \(T_\text{e}\) are the ion and electron temperatures.
The right hand side terms in the energy equations are 
collisional energy transfer \(\left(p_\text{i}-p_\text{e}\right)/\tau_\text{eq}\) between ions and electrons;
heating from Alfvén wave dissipation \(Q_\text{w} = Q^+_\text{w} + Q^-_\text{w}\) and the fraction of ion heating \(f_\text{i}\); 
work \({\left(\rho GM /r^3 \right)\vec{r} \cdot \vec{u}}\) against the star's gravitational potential;
energy \(Q_\text{rad}\) lost from the system as radiation from an optically thin plasma;
and
electron heat conduction \(\nabla\vec{q}_\text{e}\).
\end{subequations} 

\subsubsection{Ion-electron heat exchange}
The exchange of heat in equation~\eqref{eq:pressure_ion} from ions to electrons and vice versa in equation~\eqref{eq:pressure_electron}, \(\pm\left(p_\text{i}-p_\text{e}\right)/\tau_\text{eq}\), is inversely proportional to the temperature equilibrium time scale~\citep{1981phki.book.....L}; following eq.~11.46 in \citet{goldston1995introduction} the expression for \(\tau_\text{eq}\) in a pure hydrogen plasma is
\begin{equation}
\tau_\text{eq} = \frac{3\sqrt{2} \pi^{3/2}  \epsilon_0^2 m_\text{i} \left(k_\textsc{b}T_\text{e}\right)^{3/2}}{\sqrt{m_\text{e}} e^4 N \ln \Lambda}
\end{equation}
where \(N\) is the number density, \(\epsilon_0\) is electric constant, \(e\) is the elementary charge, \(m_\text{i}\) and \(m_\text{e}\) is the ion and electron mass, and \(k_\textsc{b}\) is Boltzmann's constant. 
The quantity \(\ln \Lambda\) is the Coulomb logarithm \citep[e.g.\@ eq.\@ 11.12--11.17 in][]{goldston1995introduction},
\begin{equation}
\Lambda
 = 
12\pi n \lambda_\textsc{d}^3, 
\quad 
\lambda_\textsc{d}
 =
\sqrt{{\epsilon_0 k_\textsc{b} T_\text{e}}\left/{N_\text{e}e^2}\right.} .
\end{equation}
The constant value \(\ln \Lambda=20\), which corresponds to coronal conditions, is used throughout both the \batsrus{} and \awsom{} models.

\subsubsection{Alfvén wave reflection and dissipation}
The rate of Alfvén wave reflection in equation~\eqref{eq:alfven_waves}, i.e.\ energy exchange between \(w^+\) and \(w^-\) is 
\begin{subequations} 
\begin{equation}
	R = \min\left(R_\text{imb},\max\left(\Gamma^\pm\right)\right) \cdot
	\begin{cases}
	1-\Omega\sqrt{\frac{w^-}{w^+}},\quad \Omega^2 w^- \leq w^+ \\
	\Omega\sqrt{\frac{w^+}{w^-}}-1,\quad w^- \geq \Omega^2 w^+ \\
	0, \quad \text{otherwise} \\
	\end{cases}
\end{equation}
where the reflection rate in strongly imbalanced turbulence,
\begin{equation}
	R_\text{imb} = 
		\sqrt{
			\left(\vec{B}\cdot\left(\nabla \times \vec{u}\right)\right)^2/\vec{B}^2
			+ \left(\left(\vec v_\Alfven \cdot \nabla \right) \log {|\vec v_\Alfven|}\right)^2 
		}.
\end{equation}
and \(\Omega\) is a threshold of wave imbalance; there is no reflection when \(\Omega^{-2} < w^+ / w^- < \Omega^2\).
Additionally, the reflection rate \(R\) is bounded above by the largest of the dissipation rates \(\Gamma^\pm\).

The Alfvén wave dissipation term~\citep{2014ApJ...782...81V} is \(Q^\pm_\text{w} = \Gamma^\pm w^\pm\) where 
\(\Gamma^\pm= \left(2/{L_\perp} \right)\sqrt{{w^\mp}/{\rho}}\) and, as in~\citet{JGRA:JGRA7915}, \(L_\perp \propto 1 \left/ \sqrt{B} \right.\). This gives rise to a proportionality constant \(L_\perp \sqrt{B}\); it is a free parameter of the \awsom{} model. We use the value \(L_\perp \sqrt{B}=\SI{1.5e5}{\meter\tesla^{1/2}}\) as in \citet{2018LRSP...15....4G}.

The fraction of Alfvén wave energy that is apportioned to ion heating \(f_\text{i}\) in equation~\eqref{eq:pressure_ion} and equation~\eqref{eq:pressure_electron} can be set from kinetic considerations~\citep{2011ApJ...743..197C,2014ApJ...782...81V}; 
we follow~\citet{2018LRSP...15....4G} and use the value \(f_\text{i} = 0.6\) throughout the domain. 

\end{subequations}

\subsubsection{Radiative loss}
The optically thin radiative cooling is given by 
\(Q_\text{rad} = N_\text{i}N_\text{e}\Lambda\left(T_\text{e}\right)\) where the rate of cooling curve \(\Lambda(T_\text{e})\) is calculated using the CHIANTI database~\citep{2013ApJ...763...86L}. In the calculations of radiative losses we have used Solar coronal elemental abundances, even though the Hyades cluster has a somewhat higher average metallicity of \([\text{Fe}/\text{H}]=0.13\)~\citep{2013AJ....146..143M}; we expect the impact of using Solar abundances to be very minor.
(The effects of the increased metallicity is taken into account when creating the magnetic maps in Fig.~\ref{fig:magnetograms-zdi}, see~\citetalias{2018MNRAS.474.4956F}).

\subsubsection{Electron heat conduction}
The electron heat conduction term \(\nabla \vec{q}_\text{e}\) term in equation~\eqref{eq:pressure_electron} comprises two terms \(\vec{q}_\text{e}=\vec{q}_\text{e}^\textsc{h} + \vec{q}_\text{e}^\textsc{s}\) where the 
\citet{1978RvGSP..16..689H} collisionless heat flux \(\vec{q}_\text{e}^\textsc{h}\) is analogous to Jeans loss; a portion of the highest-energy electrons escape the star's gravitational field. In our model \(\vec{q}_\text{e}^\textsc{h}\) is proportional to the flow velocity \(\vec{u}\) and the electron pressure,
\begin{equation}
    \vec{q}_\text{e}^\textsc{h} = \frac{3}{2}\alpha n_\text{e}k_\textsc{b}T_\text{e} \vec{u} = \frac{3}{2}\alpha p_\text{e} \vec{u} , \quad \alpha=1.05.
\end{equation}

The~\citet{1953PhRv...89..977S} heat flux \(\vec{q}_\text{e}^\textsc{s}\) represents heat diffusion parallel to the magnetic field lines. It is used on the form given in~\citet{2014ApJ...782...81V}
\begin{align}
\vec{q}_\text{e}^\textsc{s} = -\kappa_\text{e}T^{5/2}_\text{e} (\vec{B}\vec{B}/B^2)
\nabla T_\text{e}, 
&& \kappa_\text{e}\approx \SI{9.2e-12}{\watt\per\meter\per\kelvin^{7/2}}.
\end{align}
A microphysics version of \(\vec{q}_\text{e}^\textsc{s}\) is given in~\citet{2018LRSP...15....4G} which shows that \(\kappa_\text{e}\propto\Lambda\), where \(\Lambda\) is the Coulomb logarithm. 

The two described heating terms are smoothly joined by the scaling function \(f_\textsc{s}\), which represents the fraction of Spitzer heat flux,
\begin{equation}
\vec{q}_\text{e} = f_\textsc{s}\, \vec{q}_\text{e}^\textsc{s} + \left(1-f_\textsc{s}\right) \vec{q}_\text{e}^\textsc{h}, 
\quad 
f_\textsc{s} = 1\left/\left({1+\left(r/R_\textsc{h}\right)^2}\right)\right. 
\end{equation}
so that the Spitzer heat flux \(\vec{q}_\text{e}^\textsc{s}\) dominates at close distances, while the Hollweg heat flux \(\vec{q}_\text{e}^\textsc{h}\) dominates when \(r\gg R_\textsc{h}\). We follow~\citet{2014ApJ...782...81V} by setting \(R_\textsc{h}=5R_\Star\); this value is used in the \swmf{} and \awsom{} Solar models. 

\subsection{Numerical model and boundary conditions}\label{sec:numerical_model}

For each model, the simulation domain consists of two partially overlapping three-dimensional regions: an inner region using a spherical grid, and an outer region using a rectilinear grid. 
The solution in the outer region is driven by the inner solution; in a spherical shell from \(40\) to \(45\) stellar radii the inner region solution is copied to the outer region.

\begin{figure*}
    \centering
        \includegraphics{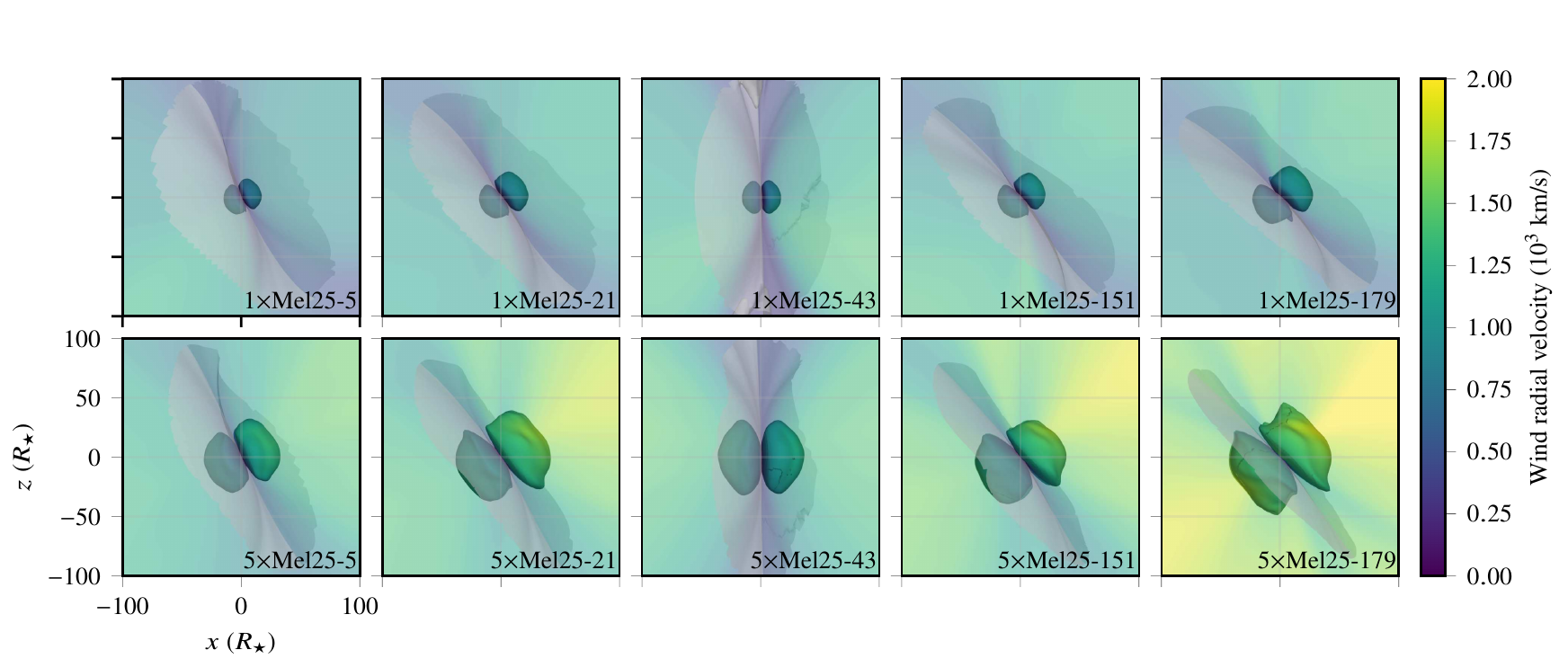}
        \caption{
        Alfvén surface and current sheet for the 
        Hyades stars. The top row corresponds to the unscaled \(B_\ZDI\) models, while the bottom row corresponds the \(5B_\ZDI\) models, where the magnetogram has been scaled by a factor of \(5\). 
        The \(z\) axis coincides with the stellar axis of rotation, and we show the inner current sheet edge-on to emphasise the two-lobed structure of the Alfvén surface. The Alfvén surface and the plane of sky is coloured according to the wind radial velocity \(u_r\). The current sheet is truncated at \(100 R_\Star\).
        The scaled magnetograms exhibit larger, more asymmetrical Alfvén surfaces as well as higher radial velocities.
        }\label{fig:mel-ur-alfven}
    \end{figure*}

\begin{figure}
        \centering
        \includegraphics{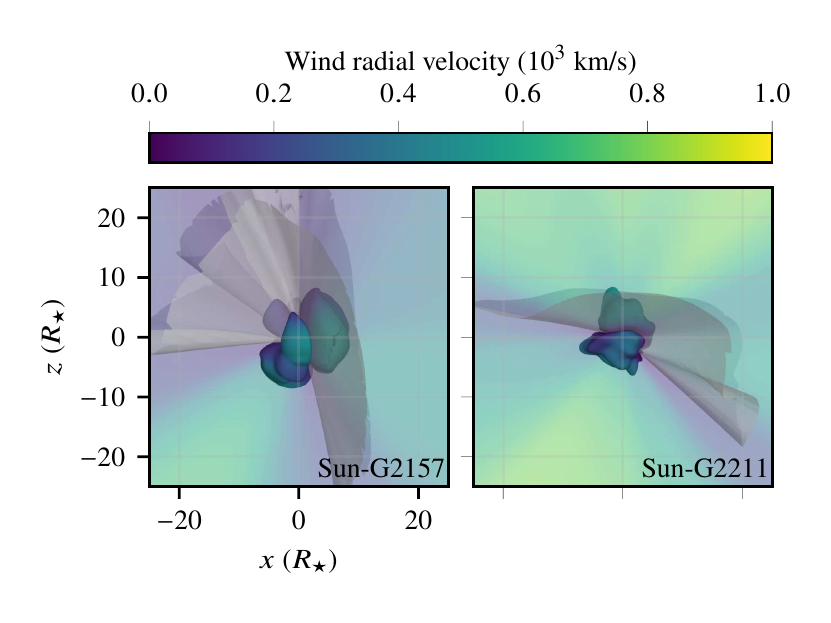}
        \caption{
        Alfvén surface and current sheet for the Solar maximum (left) and
        Solar minimum (right). The features and colouring are the same as in Fig.~\ref{fig:mel-ur-alfven}.
        Compared to the Hyades stars the Solar cases exhibit more asymmetry in the Alfvén surface as well as more pronounced kinks in the current sheet (truncated at  \(25 R_\Star\)).
        The Solar maximum case exhibits a larger Alfvén surface with a completely different orientation compared to the Solar minimum case.
        }\label{fig:sun-ur-alfven}
    \end{figure}

We use a combination of geometric mesh refinement near the stellar surface, and automatic mesh refinement around regions where \(B_r\) changes sign. These refinements permits us to obtain greater detail in the transition region and in the farther away regions where \(B_r\) changes sign; in these regions the character of the solution often changes rapidly. The region where \(B_r\) changes sign typically forms a sheet, see Fig.~\ref{fig:mel-ur-alfven} and Fig.~\ref{fig:sun-ur-alfven}.

The fully three-dimensional simulation is stepped forward in time until a 
steady or quasi-steady
state is reached, 
in which the magnetic and hydrodynamic forces balance throughout the domain of the simulation.

A key feature of the \awsom{} model is resolving the transition region, the narrow region where the plasma heats up to coronal temperatures. This is accomplished by 
\begin{enumerate}
\item making the mesh irregular in \(\Delta r\) as in~\citet{2013ApJ...778..176O}, and very fine in the transition region, and
\item applying a transformation \citep{2013ApJ...764...23S} that numerically broadens the transition region so that it covers over multiple grid cells in the radial direction. 
\end{enumerate}
The combination of these two techniques result in a physically realistic energy profile for the wind plasma in the Solar case.

In an attempt to control for the uncertainty associated with the surface magnetic field strength measured by ZDI, we conduct two series of simulations which we denote \(B_\ZDI\) and \(5B_\ZDI\). Both \(B_\ZDI\) and \(5B_\ZDI\) comprise one wind model of each star in Table~\ref{tab:observed_quantities}, driven by the magnetic maps in Fig.~\ref{fig:magnetograms-zdi}. 
All the model inputs and parameters are identical between the \(B_\ZDI\) series and the \(5B_\ZDI\) series, except that the surface radial field is scaled by a factor of 5 in the \(5B_\ZDI\) series of models. By considering these two series of models we can study the influence of the magnetic field strength on individual stars making comparisons between the \(B_\ZDI\) series and the \(5B_\ZDI\), and we can also combine the two series into a single pooled series to study the field strengths influence on a population of stars.

When discussing the wind models of individual stars, we denote the simulations by the case names Mel25-\(n\) and \(5\times\)Mel25-\(n\) for \(B_\ZDI\) and \(5B_\ZDI\) respectively. These names are used to identify each model result throughout this paper. The full names of the stars are found in Table~\ref{tab:observed_quantities}. The wind model \(5\times\)Mel25-43, for example, refers to the model in which the magnetogram of the star Cl Melotte 25 43 is scaled by a factor of \(5\), and is a model in the \(5B_\ZDI\) series. 

In addition to the stellar cases we conduct two Solar simulations corresponding to Solar maximum and Solar minimum, driven by the magnetic maps in Fig.~\ref{fig:magnetograms-sun}. The Solar cases are designated by their Carrington rotation number as Sun-G2157 (Solar maximum) and Sun-G2211 (Solar minimum). As the Solar magnetograms are not obtained through ZDI, they do not have the same associated uncertainty, and therefore we do not apply any scaling to them.

For the Solar wind models, we use the Solar values of 
mass, \(M_\Sun=\SI{1.99e30}{\kilogram}\), 
radius, \(R_\Sun=\SI{6.96e8}{\meter}\) and 
period of rotation, \(P_\text{rot}=\SI{25.4}{\day}\).
For each stellar wind model, these values are taken from Table~\ref{tab:observed_quantities}. 

At the inner domain boundary (i.e.\ in the chromosphere), the temperature and number density is set to the constant values \(T=\SI{5e4}{\kelvin}\) and \(n=\SI{2e17}{\per\cubic\meter}\). Here we have followed the example of e.g.~\citet{2016A&A...588A..28A} and used Solar values for these parameters as the stars we model are in the unsaturated X-ray regime 
 \citep[see][]{2014ApJ...794..144R,2015A&A...578A.129J} and should have coronae governed by similar physical conditions as the Sun.
 
 The outgoing Alfvén wave energy density at the inner boundary, \(w^\pm\) in equation~\eqref{eq:alfven_waves}, is set to 
 \(w = (\Pi_\Alfven/B)_\Sun\sqrt{\mu_0 \rho} \)~\citep{2014ApJ...782...81V}; %
 where \((\Pi_\Alfven/B)_\Sun\) is a free parameter (recall that \(\Pi_\Alfven \propto B\) in the \awsom{} model). We use the calibrated Solar value \((\Pi_\Alfven/B)_\Sun=\SI{1.1e6}{\watt\per\square\meter\per\tesla}\)~\citep{2018LRSP...15....4G} for the Hyades stars as well as for the Solar cases. The use of the Solar value for the stellar magnetograms is discussed in Section~\ref{sect:effect-of-zdi-limitations}.
  
In both the stellar and Solar cases, the radial component of the boundary magnetic field is fixed to the local magnetogram value in Fig.~\ref{fig:magnetograms-zdi} and Fig.~\ref{fig:magnetograms-sun}, i.e.\ \(\vec{B}_\ZDI\cdot \uvec{r}\) or \(5\vec{B}_\ZDI\cdot \uvec{r}\), depending on the model series. The non-radial components of the magnetic field are free to vary as the MHD solution evolves towards a steady state. The non-radial magnetic field components at the inner boundary are thus determined by the physics of the numerical model, rather than by phenomena occurring in deeper stellar layers.


\section{Results}\label{sec:results}
In this part we present the wind models resulting 
from our simulations (Section~\ref{sect:model-plots}), followed by a set of parameters calculated 
from the models (Section~\ref{sec:model-derived}) 
which we will proceed to discuss in Section~\ref{sec:Discussion}.

\subsection{Overview of results and plots}\label{sect:model-plots}
This section gives an overview of the main features of the 3D wind models, focusing on 
the coronal magnetic field, 
the wind speed and the Alfvén surface, 
and 
the wind pressure in the equatorial plane.

\subsubsection{Description of the wind speed and Alfvén surface}\label{eq:alfven_surface}
As the wind flow accelerates with increasing stellar distance, the wind speed eventually exceeds the local wave speed \(v_\Alfven = B / \sqrt{\mu_0 \rho}\), and wind disturbances can no longer propagate upstream towards the star. The Alfvén surface \(S_\Alfven\) is the surface where this transition first occurs.
The Alfvén surface 
\cite[or the one-dimensional equivalent average Alfvén radius \(R_\Alfven\), see e.g.\@][]{1967ApJ...148..217W} 
is a key quantity when calculating stellar angular momentum losses and spin-down and we use \(S_\Alfven\) for this purpose in Section~\ref{sec:angmom_loss}. 
In regions of superalfvénic (\(u > v_\Alfven\)) wind, shocks and discontinuities may arise in the solution, typically where a region of fast wind meets a region of slower wind or a planetary magnetosphere. 

Fig.~\ref{fig:mel-ur-alfven} and Fig.~\ref{fig:sun-ur-alfven} show the shape of the Alfvén surface and the speed of the stellar wind for the stellar (Hyades) models and the Solar models. 
To emphasise the two lobed structure of the Alfvén surface we have rotated the coordinate system around the \(\vec z\) axis (which coincides with the stellar axis of rotation) to give a side-on view of the Alfvén surface. 
In the case of the Hyades models the Alfvén structure exhibits a two-lobed structure typical of a dipolar magnetic field. 
Comparing the top row and bottom row of Fig.~\ref{fig:mel-ur-alfven}, the average Alfvén radius is about twice as large for the models in the \(5B_\ZDI\) series. The Alfvén surface and the \(xz\)-plane are coloured according to the local wind radial velocity.  The wind velocities are higher for the \(5B_\ZDI\) series, reaching values up to 
\SI{2e3}{\kilo\meter\per\second}. 

We also plot the current sheet \citep{1971CosEl...2..232S}, here characterised by \(B_r=0\). In all the stellar cases the inner current sheet, shown as a grey translucent surface, is flat (this is expected until the stellar rotation shapes the outer current sheet into a spiralling structure) and a region of lower wind velocity surrounds the current sheet. The cases Mel25-5 and Mel25-43 have very smooth Alfvén surfaces, while Mel25-21, Mel25-151 and in particular Mel25-179 have more irregular shapes. 
We also note that the amplified \(5B_\ZDI\) surface magnetic fields result in overall more irregular Alfvén surfaces. 

Mel25-43 distinguishes itself by having the Alfvén surface lobes inclined by very nearly \SI{90}{\degree}. 
    The dipole component of the corresponding magnetogram in Fig.~\ref{fig:magnetograms-zdi} has a near \SI{90}{\degree} inclination. The dipole component of the magnetic field tends to determine the inclination of the Alfvén surface lobes for the Hyades stars where the effective \(\ell_\text{max}\sim 4\).

\begin{figure*}
    \centering
    \includegraphics{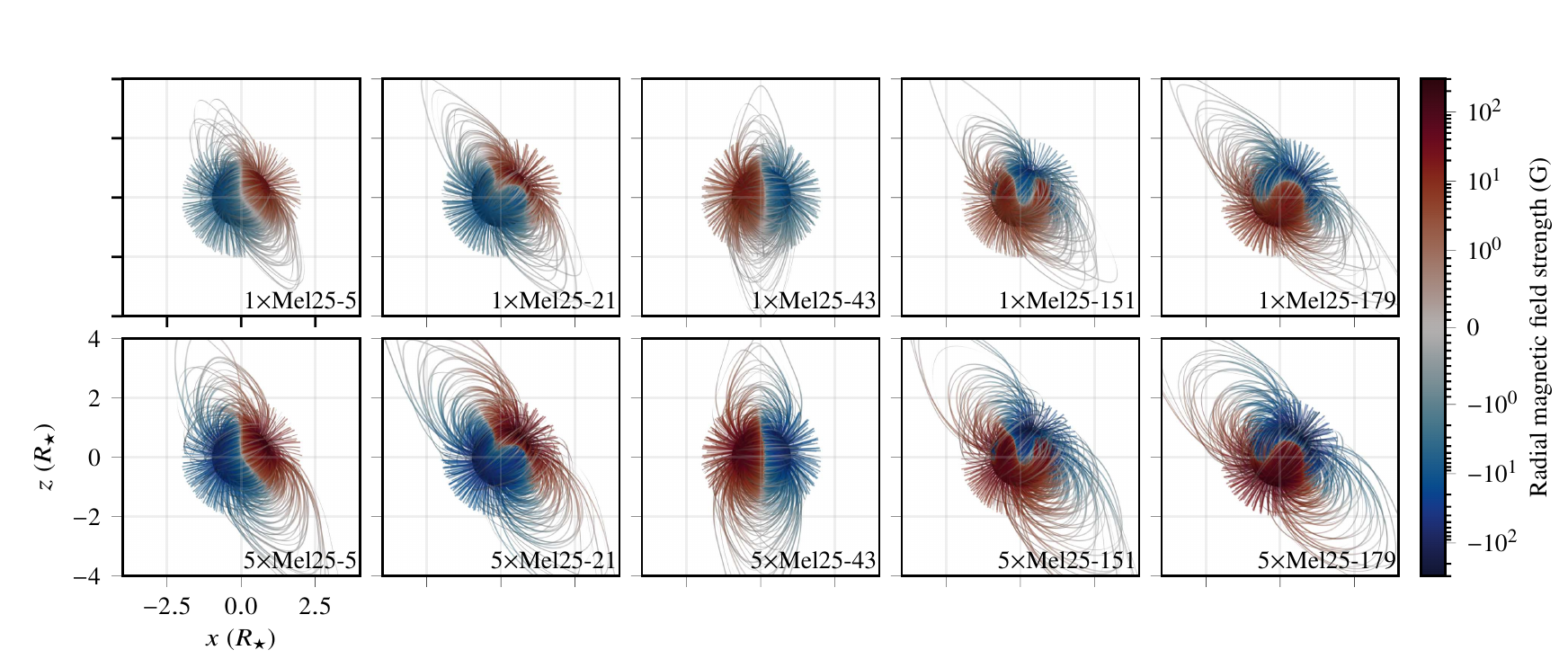}
    \caption{
        Dipole-dominated field line structures for the Hyades stars. 
        This plot shows the open and closed magnetic field lines for the models in Fig.~\ref{fig:mel-ur-alfven}. 
        The top row corresponds to the unscaled \(B_\ZDI\) models, while the bottom row corresponds the \(5B_\ZDI\) models, where the magnetogram has been scaled by a factor of \(5\).
        The open field lines are truncated at two stellar radii. 
        The stellar surface and the field lines are coloured according to the local radial magnetic field strength. Note that the colour scale is linear from \SIrange{-1}{1}{\gauss} and logarithmic outside of this range.
        With the exception of the dipole field strength and inclination, the structure of open and closed field lines appears largely independent of the surface magnetic field medium-scale structure. The \(5B_\ZDI\) series (bottom row) exhibit smaller regions of open field lines, and larger closed magnetic field line loops, see also Table~\ref{tab:main-table}.
    }\label{fig:mel-br-fieldlines}
\end{figure*}
\begin{figure}
    \centering
    \includegraphics{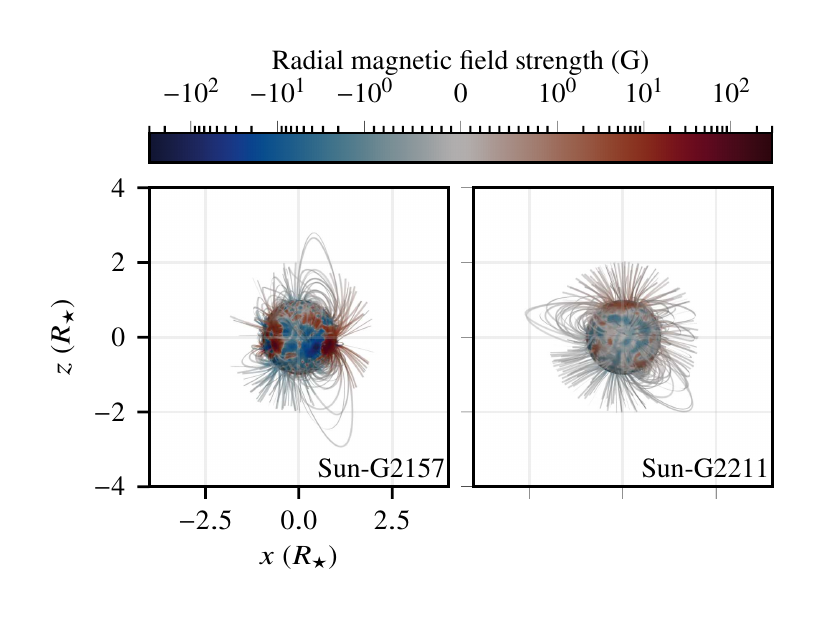}
    \caption{
        Complex field line geometry for the Solar models.
        This plot shows the open and closed magnetic field lines at Solar maximum (left) and Solar minimum (right). The features and colouring are the same as in Fig.~\ref{fig:mel-br-fieldlines}.
        The Solar models have a more complex structure of open and closed field lines, both at Solar maximum and at Solar minimum.  
}\label{fig:sun-br-fieldlines}
\end{figure}

In comparison to the Hyades stars, the Solar models in Fig.~\ref{fig:sun-ur-alfven} have significantly more irregular Alfvén surfaces and inner current sheet structures. The Alfvén surfaces also lie significantly closer to the star, about half the distance of the \(B_\ZDI\) series. 
The Solar maximum case (Sun-G2157) is the only case in this dataset that shows a four-lobed structure (one very small lobe is obscured in Fig.~\ref{fig:sun-ur-alfven}) and a nearly perpendicular fold in the inner current sheet. The Solar minimum (Sun-G2211) case has a two-lobed structure, but the Alfvén surface still has a more irregular structure than the stellar cases. 

The differences between Fig.~\ref{fig:mel-ur-alfven} and Fig.~\ref{fig:sun-ur-alfven} are to be expected; they mirror the difference in complexity between the ZDI reconstructed stellar magnetograms and the Solar magnetograms in Fig.~\ref{fig:magnetograms-zdi} and Fig.~\ref{fig:magnetograms-sun}.

\subsubsection{Description of magnetic field}

\begin{table}
    \centering
    \caption{
        Aggregate surface magnetic field values. The \(|B_r|\) and \(\max |B_r|\) value are the surface average and maximum absolute radial field values from the radial magnetic fields driving the wind models, and shown in Fig.~\ref{fig:magnetograms-zdi} and Fig.~\ref{fig:magnetograms-sun}. \(|\vec B|\) is the average surface field strength of the solution after the non-radial magnetic components have stabilised.
        `Dip.' `Quad.' and `8+' refer to the final fraction of magnetic energy in dipolar, quadrupolar and higher modes. Note that the Solar solution has most energy in higher modes, while the stellar solutions have most energy in the dipolar and quadrupolar modes. Some of the uncertainties associated with ZDI are discussed in Sections~\ref{sec:ZDI-uncertainties}--\ref{sec:ZDI-missing-field}. 
        While \(B_r\) is a model boundary condition, the tabulated \(|\vec B|\) values and percentages are not expected to have a direct correspondence with the photospheric values in \citetalias{2018MNRAS.474.4956F}.
    }
    \sisetup{
        table-figures-decimal=1,
             table-figures-integer=2,
             table-number-alignment=center,
             round-mode=places,
             round-precision=1}
    \begin{tabular}{lSSSSSS}
        \toprule
            Case & {\(|B_r|\)} & {\(\max |{B_r}|\)}  & {\(|\vec B|\)} & {Dip.} & {Quad.} & {8+} \\
             & {\((\si{\gauss}) \)}
             & {\((\si{\gauss}) \)}
             & {\((\si{\gauss}) \)}
             & {\((\si{\percent}) \)}
             & {\((\si{\percent}) \)}
             & {\((\si{\percent}) \)} \\
        \midrule
        Mel25-5                  & 5.9882           & 23.4859          & 8.88             & 61.04            & 25.59            & 13.36           \\
        Mel25-21                 & 9.9963           & 40.7272          & 14.32            & 69.71            & 16.36            & 13.93           \\
        Mel25-43                 & 5.8210           & 18.0203          & 8.75             & 70.98            & 21.71            & 7.30            \\
        Mel25-151                & 10.8717          & 54.8708          & 16.55            & 39.59            & 25.68            & 34.73           \\
        Mel25-179                & 17.0796          & 58.8239          & 24.13            & 70.90            & 17.12            & 11.98           \\
        \midrule
        $5\times$Mel25-5         & 29.8616          & 117.2837         & 42.53            & 61.47            & 25.09            & 13.43           \\
        $5\times$Mel25-21        & 49.8997          & 203.4868         & 70.20            & 69.71            & 16.18            & 14.11           \\
        $5\times$Mel25-43        & 29.0291          & 89.9841          & 40.80            & 69.00            & 25.25            & 5.75            \\
        $5\times$Mel25-151       & 54.3017          & 274.2335         & 80.56            & 39.32            & 25.64            & 35.04           \\
        $5\times$Mel25-179       & 85.3192          & 293.9951         & 119.49           & 71.10            & 16.99            & 11.92           \\
        \midrule
        Sun-G2157                & 6.6135           & 156.0516         & 12.31            & 0.33             & 1.94             & 97.73           \\
        Sun-G2211                & 0.7250           & 3.9404           & 2.28             & 8.68             & 7.44             & 83.87           \\
        \bottomrule 
    \end{tabular}\label{tab:magnetic_averages}
\end{table}

\begin{figure*}
    \centering
    \includegraphics[width=\textwidth]{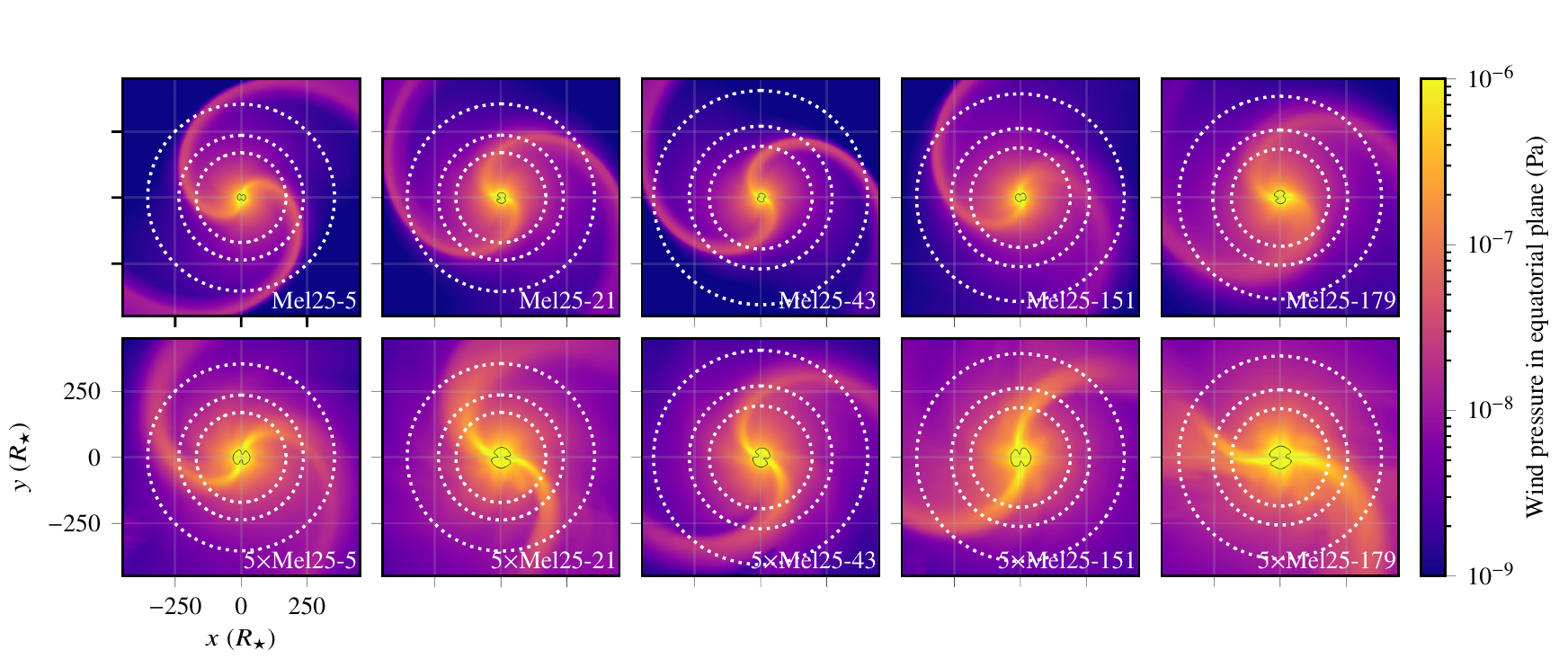}
    \caption{Wind pressure in the star's equatorial plane from the corona to the inner astrosphere. The dotted circles correspond to the current day average distances of Venus, Earth, and Mars. 
    Each model exhibits a two-armed spiral structure which reflects the two-lobed structure of the Alfvén surface in Fig.~\ref{fig:mel-ur-alfven}. The black line shows the intersection of the Alfvén surface and the equatorial plane. 
        The overdense regions arise when wind streams of different velocities interact and are called corotating interacting regions (CIRs).
    The top row corresponds to the unscaled \(B_\ZDI\) models, while the bottom row corresponds the \(5B_\ZDI\) models, where the magnetogram has been scaled by a factor of \(5\).
    We observe that the lower row of models exhibit generally higher wind pressure values.  
        In the \(5B_\ZDI\) series
    the spiral structures also have more of a bar-like appearance and are not as tightly wound. 
}\label{fig:Mel-Pwind-IH-equatorial}
\end{figure*}
\begin{figure}
    \centering
    \includegraphics{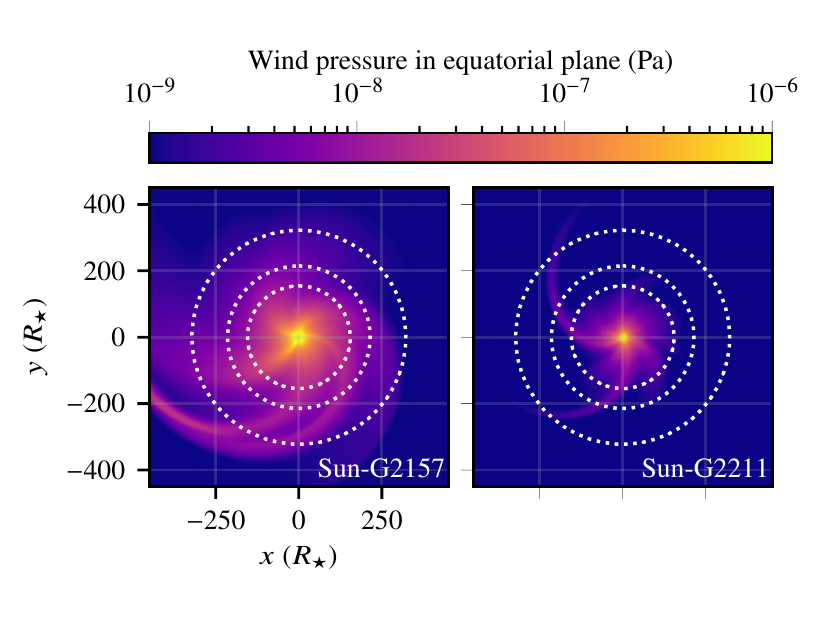}
    \caption{
        Wind pressure in the Sun's equatorial plane from the corona to the inner astrosphere at Solar maximum (left) and Solar minimum (right). The features and colouring are the same as in Fig.~\ref{fig:Mel-Pwind-IH-equatorial}. At Solar maximum a coalescing `interaction region' of two overdense CIRs is seen around \((x, y) = (-280, -230)\). In contrast to the dipole-dominated fields of the Hyades stars where only two spiral arms are generated (in Fig.~\ref{fig:Mel-Pwind-IH-equatorial}) the complex Solar magnetic fields create multiple spiral arms.
}\label{fig:Sun-Pwind-IH-equatorial}
\end{figure}

Fig.~\ref{fig:mel-br-fieldlines} and Fig.~\ref{fig:sun-br-fieldlines} show the structure of the magnetic field lines for the stellar cases and Solar cases respectively. To avoid confusing the plots, open magnetic field lines are truncated at two stellar radii, while closed field lines are not truncated. The stellar surfaces are coloured according to their radial magnetic field strength, as are the magnetic field lines themselves. To capture both the strong and weak magnetic field strength in a single plot, we employ a colour scale where the range from \SIrange{-10}{10}{\gauss} uses a linear scale, while a logarithmic scale is used outside of this range. This is also indicated by the position of the minor tick marks in the colour scales of Fig.~\ref{fig:mel-br-fieldlines} and Fig.~\ref{fig:sun-br-fieldlines}. The orientation of the plots are the same as in Fig.~\ref{fig:mel-ur-alfven} and Fig.~\ref{fig:sun-ur-alfven}. 

Comparing the top and bottom row in Fig.~\ref{fig:mel-br-fieldlines}, the plots show that the amplified \(5B_\ZDI\) models have a larger region of closed field lines. The magnetic fields only resemble dipolar fields in a superficial way; the closed field lines are stretched into typical `helmet' streamer shapes. This is a typical features of MHD wind models, as seen in e.g.~\citet{2009ApJ...699..441V}. The surface magnetic field of Mel25-43 appears almost entirely dipolar, while the other cases show kinks in the line of zero radial field strength (zero corresponds to grey on the colour scale). These features do not, however, appear to affect the structure of open and closed field lines except by a small widening of the closed field region.

In comparison with the Hyades models in Fig.~\ref{fig:mel-br-fieldlines}, the Solar models in Fig.~\ref{fig:sun-br-fieldlines} exhibit much more complicated coronal magnetic fields characterised by multiple regions of closed field lines, and many field lines closing to nearby regions of opposite polarity within a very short distance of the Solar surface. 
The large difference in field strength between Solar maximum (Sun-G2157) and Solar minimum (Sun-G2211) is evident when plotted using the same colour scale as in Fig.~\ref{fig:sun-br-fieldlines}.

Table~\ref{tab:magnetic_averages} gives aggregate magnetic quantities of the relaxed surface magnetic field; recall that while the radial magnetic field \(B_r\) is held fixed, the non-radial surface magnetic field \(
    \vec B_\perp = B_\theta \uvec{\theta} + B_\varphi \uvec{\varphi}
\)
is found as part of the model output. The values are calculated by fitting a full set of spherical harmonics coefficients, as in  \citetalias{2018MNRAS.474.4956F}, to the MHD solution and calculating the aggregate quantities from the spherical harmonics coefficients. In terms of these aggregate quantities, we observe that the MHD solution resembles the potential part of the ZDI maps in \citetalias{2018MNRAS.474.4956F}. Even though currents are permitted in the solution, the observed ZDI non-potential magnetic field is not reproduced in these models. 
The non-potential field is not thought to have a significant influence on the wind, see \citet{2013MNRAS.431..528J}. 
Of particular note is the fact that the Solar models have \SI{98}{\percent} (Solar maximum) and \SI{84}{\percent} (Solar minimum) of their magnetic energy in octupolar and higher order terms of the spherical harmonics coefficients, corresponding to surface features smaller than \( \mysim  \SI{60}{\degree}\) on the Solar disc. The Solar magnetic fields are highly complex compared to the stellar models, most of which have less than \SI{15}{\percent} of their energy in octupolar and higher order terms. Mel25-151 occupies a middle ground with \SI{35}{\percent} of its energy in octupolar and higher order terms.

\subsubsection{Description of wind pressure out to 1 au}

Fig.~\ref{fig:Mel-Pwind-IH-equatorial} and Fig.~\ref{fig:Sun-Pwind-IH-equatorial} show the total wind pressure from the stellar corona and into the inner astrosphere past the would-be orbits of Venus, Earth, and Mars. The would-be orbits of the three planets around the stellar models (and Solar models) are indicated as white dotted lines. When visible at this scale, the Alfvén surface is shown as a black curve.

The total wind pressure comprises thermal pressure, magnetic pressure, and ram pressure,
\begin{equation}\label{eq:wind-pressure}
    P_\text{w}= p+ |\vec B|^2 / (2 \mu_0) + \rho |\vec u+\vec v|^2
\end{equation}
\citep[see e.g.\@][]{2011MNRAS.414.1573V}, where \(\vec u\) is the wind velocity and \(\vec v\) is the orbital velocity of the planet. 
Near the star, all the components of equation~\eqref{eq:wind-pressure} may be significant. In our models the magnetic term only contributes \SI{10}{\percent} or more within \SIrange{10}{15}{\rSun}. The \(p\) term is greater than \(\rho |\vec u|^2\) inside \( \mysim  2 \rSun\) but never makes a significant contribution to the total wind pressure. For close-in exoplanets, \(|\vec v|\) may be comparable to \(|\vec u|\), but as orbital speeds fall with distance from the star, the orbital velocity contribution is below \SI{1}{\percent} for Venus-, Earth-, and Mars-like orbits. %
In these outer regions of the model we also find that due to the high radial velocity of the stellar wind, 
\(u_r \sim \SI{e3}{\kilo\meter\per\second}\), 
\(P_\text{w}\) is entirely dominated by \(\rho u_r^2\); both \(p\) and \(|\vec B|^2/(2\mu_0)\) are less than \SI{0.1}{\percent} of the total pressure at this distance.

In the stellar models of Fig.~\ref{fig:Mel-Pwind-IH-equatorial} each model produces a two-armed spiral structure. The spiral shape is a consequence of the stellar rotation (see Table~\ref{tab:observed_quantities} for the stellar rotation periods). We observe that the amplified \(5B_\ZDI\) models (bottom row) give rise to higher values of wind pressure throughout the model. A bar-like structure also appears close to the star; inside the Alfvén radius the magnetic forces dominate over inertial forces and the magnetic field is able to force the wind to co-rotate with the star. This co-rotation radius appears larger for the \(5B_\ZDI\) models; this is expected as a consequence of the larger Alfvén radius. Outside the Alfvén surface (comparing to Fig.~\ref{fig:mel-ur-alfven} and \ref{fig:sun-ur-alfven}) the spiral structures of the \(5B_\ZDI\) series appear less `wound up' as a consequence of the higher wind speeds. 

The arm-like structures are regions
where wind originating in different parts of the corona coalesce. In our Solar system these have been observed by spacecraft \citep{1971JGR....76.3534B} 
and are called corotating interaction regions (CIRs), see e.g. the review by \citet{1996ARA&A..34...35G}. 
The presence of CIRs in the Solar wind are thought to influence the rate of cosmic ray detections on Earth. A model of this may, however, require the inclusion of transient phenomena like flares \citep{1993JGR....98....1B}; we return to this topic in Section~\ref{sec:cosmic-ray-intensity}. In our models the 
CIRs are associated with polarity changes of the magnetic field (not shown).

The Solar models in Fig.~\ref{fig:Sun-Pwind-IH-equatorial} again show more complexity and several more CIRs. The wind pressure at Solar maximum (Sun-G2157) is almost an order of magnitude greater than the wind pressure at Solar minimum (Sun-G2211). For the Solar maximum case we also observe the collision of interaction regions around \((x, y) = (-280, -230)\) in the lower left region of the plot.

The collision of the faster wind and the slower wind leads to the formation of the spiral  CIRs that we observe in Fig.~\ref{fig:Mel-Pwind-IH-equatorial} and Fig.~\ref{fig:Sun-Pwind-IH-equatorial}. In our models, the wind pressure variations are dominated by density variations of a factor of \( \mysim  10\) or more, while the velocity varies by a factor of \( \mysim  2\). 
Tracing the planetary orbits indicated by dotted white lines (also in Fig.~\ref{fig:Mel-Pwind-IH-equatorial} and Fig.~\ref{fig:Sun-Pwind-IH-equatorial}), we observe that the wind pressure variation in a single orbit remains around a base value, but exhibits one or more spikes as a faster, less dense region of wind encounters a slower region of wind. Over one Earth orbit, the average pressures at Solar maximum 
and at Solar minimum 
correspond well to observed values, see Section~\ref{sec:wind-pressure-planets}. 
The region in which the wind pressure rapidly increases while the velocity drops can be quite narrow, leading to a sharp rise in wind pressure. 

\subsection{Quantities derived from the wind maps}\label{sec:model-derived}
Having the full three-dimensional wind solutions make it possible to calculate a large range of wind-related quantities, some of which we present in Table~\ref{tab:main-table}. A brief description of each of the calculated quantities is given in this section.

\subsubsection{Unsigned open magnetic flux}\label{sec:open_magnetic_flux}

The unsigned magnetic flux across a surface \(S\) is given by the expression
\begin{equation}
    \Phi = \oint\nolimits_S |\vec B\cdot \vec n|\, \mathrm{d}S .
\end{equation}
We calculate the unsigned flux \(\Phi_0\) at the stellar surface, where \(\Phi_0 = 4\pi R^2 |B_r|\). 
We also calculate \(\Phi_\text{open}\), the unsigned flux contribution from open field lines. (This parameter is often called the `open magnetic flux'.)
Numerically we accomplish this by picking the minimum value of \(\Phi_\text{open}\) calculated on concentric spherical shells past the Alfvén surface. We typically find this minimum at \(\mysim 2\) times the average Alfvén radius (see Section~\ref{sec:alfven_radius}). Past the last set of closed field lines this value should be constant; we observe this in our simulations within an accuracy of about \SI{1}{\percent}, indicating that the exact radius is somewhat unimportant as long as it is past the average Alfvén radius. In Solar modeling based on a potential magnetic field (PFSS, see Section~\ref{pfss-and-tangential-field}) the radius outside of which all field lines are open is often set to 2.5 by convention \citep[e.g.\@][]{2006ApJ...653.1510R}; in our Solar models this would lead to a modest overestimation of \(\Phi_\text{open}\) by \SIrange[]{10}{20}{\percent}.
The values of \(\Phi_0\) and \(\Phi_\text{open} / \Phi_0\) calculated from our models are reported in Table~\ref{tab:main-table}. 

In contrast to the observation-based estimates of \citet{2019ApJ...883...67F} our values of 
are low by a factor of 2--3. This Solar `open flux problem' is also present in many other PFSS and MHD models; see the discussion in \citet{2017ApJ...848...70L}. A recent validation study by \citet{2019ApJ...887...83S} found that the \awsom{} model can underestimate the local magnetic field strength~\(B\); 
A magnetogram scaling of \SIrange[]{1.5}{3.0}{} is often applied to remedy this problem \citep{2021arXiv210205101R}. 
In this work we use unscaled magnetograms as our work is focussed on the stellar magnetograms and how the stellar wind solutions are affected by magnetogram scaling; the Solar magnetograms are included for comparison with other ideal MHD and \awsom{} simulations. 

The amount of variation between our two Solar models also appears somewhat greater than the observation-based values of~\citet{2019ApJ...883...67F}. The larger amount of variation between Solar minimum and Solar maximum might be explained by our choice of Solar magnetograms (see Section~\ref{sec:Solar-magnetograms}) with the smallest and largest \(|B_r|\) in Solar cycle 24; as they are likely to produce more extreme conditions than `average' Solar minimum and Solar maximum conditions.

\subsubsection{Open surface fraction}
A quantity related to \(\Phi_\text{open} / \Phi_0\) is the open surface fraction \(S_\text{open}/S\), the fraction of the stellar surface where the emerging magnetic field lines extend past the Alfvén surface. 
We calculate this quantity by tracing a large amount of magnetic field lines from the stellar surface until they either loop back onto the stellar surface, or cross the Alfvén surface. To sample points near-uniformly on the stellar surface we use a Fibonacci sphere algorithm 
\citep{2006QJRMS.132.1769S}.
We observe that \(S_\text{open}/S\) is smaller than \(\Phi_\text{open} / \Phi_0\) as the surface regions with strong radial magnetic fields tend to be open, see Fig.~\ref{fig:mel-br-fieldlines} and Fig.~\ref{fig:sun-br-fieldlines}. The field lines used in calculating \(S_\text{open}/S\) are the same as shown in the two figures.

\subsubsection{Current sheet inclination}
The current sheet inclination is a measure of the inclination of the inner current sheet with respect to the stellar axis of rotation.
We calculate the current sheet inclination \(i_{B_r=0}\) by fitting a plane to the inner current sheet, and then computing the angle between the normal vector of the plane and the \(\uvec z\) axis.
$i_{B_r=0}$ is a proxy for the dipole inclination when the Alfvén surface exhibits a two-lobed structure, such as in the stellar cases (Fig.~\ref{fig:mel-ur-alfven}). For more complex magnetic fields (Fig.~\ref{fig:sun-ur-alfven}) the current sheet inclination still gives an indication of the overall orientation of the wind structure and its inclination to the stellar axis of rotation. The inner current sheet and the Alfvén surface can be seen in Fig.~\ref{fig:mel-ur-alfven} and Fig.~\ref{fig:sun-ur-alfven}, where the current sheet is appears as a grey translucent structure.

\subsubsection{Axisymmetric open flux fraction}
The non-axisymmetric open magnetic flux has been linked to the intensity of cosmic rays reaching Earth \citep{2006ApJ...644..638W}. We follow \citet{2014MNRAS.438.1162V}, calculating the axisymmetric open flux
\begin{equation}
\Phi_\text{axi} = \oint\nolimits_S |\vec B_\text{axi} \cdot \vec n| \,\mathrm{d}S, 
\text{ where }
\vec B_\text{axi} = \frac{1}{2\pi}\oint \vec B \,\mathrm{d}\varphi
\end{equation}
over a spherical surface \(S\) past the Alfvén surface, similar to the calculation of the open magnetic flux. In this formulation \(\vec B_\text{axi}\) is the average value of \(\vec B\) for a given radius and polar angle. As with the open magnetic flux, this quantity varies by about \SI{1}{\percent} past the last set of closed field lines.

\begin{table*}
    \centering
    \caption{
        Overview of the wind aggregate quantities considered.
        \(\Phi_0\) is the absolute amount of magnetic flux exiting the stellar surface, 
        \(\Phi_\text{open}/\Phi_0\) is the fraction of \(\Phi_0\) contained in open field lines,
        \(S_\text{open}/S\) is the fraction of the stellar surface area where the field lines are open, 
        \(i_{B_r=0}\) is the angle between the inner current sheet and the stellar axis of rotation; this corresponds to the dipole inclination for a dipolar magnetic field,
        \(\Phi_\text{axi}/\Phi_\text{open}\) is the axisymmetric open flux fraction,
        \(R_\Alfven\) is the distance to the Alfvén surface averaged over the stellar surface,
        \(|\vec r_\Alfven\times \uvec z|\) is the torque arm length at the Alfvén surface averaged over the stellar surface,
        \(\dot M\) is the stellar mass loss, 
        \(\dot J\) is the stellar angular momentum loss, 
        \(P_\Wind^\Earth \) is the average wind pressure for an Earth-like planet at \(\SI{1}{\astronomicalunit}\), and
        \(R_\text{mag}/R_\text{p}\) is the magnetospheric stand-off distance in Earth radii for an Earth-like planet.
    }
    \begin{tabular}{lccccccccccc} 
\toprule
Case 
& \(\Phi_0\)
& \(\Phi_\text{open}\)
& \(S_\text{open}\)
& \(i_{B_r=0}\)
& \(\Phi_\text{axi}\)
& \(R_\Alfven\) 
& \(|\vec r_\Alfven\times \uvec z|\)
& \(\dot M\) 
& \(\dot J\) 
& \(P^\Earth_\Wind\)
& \(R_\text{m}\) 
\\ %
& \(\left(\si{\weber}\right)\) %
& \(\left(\Phi_0\right)\)
& \(\left(S\right)\)
& \(\left(\si{\degree}\right)\)
& \(\left(\Phi_\text{open}\right)\)
& \(\left(R_\Star{}\right)\)
& \(\left(R_\Star{}\right)\)
& \(\left(\si{\kilogram\per\second}\right)\)
& \(\left(\si{\newton\meter}\right)\)
& \(\left(\si{\pascal}\right)\)
& \(\left(R_\text{p}\right)\)
\\ 
\midrule
Mel25-5                  & \num{3.0e+15}    & \num{0.41}       & \num{0.25}       & \num{65.3}       & \num{0.32}       & \num{13.4}       & \num{10.7}       & \num{3.1e+09}    & \num{7.5e+23}    & \num{5.4e-09}    & \num{8.5}       \\
Mel25-21                 & \num{5.0e+15}    & \num{0.36}       & \num{0.23}       & \num{55.4}       & \num{0.43}       & \num{17.4}       & \num{13.7}       & \num{3.6e+09}    & \num{1.5e+24}    & \num{7.5e-09}    & \num{8.0}       \\
Mel25-43                 & \num{2.2e+15}    & \num{0.40}       & \num{0.22}       & \num{88.3}       & \num{0.02}       & \num{14.1}       & \num{11.4}       & \num{2.0e+09}    & \num{3.9e+23}    & \num{3.5e-09}    & \num{9.1}       \\
Mel25-151                & \num{4.4e+15}    & \num{0.31}       & \num{0.19}       & \num{52.0}       & \num{0.47}       & \num{16.8}       & \num{13.3}       & \num{2.5e+09}    & \num{7.2e+23}    & \num{6.3e-09}    & \num{8.2}       \\
Mel25-179                & \num{7.3e+15}    & \num{0.30}       & \num{0.17}       & \num{46.6}       & \num{0.53}       & \num{21.1}       & \num{16.6}       & \num{4.1e+09}    & \num{1.9e+24}    & \num{1.0e-08}    & \num{7.6}       \\
\midrule
$5\times$Mel25-5         & \num{1.5e+16}    & \num{0.26}       & \num{0.14}       & \num{64.1}       & \num{0.32}       & \num{26.7}       & \num{21.2}       & \num{6.0e+09}    & \num{4.8e+24}    & \num{1.5e-08}    & \num{7.1}       \\
$5\times$Mel25-21        & \num{2.5e+16}    & \num{0.24}       & \num{0.14}       & \num{56.2}       & \num{0.41}       & \num{33.4}       & \num{26.5}       & \num{7.2e+09}    & \num{8.4e+24}    & \num{1.9e-08}    & \num{6.9}       \\
$5\times$Mel25-43        & \num{1.1e+16}    & \num{0.27}       & \num{0.15}       & \num{88.7}       & \num{0.02}       & \num{29.8}       & \num{23.9}       & \num{4.5e+09}    & \num{3.3e+24}    & \num{9.6e-09}    & \num{7.7}       \\
$5\times$Mel25-151       & \num{2.2e+16}    & \num{0.21}       & \num{0.12}       & \num{54.2}       & \num{0.42}       & \num{32.2}       & \num{25.7}       & \num{5.4e+09}    & \num{4.4e+24}    & \num{1.6e-08}    & \num{7.1}       \\
$5\times$Mel25-179       & \num{3.6e+16}    & \num{0.19}       & \num{0.11}       & \num{48.1}       & \num{0.50}       & \num{37.5}       & \num{29.6}       & \num{6.6e+09}    & \num{8.2e+24}    & \num{2.5e-08}    & \num{6.5}       \\
\midrule
Sun-G2157                & \num{4.0e+15}    & \num{0.12}       & \num{0.05}       & \num{70.2}       & \num{0.38}       & \num{6.2}        & \num{4.8}        & \num{4.1e+09}    & \num{1.5e+23}    & \num{5.2e-09}    & \num{8.5}       \\
Sun-G2211                & \num{4.4e+14}    & \num{0.38}       & \num{0.19}       & \num{22.0}       & \num{0.83}       & \num{5.1}        & \num{3.9}        & \num{5.5e+08}    & \num{1.3e+22}    & \num{6.8e-10}    & \num{11.9}      \\
\bottomrule
\end{tabular}

\label{tab:main-table} 
\end{table*}

\subsubsection{Alfvén radius}\label{sec:alfven_radius}
The average Alfvén radius \(R_\Alfven\) is the average radial distance to the Alfvén surface; this average is taken over the stellar surface.
At the Alfvén surface the motion of the wind particles can no longer remove angular momentum from the star and vice versa the star can no longer force the wind to co-rotate with the star. The co-rotation radius concept is central to formulations of angular momentum loss such as the one of \citet{1967ApJ...148..217W} and its extensions, including the one we are using to calculate angular momentum loss in this work. The Alfvén surface is shown in Fig.~\ref{fig:mel-ur-alfven} and Fig.~\ref{fig:sun-ur-alfven}.

As only the distance to the Alfvén surface in the \(xy\) plane is relevant when calculating torque around the \(\uvec z\) axis we also calculate the `torque-averaged Alfvén distance' \(|\vec{r}_\Alfven \times \uvec{z}|\); this is the average length of the Alfvén surface's torquing arm around the \(\uvec z\) axis.

\subsubsection{
    Mass loss 
}\label{sec:mass_loss}
We calculate the wind mass loss \(\dot M\) by integrating the mass flux over a closed surface centred on the star,
\begin{equation}
\dot M 
= \oint\nolimits_S \rho \vec u \cdot \vec n \, \mathrm{d}S.
\end{equation}
For a true steady state the value of \(\dot M\) should take the same value for any surface \(S\) as long as the surface encloses the star. If the solution state is only quasi-steady this need not be the case. We observe \(\dot M\) values consistent to an accuracy of \SI{1}{\percent} or better.
The mass loss values are tabulated in Table~\ref{tab:main-table}. 

    While our calculated mass loss values are well matched with other results obtained with the \awsom{} model (see Section~\ref{sec:3d_comparison}), the difference of a factor of \(\mysim10\) between 
    Solar minimum and Solar maximum differs from observational results.  In a study of Solar wind mass loss based on spacecraft observations,  
    \cite{2011MNRAS.417.2592C} found little mass loss variation over the Solar cycle, instead the observed mass loss values was found to exhibited short-term variations of a factor 2--5 around an average value of \SI{1e9}{\kilogram\per\second}. %
    A recent study by \cite{2019MNRAS.486.4671M} found yearly averaged mass loss values in the range \SIrange{1.0e9}{1.5e9}{\kilogram\per\second}. 

    Assuming that the observed short-term variations in wind speed and density are due to spatial variations, there appears to be a larger spread between the \(\dot M\) values at Solar minimum and Solar maximum in our models than what is supported by observations. However, as the Solar magnetograms in this study represent the highest and largest average surface field strength values, a larger variation in \(\dot M\) would be expected than for generic Solar minimum/maximum conditions. This could be explored by creating wind models for a larger sample of Solar magnetograms.

\subsubsection{Angular momentum loss}\label{sec:angmom_loss}
To find the stellar angular momentum loss \(\dot J\), we follow~\citet{1999stma.book.....M} and~\citet{2014MNRAS.438.1162V}. 
In our models the stars rotate around the \(z\) axis, 
so that \(\vec \Omega \parallel \uvec z\). 
We are interested in the angular momentum around the \(z\) axis \(\dot J = \uvec z \cdot \dot{\vec{J}}\).
At the Alfvén surface \(S_\Alfven\) (see Section~\ref{eq:alfven_surface}) the angular momentum loss around the \(z\) axis is given by 
\begin{equation}
\label{eq:angmom_loss}
\dot J 
= \oint\nolimits_{S_\Alfven}
\left(\vec r \times \vec n\right)_3 \left( p+ \frac{B^2}{2 \mu_0 } \right)
+ ({\vec V \cdot \vec n}) \varpi^2 \Omega \rho 
\, {\rm d} S_\Alfven  
\end{equation}
where 
\(
\varpi 
\)
is the projection of \(\vec r\) into the \(xy\)-plane, \(\vec V = \vec u - \vec \Omega  \times \vec r\) is the flow velocity in the rotating coordinate system centred on the star, \(\vec \Omega\) is the stellar rotational angular velocity calculated from rotation periods in Table~\ref{tab:observed_quantities}, and the \(z\) component is denoted by \((\cdot)_3\). The first term under the integral sign is the pressure moment (thermal and magnetic), while \(({\vec V \cdot \vec n}) \varpi^2 \Omega \rho\) is an effective corotation term. 

We note that by retaining only the corotation term we recover the~\citet{2014ApJ...783...55C} equation for angular momentum loss. In our models we find that 
the pressure moment makes a negligible contribution to the total angular momentum at the Alfvén surface, justifying the use of this simplified equation. When the effective corotation term dominates we have \(\dot J\propto \Omega\), so that the Solar models should have smaller values of \(\dot J\) due to the Sun's period of rotation being significantly longer than the stellar periods in Table~\ref{tab:observed_quantities}.

There are two more terms in the general expression for \(\dot J\); these terms vanish when integrating over the Alfvén surface \(S_\Alfven\), 
\begin{equation}
0 
= \oint\nolimits_{S_\Alfven}
(\vec r \times \vec V)_3 ({\vec V \cdot \vec n})\rho
- (\vec r \times \vec B)_3 \left( \frac{\vec B \cdot \vec n}{\mu_0 } \right) 
\, {\rm d} {S_\Alfven},
\end{equation}
but they must be accounted for when integrating over a different surface (such as a sphere). We use these two terms to estimate discretisation errors in the calculation of \(\dot J\) and find that the error is a few percent of \(\dot J\). 

Based on in-situ observations by the \emph{WIND} spacecraft,
\citet{2019ApJ...885L..30F} calculated an average value \(\dot J_\Sun = \SI{3.3e23}{\newton\meter}\) in Solar cycle 23--24, and a spread of  \(\mysim4\)
between Solar maximum (November 2014) and Solar minimum (December 2018) when applying a year-long smoothing window, and more when considering each individual Carrington rotation.
The reported average value is \(\mysim2\) times larger than our \(\dot J\) value at Solar maximum, and \(\mysim 25\) times larger than our Solar minimum value. 

As with the calculated mass loss, we note that our calculated mass loss values are well matched with other results obtained with the \awsom{} model; we return to these issues in Section~\ref{sec:3d_comparison}.

\subsubsection{Wind pressure for planets}\label{sec:wind-pressure-planets}
To calculate the wind pressure at an Earth-like orbit \(P_\Wind^\Earth\) we sample many orbits with Earth-like orbital distance and inclination while varying the longitude of the ascending node. This permits us to sample the full range of positions that the Earth-like planet would occupy. The wind pressure at \SI{1}{\astronomicalunit} is calculated using equation~\eqref{eq:wind-pressure}. The values are tabulated in  Table~\ref{tab:main-table}. 

For the Solar values it is possible to make a comparison with in situ based space measurements.
The average pressures at Solar maximum, 
\(\langle P^\Earth_\Wind\rangle = \SI{5.2}{\nano\pascal}\),
and at Solar minimum, 
\(\langle P^\Earth_\Wind\rangle = \SI{0.68}{\nano\pascal}\),
correspond reasonably well with the flow pressure values provided by the OMNIWeb\footnote{\url{https://omniweb.gsfc.nasa.gov/} (Accessed in January 2021).} service, see \citet{2005JGRA..110.2104K}. Monthly rolling averages of the values in the OMNI dataset range from \(\mysim 0.1\) to \(\mysim \SI{10}{\nano\pascal}\), while hourly variations  go as high as \(\mysim \SI{100}{\nano\pascal}\).
We note that the steady state wind represents only a snapshot in time of the true stellar wind; the true wind is likely to exhibit greater variation than any snapshot, even when sampling across the full range of possible Earth positions.
\subsubsection{Magnetospheric stand-off distance}

A planet's magnetosphere \citep{1959JGR....64.1219G} is the region of space where the Solar wind is disrupted by the planet's presence; for a magnetised planet (i.e.\ an Earth-like planet) the region typically takes on a teardrop or comet-like shape characterised by the formation of a shock on the dayside of the planet and a long tail forming on the planet's nightside. The magnetospheric stand-off distance---the distance from the planet's centre to the dayside shock---gives an indication of the size of the planetary magnetosphere and whether atmospheric erosion is likely to occur. 

A pressure balance argument \citep{1931TeMAE..36...77C, 2009ApJ...703.1734V} lets us estimate the magnetospheric stand-off distance \(R_\text{m}\) (the distance to the sub-Solar magnetopause) if the Earth were to orbit each of our stellar models at \SI{1}{\astronomicalunit}. We let \(R_\text{m}\) be the distance at which the wind pressure \(P_\text{w}\) matches the magnetosphere pressure \(P_\text{p}\) from the planet. For the Earth, the dominant term is the planetary magnetic pressure, \(P_\text{p} = B^2/(2\mu_0)\). Assuming the planetary magnetic field is dipolar, the magnetic field strength scales with height as as \(B/B_0=(R / R_\text{p})^{-3}\). Equating the wind pressure and the planetary pressure gives an expression for the magnetospheric stand-off distance
\begin{equation}\label{eq:magnetospheric_stand_off_distance}
    \frac{R_\text{m}}{R_\text{p}} = \left(\frac{P_\text{p}}{P_\text{w}}\right)^{1/6} = \left(\frac{B_0^2/(2\mu_0)}{\rho u^2}\right)^{1/6}.
\end{equation}
It is evident from the \(1/6\) exponent that the magnetospheric stand-off distance is only weakly affected by wind pressure changes: When the wind pressure varies by an order of magnitude, \(R_\text{m}\) varies only by \(10^{1/6}\approx1.5\). 
Fig.~\ref{fig:orthogonal_trend} shows the resulting magnetospheric stand-off distance for a planet with a current-day Earth-like dipolar magnetic field of \SI{0.7}{\gauss}. 
This value includes a factor of \(2\) in order to account for magnetospheric currents, see e.g.~\citet{1964JGR....69.1181M}.
The value of \(\SI{0.35}{\gauss}\) is calculated using the Earth's magnetic dipole moment and its axial tilt.
For the Solar cases in Table~\ref{tab:main-table} of \(R_\text{m}/R_\text{p}=8.5\) (Solar maximum) and 12 (Solar minimum) we have good general agreement with observations \citep{2010JGRA..115.4207L} and MHD models \citep{2020GeoRL..4786474S}. The full range of instantaneous values shown by the boxplots in Fig.~\ref{fig:orthogonal_trend} of \(R_\text{m}/R_\text{p}=6.9\)--10 (Solar maximum) and 8.6--15 (Solar minimum) may be a bit wide although values as low as 4.6 have been reported during Solar storms \citep{PhysRevLett.117.171101}.

In a study of the close-in exoplanets of M-dwarf stars, \citet{2007AsBio...7..185L} %
suggested that the magnetospheric distance must exceed 2 planetary radii in order to be protected against atmospheric erosion. As all the average \(R_\text{m}\) values in Table~\ref{tab:main-table} exceed \(6 R_\Earth\) we do not expect significant atmospheric erosion for Earth-like planets orbiting the Hyades stars in our study at a distance of \SI{1}{\astronomicalunit}.

\section{Discussion}\label{sec:Discussion}

In this section we examine the effect of magnetic scaling on our model results in Section~\ref{sec:mag_scale_effects}.

We compare our results to published three-dimensional wind models in Section~\ref{sec:3d_comparison}
and to published scaling laws in Section~\ref{sec:scaling-law-comparison}. 
We review the impact of scaling the ZDI magnetic fields in Section~\ref{sect:effect-of-zdi-limitations}. 
Section \ref{sec:magnetic_cycles} considers the role of magnetic varibility and cycles for our results. 
Effects of the modelled wind on a young Earth is considered in Section~\ref{sec:young-solar-system}.

\subsection{Effect of magnetic scaling}\label{sec:mag_scale_effects}
MHD studies such as the ones of \citet{2015MNRAS.449.4117V} and \citet{2018ApJ...856...53P} have investigated the relationship between age, wind and planetary impacts. The stellar rotation rate \(P_\text{rot}\) and average field strength \(|\vec B|\) are however both dependent on stellar age \citep[although the age-spin relation is bimodal, see][]{2003ApJ...586..464B}, and by design such studies cannot provide as much information on the direct influence of magnetic field strength on the wind. In addition to the correlation between  \(P_\text{rot}\) and \(|\vec B|\), very rapid rotation increases not only angular momentum loss, but also mass loss via the centrifugal force \citep{1976ApJ...210..498B}.

By having two models for each star, that vary only in their absolute radial magnetic field strength, we are able investigate the effect of the scaling of the magnetic field on the model results.

\begin{figure}
    \centering
    \includegraphics[width=\linewidth]{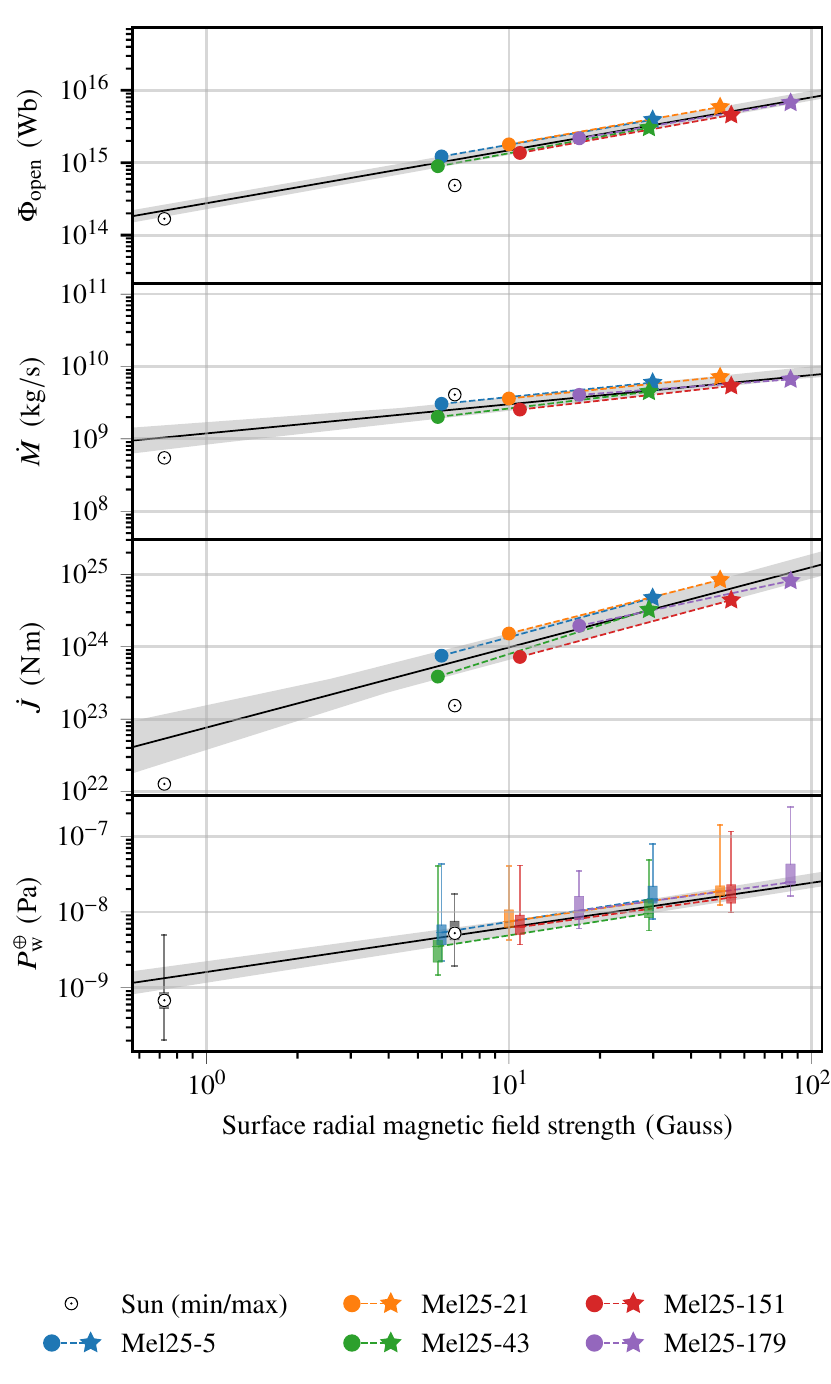}
    \caption{ 
    Effect of magnetic scaling, and lower bound on residual variation due to magnetic geometry.
    This plot shows the open magnetic flux, mass loss, angular momentum loss, and total wind pressure at \SI{1}{\astronomicalunit} plotted against the mean surface radial magnetic field strength.
    For each star we plot the value in the \(B_\ZDI\) and \(5B_\ZDI\) series as a barbell (a dot and a star connected by a dashed curve \(\hat y_i(x) \propto |B_r|^{\alpha_i}\). The dot represent the values in the \(B_\ZDI\) series and the star represents the values in the \(5B_\ZDI\) series.
    The midpoint of the dashed barbell lines \(\hat y(x)\propto |B_r|^\alpha\) is also plotted (black line). The shaded region represents the span of the barbell curves \(\hat y_i(x) \propto |B_r|^{\alpha_i}\).
    The Sun values are represented by the Sun symbol `$\Sun$'; the left symbol is the value at Solar minimum. 
    In the bottom panel we use boxplots to indicate the variation with orbital phase and Solar rotation in each model.
    }\label{fig:orthogonal_trend}
\end{figure}

In Fig.~\ref{fig:orthogonal_trend} we plot a `barbell' comprising a line segment that connects the simulation results of the \(B_\ZDI\) case and the \(5B_\ZDI\) case for each star. On the barbell, the end representing the \(B_\ZDI\) case is drawn as a dot, and the end representing the \(5B_\ZDI\) case is drawn as a star. From Fig.~\ref{fig:orthogonal_trend} it may be seen that the five barbells all have similar slopes. 
Letting the index \(i\) represent the star, and \(\hat y_i(x)\) be a parametrisation of the dashed line of each barbell, the equation for each of the line segments are
\begin{subequations}
\begin{equation}
    \log_{10} \hat y_i(x) = \alpha_i \log_{10} x + \log_{10}\beta_i, \text{ so that } \hat y_i(x) \propto x^{\alpha_i}.
\end{equation}
We report the midpoint and range of the barbell slopes \(\alpha_i\) in the \(\alpha\) column of Table~\ref{tab:orthogonal_correlations}. 
In Fig.~\ref{fig:orthogonal_trend} we draw a shaded area 
between the curves \(\min_i \hat y_i(x)\) and \(\max_i \hat y_i(x)\) to indicate the range of fitted \(\hat y_i(x)\) curves,
and a black line \(\hat y(x) = \beta x^\alpha\) using the midpoint values of \(\alpha_i\) and \(\beta_i\). 

The height of the shaded area in the region on the \(x\) axis occupied by the \(B_\ZDI\) and \(5B_\ZDI\) models (but not extending to the Sun models) represents the observed variation around the middle curve \(\hat y(x) = \beta x^\alpha\) in our models.  
We report the value
\begin{equation}
    \frac{ y_\text{max}}{ y_\text{min}} = \max_{x \in X_\Star} \,
    {
        \frac{\max \hat y_i(x)}{\min \hat y_i(x)}
    },
    \quad
    X_\Star = (\min |B_r|_\Star, \max |B_r|_\Star)
\end{equation}
in Table~\ref{tab:orthogonal_correlations}.
The \(y_\text{max}/y_\text{min}\) this represents the maximal vertical variation in the shaded area in the region of the plot occupied by the barbells. The range of this variation provide a measure of the influence of magnetic geometry on the wind models.

The height of the shaded area indicates the magnitude of the effect of magnetic geometry, but the
coefficients of determination (\(r^2\) values) are a better measure of the explanatory power of the \(\hat y(x)\) curve, as \(r^2\) is independent of the value of \(\alpha\). We let \(j\) range over the models in both the \(B_\ZDI\) and the \(5B_\ZDI\) series, so that \(x_j\) and \(y_j\) represent the result of an individual model, and report the coefficients of determination of the \(\hat y(x)\) curve calculated in the regular way:
\begin{equation}
    r^2 = 1 - 
    \left.
        \sum{\big(\log_{10} y_j / \bar y \big)^2} 
    \middle/ \,
        \sum{\big(\log_{10} y_j / \hat y(x_j) \big)^2}
    \right.
    ;
\end{equation}
we have used \(\log y_j - \log \bar y = \log y_j/\bar y\) and \(\log y_j - \log \hat y(x_j) = \log y_j/\hat y(x_j)\). The \(r^2\) value may be interpreted as the proportion of the variation in the \(y_j\) values that are predicted from the \(x_j\) values. 

We emphasise that quantities \(\alpha\), \(y_\text{max} / y_\text{min}\) and \(r^2\) are purely descriptive, and independent of assumptions about the underlying statistical distribution of the model result and/or their variation. 
This approach does not over-estimate the influence of the magnetic geometry due to the widening of confidence intervals when only few data points is available; instead, this approach gives a lower bound on the effect of magnetic geometry compared to an approach based on regression analysis based approach (see Appendix~\ref{sec:correlations}).
\end{subequations}
\begin{table}
    \centering
    \caption{
        Effect of magnetic scaling and lower bound on residual variation due to magnetic geometry. Power laws on the form \(\hat y \propto |B_r|^\alpha\) have significant explanatory power over the physical quantities reported in Table~\ref{tab:magnetic_averages} and Table~\ref{tab:main-table}. 
        The \(\alpha\) coefficient represents the midpoint of the \(\alpha_i\) values found by drawing power law relations on the form \(\hat y_i \propto |B_r|^{\alpha_i}\) from the stellar models of the \(B_\ZDI\) series to the corresponding model in the \(5B_\ZDI\) series. 
        The \(\frac{y_\text{max}}{y_\text{min}}\) values are a measure of the residual amount of variation not attributable to the \(|B_r|\) parameter.
        The coefficients of determination \(r^2\) also measure how much of the variation in the dependent variable may be attributed to variations in the independent variable, but in contrast to the \(\frac{y_\text{max}}{y_\text{min}}\) parameters, the \(r^2\) values are independent of the \(\alpha\) values. The mass loss \(\dot M\) is the value that has the least amount of variation attributable to \(|B_r|\), while the angular momentum loss \(\dot J\) has the largest range values not attributable to \(|B_r|\). 
        }\label{tab:orthogonal_correlations}
    \sisetup{
        table-figures-decimal=3,
        table-figures-integer=2,
        table-figures-uncertainty=2,
        table-number-alignment=center,
        add-decimal-zero=true,
        add-integer-zero = false,
        omit-uncertainty = false
    }
    \begin{tabular}{
    l
    S[
        table-figures-decimal=3, 
        table-figures-integer=1,  %
        table-figures-uncertainty=4,
        table-figures-exponent = 0
    ]
    S[
        table-figures-decimal=3,
        table-figures-integer=1,
        table-figures-uncertainty= 0,
        table-figures-exponent = 0,
    ]
    S[
        table-figures-decimal=3,
        table-figures-integer=1,
        table-figures-uncertainty= 0,
        table-figures-exponent = 0,
    ]
}
\toprule
{Quantity} & \multicolumn{3}{c}{Relation to $\alpha \log_{10}|B_r|+ \log_{10}\beta$} \\
\cmidrule(lr){2-4}
{} & {$\displaystyle \alpha$} & {$\frac{y_\text{max}}{y_\text{min}}$} & {$r^2$} \\

\midrule
$\log_{10}\max|B_r|$ & 1.001+-0.000 & 1.630 & 0.968 \\
$\log_{10}\Phi_0$ & 1.000+-0.000 & 1.313 & 0.985 \\
$\log_{10}|\vec{B}|$ & 0.980+-0.013 & 1.097 & 0.999 \\
$\log_{10} R_\Alfven$ & 0.412+-0.035 & 1.307 & 0.959 \\
$\log_{10} |\vec{r_\Alfven} \times \uvec{z}|$ & 0.414+-0.034 & 1.333 & 0.958 \\
$\log_{10}\Phi_\text{open}$ & 0.736+-0.020 & 1.406 & 0.956 \\
$\log_{10}\dot M$ & 0.424+-0.065 & 1.587 & 0.819 \\
$\log_{10}\dot J$ & 1.110+-0.140 & 2.391 & 0.900 \\
$\log_{10}P_\text{wind}^\Earth$ & 0.587+-0.038 & 1.656 & 0.912 \\
$\log_{10} R_\text{mag}$ & -0.098+-0.006 & 1.088 & 0.912 \\
\bottomrule
\end{tabular}
\end{table}

We now point out some conjectural conclusions that may be drawn from Fig.~\ref{fig:orthogonal_trend} and Table~\ref{tab:orthogonal_correlations}. 
\begin{enumerate}
    
    \item For the magnetic quantities \(\max |B_r|\) and \(\Phi_0\) our model enforces \(\alpha=1\), as \(|B_r|\) is a fixed boundary condition of the model, and \(\Phi_0 = 4\pi R_\Star^2 |B_r|\) (see Section~\ref{sec:open_magnetic_flux}), so that both \(\max |B_r|\) and \(\Phi_0\) scale with \(|B_r|\). This is reflected in the \(\alpha\) values in Table~\ref{tab:orthogonal_correlations}.
The large \({y_\text{max}}/{y_\text{min}} = \SI{163}{\percent}\) for \(\max |B_r|\) is a direct consequence of the magnetic geometry as the geometry directly determines 
\(\left.\max |B_r| \middle/ |B_r|\right.\) 
(values given in Table~\ref{tab:magnetic_averages}).

\item The average unsigned surface field strength \(|\vec B|\) is very nearly proportional to \(|B_r|\),
\(\alpha=0.98 \pm 0.01\), suggesting that for the Hyades models, the radial surface fields very nearly determines all three components of the surface field. The \(y_\text{max}/y_\text{min} = \SI{109.7}{\percent}\) shows that the range of variation in \(|\vec B| / |B_r|^{0.98 \pm 0.01}\) is only \SI{9.7}{\percent} across the Hyades models.

\item We observe that the scaling behaviour of the average Alfvén \(R_\Alfven\) radius and the torque-averaged Alfvén radius \(|\vec r_\Alfven \times \uvec z|\) have nearly identical values in Table~\ref{tab:orthogonal_correlations}. In both cases \(\alpha=0.41\) and about \SI{30}{\percent} of variation is attributed to the magnetic geometry.

\item The open magnetic flux \(\Phi_\text{open}\) is shown in the upper panel of Fig.~\ref{fig:orthogonal_trend}. While the shaded region in Fig.~\ref{fig:orthogonal_trend} appears narrow,  Table~\ref{tab:orthogonal_correlations} shows that the highest \(\hat y_i(x)\) values for \(\Phi_\text{open}\) are \SI{41}{\percent} greater than the smallest values in the shaded region.

\item The mass loss values \(\dot M\) are shown in the upper middle panel of Fig.~\ref{fig:orthogonal_trend}.
From Table~\ref{tab:orthogonal_correlations} we see a model trend \(\dot M \propto |B_r|^{0.42\pm 0.06}\), with a variation of \SI{58.7}{\percent} across the Hyades models. 
This gives \(\dot M\) the smallest \(r^2\) value in Table~\ref{tab:orthogonal_correlations}; only \(r^2=0.82\) of the variation in \(\dot M\) is explained by the midpoint curve \(\hat y(x)\) (black line).

\item The angular momentum loss values are shown in the lower middle panel of Fig.~\ref{fig:orthogonal_trend}. 
\(\dot J\) has the largest range in \(\alpha = 1.11 \pm 0.14\), and \(y_\text{max}/y_\text{min} = \SI{239.1}{\percent}\) shows that the magnetic geometry variations produce a spread of more than a factor of \(2\) in the angular momentum loss magnitude. This is the greatest spread observed in Table~\ref{tab:orthogonal_correlations}. The magnitude of \(\alpha\) plays a role here as the \(r^2\) value is better than the \(r^2\) value for \(\dot M\).

\item The wind pressure at \SI{1}{\astronomicalunit{}} is shown in the lower panel of Fig.~\ref{fig:orthogonal_trend}. In Table~\ref{tab:orthogonal_correlations}, the \(P_\text{wind}^\Earth\) and \(R_\text{mag}\) have identical \(r^2\) values and a factor of \(6\) separates their \(\alpha\) values as a consequence of equation~\eqref{eq:magnetospheric_stand_off_distance}.
\end{enumerate}

\subsubsection{Incorporation of Solar models}\label{sec:mag_scale_solar}

The values calculated from the two Solar models are shown in Fig.~\ref{fig:orthogonal_trend} as Sun symbols. We do not draw a barbell connecting the Solar minimum and Solar maximum models as the magnetic geometry is not the same for these two models, see Fig.~\ref{fig:sun-br-fieldlines}. The Solar models rarely sit inside the shaded area; this is another indication that the shaded area represents a lower bound on geometry-induced variation in parameters. 

In Section~\ref{sec:results} we saw that our calculated values of \(\Phi_\text{open}\) for the Solar cases are 2--3 times lower than observational values. The results in this section indicate that a Solar magnetogram scaling of about 3 would bring our calculated \(\Phi_\text{open}\) values in agreement with observations. This scaling would also increase the Solar \(\dot J\) values so that the our Solar maximum value would be within a factor of \(\mysim2\) of the observed values. Our calculated value at Solar minimum would, however, still be 8 times smaller than what is observed at Solar minimum. 

\subsubsection{Effect of  quadrupole field components and dipole axisymmetry}\label{sec:effect-qpole-axisymm}
We do not observe any significant correlation between the dependent values in 
Table~\ref{tab:orthogonal_correlations}
and the fraction of magnetic energy in the dipolar, quadrupolar, and octupolar surface field component. As our dipole energy fractions fall in the range 0.4--0.7 (see Table~\ref{tab:magnetic_averages}) we do not expect to see the dramatic variations that was found by \citet{2015ApJ...807L...6G} for pure quadrupolar and higher order fields. 

In the same vein, \citet{2019ApJ...886..120S}, see also \citet{2018ApJ...854...78F}, found that spin-down torque is dominated by the dipole component if \(\dot M\) is smaller than a critical mass \(\dot M_\text{crit}\)~\citep[eq. 3 in][]{2019ApJ...886..120S}. For the cases we have simulated in this study, including the \(5B_\ZDI\) series, the mass loss rate lies below the critical rate, hence this is consistent with the results in this study where no correlation is found between \(\dot M\), \(\dot J\), and the non-dipolar magnetic field components.

There is a noticeable correlation between the axisymmetric dipole field and \(\dot M\), \(\dot J\) and \(P_\Wind\). There is however also a correlation between \(|B_r|\) and the axisymmetric dipole field strength in the underlying dataset which prevents any conclusion to be drawn about the effect of the axisymmetric dipole fraction.

\subsection{Comparison to other 3D models}\label{sec:3d_comparison}

To provide context for the model results of Section~\ref{sec:results} and Table~\ref{tab:main-table}, we compare our mass loss and angular momentum loss values to a selection of recently published values from stellar wind models. We do not include papers whose main focus is reproducing features of the Solar wind, but note that it has long been argued \citep[e.g.\@][]{1977ApJ...213..874M} that a variable polytropic index or explicit heating terms are needed to reproduce coronal and wind properties at the same time.  \citet{2017ApJ...835..220C} argues that ideal MHD models cannot reproduce the conditions in the Solar corona and at \SI{1}{\astronomicalunit}  %
with a single set of parameters, leading first to the introduction of a variable polytropic index and later to models with explicit physical heating and cooling terms, such as the \awsom{} model, where simultaneous reconstruction of conditions in the different regions is possible. 

The results of our comparison to published stellar wind models are shown in Fig.~\ref{fig:results-context}. Values obtained from full MHD simulations are shown as coloured symbols with black outlines. In the original papers where these models are presented, dissimilar quantities are used to describe the inner boundary (i.e.\ chromospheric or coronal base) magnetic field. 
The quantities used include averages of \(|\vec B|\), \(|B_r|\), \(\vec B^2\) and \(B^2_r\), as well as the total surface flux $\Phi_0$. 
To facilitate comparisons we have used approximate power laws which we have derived from the magnetograms of \citet{2016MNRAS.457..580F} and \citetalias{2018MNRAS.474.4956F}: 
\(\langle |\vec B| \rangle = 1.4 \, \langle \left|B_r\right| \rangle^{1.0}\), 
\(\langle |\vec B| \rangle = 1.1 \, \langle       B_r^2      \rangle^{0.5}\), and 
\(\langle |\vec B| \rangle = 0.9 \, \langle  \vec B^2        \rangle^{0.5}\). 
We emphasise that these relations between the surface averages are approximate and do not capture the complexity of a full wind model. They do, however, enable rough comparisons of the literature values, which is our goal in Fig.~\ref{fig:results-context}.

In our comparison we incorporate two families of models, those obtained with \awsom{}-like models (blue symbols), and those with ideal MHD models using `already hot' coronae and polytropic indices near unity (orange and green symbols).
A red outline around these symbols indicate that the model is a Sun model, otherwise we use a black outline. Our own models are shown as white symbols with black outlines. 

\subsubsection{Alfvén wave driven models}\label{sec:aw-driven-models}
The most similar numerical models to our own are the models of 
\citet{2016A&A...588A..28A,2016A&A...594A..95A} (blue triangles), who used the \awsom{} model to model the winds and coronal structure of the Sun, and of the G and K type stars HD~1237, HD~22049, and HD~147513. In the paper two different ZDI reconstruction techniques 
were applied; we group them together in this comparison as the method of reconstructing the magnetic field does not affect the trends in \(|\vec B|\). 
Their stellar values for \(\dot M\) and \(\dot J\) show very good agreement with our values. For their Solar models we estimate the surface \(|\vec B|\) values from SOHO/MDI magnetograms truncated to \SI{120}{\gauss}, this does however place their Solar minimum \(|\vec B|\) value closer to our Solar maximum value. \citet{2014SoPh..289..769R} recommends a scaling factor \numrange{2}{3} when comparing MDI magnetograms and GONG magnetograms which would move these values leftwards in the plot. In Fig.~\ref{fig:results-context} their Solar \(\dot J\) values are a bit lower than ours, while the \(\dot M\) values show good agreement.

Recently, \citet{2018ApJ...856...53P} (blue squares) used \awsom{} to model eight Solar-type stars with ages ranging from \SIrange{.03}{4.6}{\giga\year}. This work, however, does not model the transition region; instead it uses coronal boundary conditions of \SI{2e8}{\per\cubic\centi\meter} and \SI{1.5}{\mega\kelvin}. We group these results with the Alfvén wave driven models as heating and cooling terms are incorporated similar to Section~\ref{sec:model-equations}. For \(\dot M\) the values show excellent agreement with our models. The \(\dot J\) values also agree well with our values but have a higher spread, which we attribute the large spread in periods of rotation (\SIrange{0.5}{26}{\day}) and stellar ages in their work.

\subsubsection{Comparison with ideal MHD models}
Next, we compare our results with a number of wind models using ideal MHD and polytropic indices near unity. These models do not include the transition region, instead the corona is already hot at the inner boundary.
In the ideal MHD models these stars were modelled with similar parameters: a polytropic index 
\(\gamma\approx 1.1\) \citep[see\ ][]{2011ApJ...727L..32V}, a coronal temperature of 
\( \mysim \SI{2}{\mega\kelvin}\), and a coronal base density of 
\( \mysim \SI{e9}{\per\cubic\centi\meter}\). 
The six sets of models that we compare to are listed below, along with some of their key features and differences. The first five of these use the \batsrus{} code, while the sixth one uses the \pluto{} code. 
\begin{enumerate}
    \item HD 189733, a K2V star which has a close-in hot Jupiter (orange downwards-pointing triangle), was modelled by \citet{2013MNRAS.436.2179L}. 
    The average ZDI magnetic strength is given in \citet{2010MNRAS.406..409F} as \SI{36}{\gauss}, of which \SI{77}{\percent} of the energy is is not poloidal.
    The resulting \(\dot M\) and \(\dot J\) values of this star is reported in \citet{2017MNRAS.466.1542S}.  
    \item Five hot Jupiter hosting F-K type stars studied by \citet{2015MNRAS.449.4117V} (orange upwards-pointing triangle), with rotation periods from \SIrange{7.6}{42}{\day}. 
    \item The star \(\kappa^1\) Ceti \citep{2016ApJ...820L..15D} (orange square), the \(\dot J\) value of which is reported in \citet{2017MNRAS.466.1542S}.
    \item The rapidly rotating (\(P_\text{rot}=\SI{3}{\day}\)) star F7V \(\tau\)~Bo\"{o}tis modelled by~\citet{2016MNRAS.459.1907N} 
    (orange pentagon)
    at eight different epochs. These results are located in the upper left of Fig.~\ref{fig:results-context}, we expect this because of the star's rapid rotation. 
    \item Six stars and two Solar models from \citet{2019MNRAS.483..873O} (orange hexagons). In these models the coronal base density and temperature are scaled with the stellar rotation period, which gives base density and temperature value ranges of \SIrange{6.7e8}{1.9e9}{\per\cubic\centi\meter} and \SIrange{1.5}{3.0}{\mega\kelvin}. 
    \item Five Solar-type stars as well as a Solar case \citep{2016ApJ...832..145R} (green triangles) modelled using an ideal MHD model with ratio of specific heats \(\gamma=1.05\). 
    These models extends from the coronal base where the temperature and density are set according to the scaling laws given by~\citet{2007A&A...463...11H}, \(T\propto\Omega^{0.1}\) and \(\rho \propto \Omega^{0.6}\). 
    This gives base densities from \SIrange{1.0e8}{4.2e8}{\per\cubic\centi\meter} and temperatures from \SIrange{1.5}{1.9}{\mega\kelvin}. The mass loss values agree well with our values, while the angular momentum loss values are higher than our values.
\end{enumerate}

\subsubsection{3D model comparison summary}
Direct comparison between stellar wind models can most easily be made if the studies include Sun models (red outline symbols), as the Solar models can be used to judge the impact of model choices that affect the mass loss and angular momentum loss. For mass loss, the Solar cases (red outline symbols) agree to within one order of magnitude even between the ideal MHD models and the \awsom{} models but the basis data are scant and non-conclusive as can be seen in Fig.~\ref{fig:results-context}. 

By broadening the comparison to all 3D models (black outline symbols and red outline symbols) we see from Fig.~\ref{fig:results-context} that the ideal MHD models (orange) predict significantly (1--2 orders of magnitude) higher values of mass loss, and one order of magnitude higher values for angular momentum loss than the \awsom{}-like models. We know, however, from equation~\eqref{eq:angmom_loss} that the scatter in \(\dot J\) is increased by dissimilar rates of rotation. This is true in particular for the \(\tau \) Bo\"{o}tis models (orange pentagons) where \(P_\text{rot} \sim \SI{3}{\day}\). 
For the mass loss values the \awsom{} type models (white and blue symbols) have only a small degree of scatter in comparison to the total scatter across all the model types. The angular momentum loss values contain a comparable amount of scatter for the ideal MHD models and for the \awsom{} type models. 
The correlation between \(\Omega\) and \(|\vec B|\) also affects some of the data plotted. We expect this to be the case for any study that models stars at a range of different ages, such as the models from 
\citet{2016ApJ...832..145R} (green triangles),
\citet{2015MNRAS.449.4117V} (orange triangles), 
\citet{2018ApJ...856...53P} (blue squares), and
\citet{2019MNRAS.483..873O} (orange hexagons).
The approximate scaling laws we used also increases the uncertainty of these conclusions; this is unlikely to be resolved due to the lack of standardisation in magnetogram parameter reporting in the literature.

\subsection{Comparison with known scaling laws}\label{sec:scaling-law-comparison}

The stellar wind mass loss is a key parameter in determining both the stellar angular momentum evolution and the wind pressure and density at planetary distances from the star. 
The mass loss is also notoriously difficult to determine. While direct observations of stellar winds have been made for giant stars~\citep{1995ApJ...452..407H},
and young pre-main sequence stars~\citep{2015ASSL..411...19W},
the mass loss of main sequence Solar-type stars is not well constrained. When no wind signal is found, observations only yield upper bounds on the wind strength and mass loss rates.

Methods used to date or proposed for detecting dwarf star stellar winds include observing 
the Lyman-\(\alpha\) line produced by stellar wind interacting with the interstellar medium~\citep{2001ApJ...547L..49W}, 
ultraviolet features resulting from  accretion in binary systems~\citep{1989ApJ...339L..33M,2006ApJ...652..636D}, 
free-free radio emissions from the wind itself~\citep{1992ApJ...397..225M},
and 
X-ray emissions from the regions where the stellar wind interacts with the local interstellar medium~\citep{2002ApJ...578..503W}, see the reviews in~\citet{2004LRSP....1....2W} and~\citet{2018haex.bookE..26V}. 

Of these direct observation methods, the most stringent upper limits on stellar wind mass loss come from observing the Lyman-\(\alpha\) line. From Lyman-\(\alpha\) observations of \( \mysim  10\) dwarf stars, \citet{2002ApJ...574..412W, 2005ApJ...628L.143W} found an empirical relation between stellar age~\(t\) and wind mass loss, \(\dot M\propto t^{-2.33\pm0.55}\). 
Going backwards in time from the present-day Sun, the relation predicts mass loss values up to \( \mysim  10^2 \dot M_\Sun\) before the relation breaks down at an age of \( \mysim \SI{0.7}{\giga\year}\). At ages below this threshold, stars were found to have lower values of mass loss \citep{2004LRSP....1....2W,2014ApJ...781L..33W}.

\citet{2005ApJ...628L.143W} suggested that the breakdown of this relation may be linked to the emergence of polar spots and strong dipolar fields in stars younger than this age. \citet{2016MNRAS.455L..52V}, however, did not find any evidence of strong dipolar field in young stars, but noted that young stars' magnetic fields have a significant component resulting from currents in the photosphere.

The weak observational constraints on stellar mass loss values have led to a range of mass loss predictions that can differ by an order or magnitude or more at the age of the Hyades cluster.  \citet{2007A&A...463...11H} presented a wind model in which the predicted mass loss rate does not exceed \(10\dot M_\Sun\) for main sequence stars. In this model the effects of rotation are only expected to become significant for fast magnetic rotators \citep[see][]{1976ApJ...210..498B} with \(\Omega R^2 B_r\) values be 10--20 times greater than the current day Sun. This model, however, does not explain the Lyman-\(\alpha\) derived mass loss \citep{2014ApJ...781L..33W} of some K dwarf stars such as \(\epsilon\) Eridani.

\citet{2011ApJ...741...54C} developed a method to estimate \(\dot M\) based on the fundamental properties of the star. This method was used by \citet{2014A&A...570A..99S} to model a large sample of stars from the BCool \citep{2014MNRAS.444.3517M} study, predicting mass losses up to \( \mysim  10^2 \dot M_\Sun\).

In a study where the corona is heated and supported by Alfvén waves, \citet{2013PASJ...65...98S} found \(\dot M \propto t^{-1.23}\), no break in the wind-mass loss relation, and loss rates up to \(10^3 \dot M_\Sun\) for very young stars. At the age of the Hyades, however, the model predicts a mass loss rate of \( \mysim  2\times10^2\dot M_\Sun\). \\\\

Unlike wind mass loss, stellar angular momentum loss can broadly be inferred by studying stellar ages and rotation rates \citep{1972ApJ...171..565S,2003ApJ...586..464B}. In general, stellar rotation slows with age, caused by shedding of angular momentum via stellar winds. 

Given the difficulties in observationally constraining \(\dot M\) for Solar-type stars, many models of angular momentum loss \(\dot J\) have \(\dot M\) as a free parameter. In general, angular momentum loss models assume that the stellar magnetic field forces the wind to co-rotate with the star some distance \(R_\Alfven\) into the corona. This co-rotation radius functions as a lever arm that greatly increases the amount of angular momentum lost via stellar winds~\citep{1962AnAp...25...18S}.

The one-dimensional model of \citet{1967ApJ...148..217W}, where \(\dot J = \frac{2}{3} \dot M \Omega R_\Alfven^2 \), can be extended through the incorporation of dipolar magnetic fields where some field lines loop back to the stellar surface. This gives rise to a dead zone of closed field lines where the wind is trapped, reducing the power of \(R_\Alfven\) so that \(\dot J \propto R_\Alfven\) \citep{1968MNRAS.138..359M}. Several models have followed with \(\dot J \propto R_\Alfven^n\) \citep{1984LNP...193...49M,1988ApJ...333..236K} where \(n\) is a magnetic field geometry parameter.

Many studies of dipolar~\citep{2008ApJ...678.1109M,2012ApJ...754L..26M} and more complex axisymmetric magnetic fields~\citep{2017ApJ...845...46F,2018ApJ...854...78F} employ a dimensionless `wind magnetisation parameter', which in SI units is of the form \(\Upsilon = (4\pi/\mu_0) B_\text{chr}^2 R^2 / (\dot M v_\text{esc})\), where \(v_\text{esc}\) is the stellar escape velocity. Here \(B_\text{chr}\) is a characteristic magnetic field strength but does not directly correspond to the average surface field strength. The related methodology by \citet{2015ApJ...798..116R,2015ApJ...814...99R} uses the open magnetic flux (the surface magnetic field in regions where the magnetic field is not closed) in their wind magnetisation parameter, thus accounting for the effect of a dead zone of closed field lines.  Using a three-dimensional potential field extrapolation to find the open flux from ZDI maps, \citet{2017MNRAS.466.1542S} found good agreement between these two methods. In this work we observe slightly better correlations between \(\dot J\) and \(\Phi_\text{open}\) than between \(\dot J\) and \(|B_r|\), tentatively supporting the open flux approach.

A recent model by~\citet{2020ApJ...896..123S} found that mass loss saturates at the comparatively low value of \(2 \dot M_\Sun\) and angular momentum loss values around \(\dot J=\SI{5e24}{\newton\meter}\) for a ten-day rotation period like that of the Hyades stars. 

 In another recent work
 \citet{2019ApJ...886..120S,2017MNRAS.466.1542S} compared the scaling law for mass loss from~\citet{2011ApJ...741...54C}, with a scaling law for mass loss based on reproducing the spin-down evolution of open clusters, and using the angular momentum loss model from~\citet{2017ApJ...845...46F,2018ApJ...854...78F}, thus attempting to infer \(\dot M\) from \(\dot J\). 

\begin{figure}
    \centering
    \input{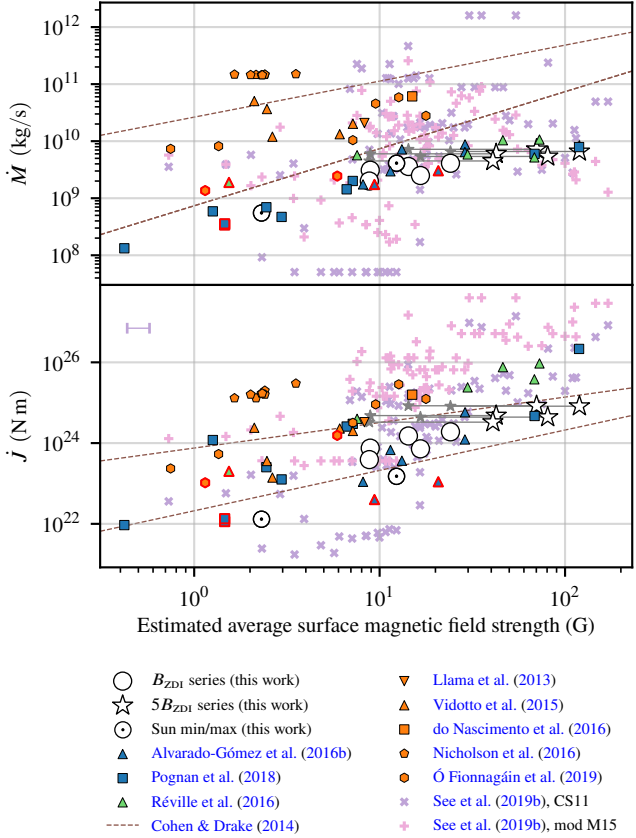}
    \caption{ 
        Comparison of our results with published mass loss and angular momentum loss values.
        The large white shape with black outlines represent the models in this work. 
        The other shapes with black or red edges represent individual MHD simulations of stellar (black) and Solar (red) models in the published works of others.
        The blue filled shapes are \awsom{} models while the yellow filled shapes are ideal MHD \batsrus{} models. The green filled shapes are \pluto{} models.
        The shapes with no edge colour represent scaling laws applied to a population of stars.
        The dashed lines represent the scaling laws from 
        \citet{2014ApJ...783...55C} 
        for Sun-like coronal densities 
        (\(n=\SI{2e8}{\per\cubic\centi\meter}\) and \(n=\SI{2e9}{\per\cubic\centi\meter}\)), 
        and 
        \(P_\text{rot}=\SI{10}{\day}\) like that of the Hyades stars. 
        The interval shown in the upper left of the \(\dot J\) panel shows the error in the \(x\) direction from our estimates for the CS11 and mod-M15 populations.
        Note that the \(x\) axis values are the values at the inner boundary of the model and are calculated from the final relaxed wind solution as described in Section~\ref{sec:numerical_model}; therefore no magnetic field is contributed by photospheric currents. 
        They grey stars connected with lines to the \(5B_\ZDI\) models show the \(\dot M\) and \(\dot J\) values of the \(5B_\ZDI\) models plotted against the \(B_\ZDI\) field strength; 
        they appear to agree well with the 
        \citet{2019ApJ...886..120S} 
        CS11 and mod-M15 models, something that may indicate that the ZDI field strength is underreported.
        As the CS11 method uses only fundamental stellar parameters to estimate \(\dot M\), rows of CS11 data points appear in the upper panel when a star is observed at multiple epochs. The row of of \(\dot M < \SI{1e8}{\kilogram\per\second}\) values represent multiple observations of 61~Cyg~A by \citet{2016A&A...594A..29B}.
    }\label{fig:results-context}
\end{figure}

In this work we compare our models with scaling laws from \citet{2014ApJ...783...55C} and two scaling laws from \citet{2019ApJ...886..120S}. These are shown in Fig.~\ref{fig:results-context} as dashed lines and a population of plus symbols and cross symbols. 

\subsubsection[Variable polytropic index model]{Variable polytropic index model of \citet{2014ApJ...783...55C}}

The first scaling law to which we make comparisons is a variable polytropic index model by \citet{2014ApJ...783...55C}, who created a grid of dipolar stellar wind models with varying 
dipolar magnetic field strengths, %
periods of rotation, and %
coronal base density \(n\). 
The base temperature, which is set in the corona, is specified from an energy argument, see \citet{2009ApJ...699.1501C}. These models reproduce the Solar wind but do not accurately reflect the energetics of the corona; in terms of model complexity, the inclusion of a variable polytropic index is an intermediate step between ideal MHD models and models with explicit heating/cooling terms.

Even though our models have the coronal density values emerge within the model, and even though our magnetic fields are not pure dipoles, we have good agreement for \(\dot J\). 
Our stellar values lie between the lines for 
\(n=\SI{2e8}{\per\centi\meter\cubed}\) and 
\(n=\SI{2e9}{\per\centi\meter\cubed}\); these are displayed as dashed lines.
The situation is not as persuasive for the \(\dot M\) values, where our values lie below the \(n=\SI{2e8}{\per\centi\meter\cubed}\) line (the lower of the two lines) by a factor of \( \mysim  2\) for the Solar cases and the non-amplified \(B_\ZDI\) series of stellar cases. For the \(5B_\ZDI\) amplified magnetic fields the difference increases to almost an order of magnitude. The predictions from the \citet{2014ApJ...783...55C} model resemble more closely the ideal MHD model trend than the trend we observe on our own models and the \awsom{}-type models (blue symbols).
For a recent comparison of various polytropic indices and explicit heating and cooling in the Solar corona, see \citet{2017ApJ...835..220C}.

\subsubsection[Populations from See et al.]{Populations from \citet{2019ApJ...886..120S}}\label{sect:see-comparison}

We also compare our results to two populations of mass loss and angular momentum loss values from \citet{2019ApJ...886..120S}, both of which are based on scaling laws. These values are shown as cross and plus symbols in Fig.~\ref{fig:results-context}. To estimate the aggregate surface field strength from the listed dipolar, quadrupolar and octupolar values used in \citet{2019ApJ...886..120S} we compare these quantities in the~\citet{2016MNRAS.457..580F} and~\citetalias{2018MNRAS.474.4956F} datasets which gives an approximate relation 
\(\langle |\vec B| \rangle = 
                  2 \,
 \langle |\vec B_\text{d}| \rangle^{\num{0.5} } \,
 \langle |\vec B_\text{q}| \rangle^{\num{0.2} } \,
 \langle |\vec B_\text{o}| \rangle^{\num{0.3} } \,
\)
which nicely preserves scaling of the magnetic field. We stress that this relation is made only to facilitate the comparison in Fig.~\ref{fig:results-context}, in which trends in 
\(|\vec B|\) are the object of study. The formal uncertainty in these estimates is shown in the top left corner of the bottom panel. The two methods, which they name CS11 and mod-M15 are different in several important regards:
\begin{enumerate}
    \item The CS11 method (crosses) uses a mass loss scaling law from~\citet{2011ApJ...741...54C} that predicts the stellar mass loss from fundamental stellar parameters (mass, radius, luminosity, period, and metallicity). 
    The \(\dot J\) values are then calculated using the method of \citet{2018ApJ...854...78F}.
    \item The mod-M15 method (plus signs) involves determining the mass loss rate such that the braking law is fitted to the spin-down evolution of open clusters, see \citet{2015ApJ...799L..23M}, and scaled the results by a factor of 25 to match Solar values \citep{2019ApJ...886..120S}.
\end{enumerate}

In Fig.~\ref{fig:results-context} it is evident that both the CS11 and mod-M15 values exhibit a large amount of scatter due to parameters other than \(B_r\). We observe that for mass loss \(\dot M\), the values of our study fall inside the population scatter for both the CS11 population and the mod-M15 population; for the CS11 scaling law our values do however appear to lie below the main trend. For \(\dot J\) our values lie just below the CS11 method, and significantly below the results of the mod-M15 method.

As the work of \citet{2019ApJ...886..120S} contain models of the same stars as in this work, we include a comparison of their values and ours in Appendix~\ref{appendix:see-comparison}. This confirms the agreement between the CS11 values and the values in this study. The mod-M15 values are the same order of magnitude as our values for \(\dot M\), while for \(\dot J\) they are 1-2 orders of magnitude higher.

\subsection{Limitations of the model and boundary conditions}\label{sect:effect-of-zdi-limitations}
The \awsom{} model itself, while physics-based, is calibrated to Solar values. We expect that the validity of this calibration is progressively reduced as we model stars more unlike the Sun; the variation of these parameters with stellar type and age can be informed by theoretical studies and observational data in the future. The efficacy of non-Alfvén wave based heating such as nanoflares and others \citep{2019ARA&A..57..157C} may be different, something which would affect our results particularly for heating mechanisms where \(\Pi_\Alfven \propto B\) does not hold. 

The magnetic maps and numerical models we use in this study represent the state of the art in Zeeman-Doppler imaging and wind modelling, yet questions remain whether some of the discrepancies between model types can be explained by systematic errors in ZDI and differences between Solar and stellar magnetograms. \citet{2017ApJ...835..220C} warned against using wind models without careful calibration against Solar cases, and \citet{2020A&A...635A.178B} warned that the uncritical use of stellar magnetograms in a Solar wind model could result in inaccurate results when stellar parameters differ significantly from the Solar parameters.

In the future the inclusion of comparisons with X-ray data values and complementary Zeeman broadening measurements can likely improve our understanding of the ZDI underreporting and the relevance of the small scale field. 
This will be aided by models like \awsom{}, where simultaneous reconstruction of coronal conditions and \SI{1}{\astronomicalunit} conditions is possible. The simultaneous reconstruction of these two regions facilitate comparison of wind properties like \(\dot M\) and \(\dot J\) with X-ray and UV fluxes from the corona.

In the remainder of this section we consider the effect of a free floating tangential magnetic field at the inner boundary,  the effect of ZDI measurements underreporting the true magnetic field strength, the effect of the missing small-scale field in ZDI, the effect of unobservable stellar regions, and the possible influence of unseen binary companions on our results.

\subsubsection{Relevance of free floating tangential field}\label{pfss-and-tangential-field}

Traditionally, Solar magnetograms coefficients are found using the potential field source surface  method \citep[PFSS, ][]{1969SoPh....6..442S,1969SoPh....9..131A,1984PhDT.........5H,1992ApJ...392..310W}, see also the review by~\citet{2017SSRv..210..249W}. 
In the PFSS method the set of radial spherical harmonics coefficients suffice to completely determine the magnetic field as the field is assumed to be purely potential between the stellar surface and a `source surface', and purely radial outside the source surface. A potential field does not permit `wind-up' of the magnetic field lines as a consequence of (rapid) stellar rotation.

While vector magnetograms of the Sun have been available with the Synoptic Optical Long-term Investigations of the Sun \citep[SOLIS; ][]{2003AGUFMSH42B0545H} and the
Helioseismic and Magnetic Imager \citep[HMI; ][]{2012SoPh..275....3P,2014SoPh..289.3483H},
the \awsom{} model is driven only by the radial magnetogram component; the non-radial magnetic field values are relaxed freely on the inner boundary \citep{2014ApJ...782...81V}. Indeed, the effect of the non-radial, non-potential field is thought to have little influence on the wind, as it is assumed that Solar-type coronae are dominated by magnetic pressure (low-\(\beta\)) and that the magnetic field may be adequately described using the PFSS method, see e.g.~\citet{2013MNRAS.431..528J}. This approach means that the non-radial magnetic field at the chromospheric base is determined by phenomena occurring inside the simulation, rather than by phenomena occurring below the base of the chromosphere and in the photosphere where the ZDI Stokes \(I\) and \(V\) profiles originate. 

We emphasise that while our models are driven only by the radial magnetic field, wind-up and other non-potential effects as seen in Fig.~\ref{fig:Mel-Pwind-IH-equatorial} are permitted in all parts of the numerical solution; the effect of magnetic and thermal pressures are given the full treatment described in Section~\ref{sec:model-equations}.

We also note that transient phenomena such as flares and coronal mass ejections are likely powered by non-potential magnetic fields. The free floating tangential field employed in our models mean that the final magnetic field configuration is a low-energy state where the energy to power these phenomena is unavailable. Transient phenomena likely become more important for younger, more magnetically active stars.

\subsubsection{Effect of ZDI underreporting}\label{sec:effect_of_zdi_underreporting}
From our results it is clear that the ZDI underreporting of magnetic field strength 
(see Section~\ref{sec:ZDI-missing-field}) would significantly affect the wind models. 
In this work the magnetic field scaling factor of 5 which separates the \(B_\ZDI\) and \(5B_\ZDI\) series is considered a worst case scenario. We consider this justified as the lowest magnetogram degrees are the ones that are best reconstructed by ZDI, and most of the magnetic energy of the Hyades magnetograms is in the dipolar and quadrupolar field (see Table~\ref{tab:magnetic_averages}). The relevance of the missing small scale field is considered separately in Section~\ref{sec:relevance-of-small-scale-field}.

A key difference between the ideal MHD models and \awsom{} models we compared our work with in Section~\ref{sec:3d_comparison}, is that  ideal MHD models separate the coronal heating effect from the influence of the surface magnetic field strength, while in the \awsom{}-type models 
the coronal heating is linked to the surface field via the Poynting flux-to-field ratio parameter \(\Pi_\Alfven/B\). 

We saw in Fig.~\ref{fig:mel-ur-alfven} that with \awsom{} an increase in the large scale field strength does not only increase the strength of the magnetic field; it also leads to a faster wind as it increases the supply of Alfvén wave energy at the inner model boundary where \(\Pi_\Alfven \propto B\). 

In Fig.~\ref{fig:results-context} we have marked the shift in horizontal position of our \(5B_\ZDI\) stellar models would undergo given that the missing field strength factor of $ \mysim  5$ is realistic (grey lines and stars).
It is suggestive to see how this would make our result fall more in the centre of the scatter for both the CS11 population and the mod-M15 population of \citet{2019ApJ...886..120S}, and the ideal MHD models (orange symbols) where the \SI{e6}{\kelvin} coronal temperature is set at the inner boundary of the model. These shifted values of the \(5B_\ZDI\) series are well matched with the CS11 model which makes no direct use of magnetic field strength in predicting \(\dot M\)~\citep{2011ApJ...741...54C}. 

\subsubsection{Relevance of small-scale field}\label{sec:relevance-of-small-scale-field}
The effect of omitting the small scale magnetic field in stellar wind models has been studied by several authors and groups, and there are some indications that the results are model-dependent; setting the model boundary temperature in the \si{\mega\kelvin} range reduces the impact of neglecting the small scale magnetic field.

In a study using an extension of the \citet{1967ApJ...148..217W} model, \citet{2005A&A...440..411H} found that non-dipolar magnetic fields can change the angular momentum loss by a factor of \( \mysim  2\), while having only a small impact on mass loss. 

Modelling M dwarf stars with ideal MHD, \citet{2014MNRAS.438.1162V} found that \(\dot M\) and \(\dot J\) is somewhat proportional to the open flux and that the complexity of the magnetic field could play an important role in stellar revolution. 

In a model where \(\dot M\) and \(\dot J\) are determined by the open magnetic flux \(\Phi_\text{open}\), \citet{2014MNRAS.439.2122L} added a `flux carpet' of small scale magnetic fields in order to align ZDI and Zeeman broadening measurements, finding that the small scale field has little effect on \(\Phi_\text{open}\). They do however warn that modelling the magnetic field as a pure dipole can lead to overestimates of \(\Phi_\text{open}\) and thus of \(\dot M\) and \(\dot J\).
A similar result was found by \citet{2015ApJ...807L...6G}  studying \(\dot M\) and \(\dot J\) with \awsom{} in the case of \emph{pure} dipolar, quadrupolar, and higher order fields. The authors found a decrease of both parameters by several orders of magnitude and propose this as an explanation of the bimodal distribution of rotational periods in young clusters.
Following up on their work, \citet{2016A&A...595A.110G} found that \(\dot M\) and \(\dot J\) are more strongly affected by the magnetogram degree \(\ell\), and less so by the order \(m\) (see equation~\ref{eq:zdi_br}) for a given degree.

Using the Wang-Sheeley-Arge \citep[WSA,\ ][]{1990ApJ...355..726W,2000JGR...10510465A} semi-empirical method, \citet{2017MNRAS.465L..25J} found variations between \SI{5}{\percent} and \SI{20}{\percent} for \(\dot M\) and \(\dot J\), when truncating Solar magnetograms at degrees \(\ell_\text{max}\) between 1 and \({\sim}60\). 

Comparing these results to the dipole-dominated mass loss values of \citet{2019ApJ...886..120S} that we mentioned in Section~\ref{sec:effect-qpole-axisymm} and several recent studies \citep{2017MNRAS.466.1542S,2018MNRAS.474..536S,2017ApJ...845...46F,2018ApJ...854...78F} it appears that using pure dipolar, etc.\ geometries requires extra caution in particular if the field strength is set from Zeeman broadening data. Zeeman broadening data reflect the total magnetic field strength of both the large- and the small-scale field; adding all this magnetic energy into e.g.\ a pure dipolar magnetic geometry would lead to an overestimate of \(\dot M\) and \(\dot J\). Conversely, a less idealised geometry with energy distributed over multiple order and degrees would result in smaller \(\dot M\) and \(\dot J\) values. 

For the \awsom{} model the addition of a flux carpet of small scale magnetic fields would increase \(|\vec B|\) and thus the Poynting flux \(\Pi_\Alfven\) at the inner boundary, which in turn would increase the supply of Alfvén wave energy.
If the ratio between the small- and the large-scale field strength is known to differ from the Solar case
the value of \(\Pi_\Alfven/B\) may be modified. This may involve complementary observations of Zeeman broadening, X-ray flux, starspot distributions, or the use of semi-empirical models of these parameters such as the trends in \citet{2014MNRAS.441.2361V}.
An example that involves tuning \(\Pi_\Alfven/B\) is the work of \citet{2018PNAS..115..260D}, %
where the authors varied the parameter to reproduce the X-ray luminosity of an M-dwarf while driving the model with a scaled Solar minimum magnetogram.

In this work we do not vary the \(\Pi_\Alfven/B\) parameter as the \awsom{} model already reproduces Solar conditions with 
\(\Pi_\Alfven/B=\SI{1.1e6}{\watt\per\square\meter\per\tesla}\)
and 
\(\ell_\text{max}=15\), i.e.\ without small Solar features with an angular length scale smaller than \SI{12}{\degree} present in the magnetogram. Instead we have considered the assumption that the proportion between the large- and small-scale magnetic fields of the Hyades stars is such that the Solar \(\Pi_\Alfven/B\) value is still applicable, which has given good agreement with literature values. Varying \(\Pi_\Alfven/B\) instead of directly varying the magnetic field is an interesting area of open research which we should investigate further when e.g.\ X-ray measurements and ZDI maps are both available.

\subsubsection{Effect of unobservable stellar regions}
For all stellar inclinations \(i\) less than \SI{90}{\degree} there is a surface region that never rotates into view. This unobservable region consists of the colatitudes past \(i + \SI{90}{\degree}\), so that the observable fraction of the stellar surface is \((\sin i + 1)/2\). For the Hyades magnetograms \(i\) varies from \SIrange{46}{65}{\degree} (see \citetalias{2018MNRAS.474.4956F}), which means that the observable surface varies from \SIrange{86}{95}{\percent} of the total surface area.

The spherical harmonics representation ensures that the magnetic field in the unobservable region is smoothly connected to the field in the visible region, but the magnetic field strength in the unobserved regions is minimised by the entropy constraint employed to find finding a unique ZDI solution unless strong assumptions are made about the symmetry of the surface field.
The consequent reduction in detail at colatitudes past \SIrange{136}{155}{\degree} may be observed in the ZDI surface magnetic maps in Fig.~\ref{fig:magnetograms-zdi}. The average field strength \(|B_r|\) is also likely to be somewhat reduced, although the dipolar component should not be affected.

The lack of detail in the unobservable region can give rise to a corresponding lack of detail in the lower hemisphere of the wind model. We do not, however, see significant differences between the upper and lower hemispheres in Fig.~\ref{fig:mel-ur-alfven} and Fig.~\ref{fig:mel-br-fieldlines} so we conclude that both this effect and the reduction in \(|B_r|\) is small for the Hyades magnetograms where the effective \(\ell_\text{max}\sim 4\), as we saw in Section~\ref{sec:ZDI-missing-field}. We expect the two effects to be more significant for younger stars with more complicated magnetic field geometries.

\subsubsection{Negligibility of unseen binary companions}
About \SI{50}{\percent} of G and K type stars in the Hyades cluster are 
binaries~\citep{2007AJ....133.1903B}. %
Of the five stars in our sample (see Table~\ref{tab:observed_quantities}), Mel25-43 and Mel25-179 may be single-line spectroscopic binary stars~\citepalias{2018MNRAS.474.4956F}. Mel25-43 was found by \citet{1998A&A...331...81P} to be a single-line spectroscopic binary (SB1) and  Mel25-179 may also be an SB1 binary \citep{1998AJ....115.1972P}. In the work of \citetalias{2018MNRAS.474.4956F}, Mel25-5, Mel25-21 and Mel25-179 
exhibit luminosity values in excess of the cluster isochrone curve; this may also be an indication of binarity. For the two most likely binaries in our study, Mel25-43 and Mel25-179, we do not see any clear consequences in the magnetograms, wind models, or trends. 

For the surface magnetic field to be affected, the distance to the binary companion would have to be on the same order of magnitude as the star's Alfvén radius. Any such close companion would be noticeable in the star's radial velocity curve as part of the ZDI mapping as described in \citetalias{2018MNRAS.474.4956F}. 
For wider binaries we do not expect binary companions to affect the stellar surface magnetic fields. As we model single stars based on the surface magnetic field, only the effect on the surface magnetic field is relevant to our study. Therefore we consider it justified to model each of the stars in this study as single stars. For future studies with comparisons involving UV/X-ray luminosity and Lyman-\(\alpha\) emissions the contributions of a binary companion and interacting astrospheres may need to be considered.

\subsection{Magnetic cycles or permanent differences?}\label{sec:magnetic_cycles}
Given the small variation in rotation rates for the Hyades stars in general~\citep{2011MNRAS.413.2218D}, and 
the nearly identical stellar parameters in our sample 
(see Table~\ref{tab:observed_quantities}), we expect that 
the variation factor of \(\mysim2.4\) in angular momentum loss rates 
and 
the variation in mass loss rates
of the Hyades stars in our sample are the result of differences in the stellar magnetic field strength and geometry. 
Stars may harbour different dynamos even if their fundamental parameters are the same:
In a survey of ZDI magnetic data, \citet{2014MNRAS.441.2361V} found variations in \(|\vec B|\) of at least an order of magnitude at a Sun-like age; see also appendix D in \citetalias{2018MNRAS.474.4956F}. In our sample of five Hyades stars we see a variation of \(\mysim4\) in \(|B_r|\); we would expect a larger sample to exhibit a larger variation. 

While we attribute the variation in angular momentum loss to differences in the state of the stellar dynamo, we expect that the state of the stellar dynamo itself may vary on multiple timescales.
Observations have shown that the variability of stellar activity in general is greater than the Solar variability \citep{1995ApJ...438..269B}; it is also likely that variations exist on timescales longer than the stellar activity observational record of \(\mysim50\) years. 
Consequently, the variations in magnetic geometry seen in the Hyades may be caused by Solar cycle-like variations, and also by temporal variations of the stellar magnetic field 
on longer timescales. The Solar angular momentum loss has been reconstructed for the last \SI{e4}{\year} \citep{2019ApJ...883...67F}; for even longer timescales observations of stellar populations appears to be the only way to infer the magnitude of variations.
The small variation in observed rotational periods sets an upper limit of about \(\mysim 10^7\) years for persistent systematic differences in angular momentum loss, otherwise the stars would not have had sufficient time to switch several times between the low- and high-torque states.

\subsubsection{Torques inferred from the the Skumanich law}

The angular velocities of stars older than \(\mysim \SI{1}{\giga\year}\) tend to adhere to the Skumanich law \(\Omega \propto t^{-1/2}\), while younger stars may exhibit diverging rotational histories \citep{2013A&A...556A..36G}. Under certain mild assumptions the Skumanich law also imposes torque values: Assuming that the stellar moment of inertia \(I\) varies little over the time span when the Skumanich law applies, so that
\(|\Omega/\dot \Omega |\ll |I /\dot I |\), 
the angular velocity loss function \(\dot J(t)\) imposed by the Skumanich law \(\Omega \propto t^{-1/2}\) is 
\begin{equation}\label{eq:skumanich-dotJ}
    \dot J(t) = \left(t\middle/t_0\right)^{-3/2} \dot J_0 , 
    \quad 
    \dot J_0=\left. -I \Omega_0\middle/\left(2 t_0\right). \right.
\end{equation}
While more rapid rotators still exist at an age of \(\mysim\SI{0.6}{\giga\year}\) \citep{2011MNRAS.413.2218D}, which is the age of the Hyades, the stars are approaching the Skumanich age range. The Skumanich law predicts that the Sun's period should be 2.7 times greater than the period of the Hyades stars; this is well matched by the Hyades' rotational periods of \(\mysim\SI{10}{\day}\) (see Table~\ref{tab:observed_quantities}). 
Assuming that the moment of inertia for the Hyades stars is similar to the Solar moment of inertia \(I_\Sun=\SI{5.8e46}{\kilogram\square\meter}\) \citep{2000asqu.book.....C}, we find that the average \(\dot J\) for the Hyades should be \SI{1.1e25}{\newton\meter}, which is about twice as high as our \(\dot J\) values for the \(5B_\ZDI\) series.

Interestingly, the Solar \(\dot J\) value from equation~\eqref{eq:skumanich-dotJ} is  \(\mysim 3\)  %
times greater than the Sun's recent observed \(\dot J\) values \citep{2019ApJ...885L..30F}. The
Skumanich law may have an upper age limit around the age of the Sun, as older stars have been observed to rotate more rapidly, and thus have smaller \(\dot J\) values, than suggested by the Skumanich law, see \citet{2016Natur.529..181V, 2021arXiv210410919H}. Alternatively, the Sun could be in a `low-torque state', see \citet{2019ApJ...883...67F} and the discussion in \cite{2018ApJ...864..125F}.

In conclusion, the Skumanich law suggests a magnetic scale factor of \(\mysim10\). However, given the mismatch between observed Solar \(\dot J\) values and the Solar \(\dot J\) values predicted by the Skumanich law, we consider that a magnetic scale factor of 5 is reasonable and provides good comparison between Solar and stellar cases.

\subsection{The young Solar system}\label{sec:young-solar-system}

At the age of the Hyades we expect the Sun to have had a higher rate of rotation and a stronger magnetic field and, as we have seen, a stronger Solar wind. While the overall luminosity should be reduced, the X-ray and UV flux is expected to be \( \mysim 40\) and \( \mysim 10\) times higher than present day values \citep{2007LRSP....4....3G}. This can heat the planet's outer atmosphere, which again can enhance atmospheric evaporation and erosion due to Solar/stellar wind.  
Flare activity is also expected to be significantly higher for the Sun at the age of the Hyades.
All these factors would have affected the atmosphere and radiative environment for the Earth and the other Solar system planets.

In this paper we have considered an Earth-like planet with a modern day Earth-like magnetic field. The earliest records of a modern-day magnetic field on ancient Earth are however from a Solar system age of \SI{1}{\giga\year}, when the magnetic field may have been \SI{50}{\percent} of its current-day value \citep{2010Sci...327.1238T}. Such a
reduction in the planetary magnetic field strength would reduce the magnetospheric stand-off distance by \SI{20}{\percent} according to equation~\eqref{eq:magnetospheric_stand_off_distance}; this does not significantly affect the findings in this section. The atmosphere of the early Earth would also have been different from today's atmosphere; this is however beyond the scope of this work.

\subsubsection{Ram pressure and magnetospheric stand-off distance}
The wind pressure and magnetospheric stand-off distance calculations in Section~\ref{sec:model-derived} are not drastically different from the current day Solar values. With the wind pressure \(P_\Wind^\Earth\) reaching values of \( \mysim  10\) times the Solar value for our most active member of the \(5B_\ZDI\) series, the ram pressure the magnetospheric stand-off distance for an Earth-like planet is reduced from \( \mysim  10 R_\Earth\) to \( \mysim  7 R_\Earth\); this shows the efficacy with which the Earth's magnetic field protects against the Solar wind. In their model with very powerful winds (orange square in Fig.~\ref{fig:results-context}), \citet{2016ApJ...820L..15D} still found a magnetospheric stand-off distance of \(5 R_\Earth\); all of these values exceed the \(2 R_\Earth\) threshold for atmospheric erosion proposed by \citet{2007AsBio...7..185L}, indicating that atmospheric erosion is not occurring at significant rates for Earth-like planets in the Hyades. This view is supported by the detailed MHD simulations of atmospheric erosion by stellar wind by \citet{2017ApJ...837L..26D,2018PNAS..115..260D}, which showed that a strong Earth-like magnetic field can protect against erosion even in the potentially hostile space environment surrounding an M-dwarf, but that without this magnetic protection an Earth-like planet would lose its atmosphere in \(<\SI{1}{\giga\year}\).

The auroral opening for a planet with an Earth-like magnetic field has been estimated by \citet{2011MNRAS.414.1573V}. For a spin-aligned field 
the colatitude \(\alpha_0\) of the ring separating open and closed field lines would be given by
        \(
        \alpha_0 = \arcsin 
            \sqrt{R_\text{p}/R_\text{m}}
        \).
The value of \(\alpha_0\) ranges from about \SI{27}{\degree} to \SI{15}{\degree} for values of \(R_\text{m}/R_\text{p}\) from 5 to 15. The auroral openings can permit energetic particles into the planetary atmosphere and affect its composition \citep{2016NatGe...9..452A}, aiding in nitrogen fixation and creating greenhouse gases. The increased angular diameter of the auroral opening would increase the effect of these processes.

\subsubsection{Axisymmetric open flux and cosmic ray intensity}\label{sec:cosmic-ray-intensity}
The variation of galactic cosmic ray flux observed at the Earth and by interplanetary spacecraft is correlated with the Sun's axisymmetric open magnetic flux \(\Phi_\text{axi}\) \citep{2006ApJ...644..638W}. This is thought to be through the formation of 
shell-like global, merged CIRs (see Fig.~\ref{fig:Sun-Pwind-IH-equatorial} for a region where two CIRs interact) in the outer heliosphere; these regions of higher magnetic field strength can scatter incoming cosmic rays. 

There is a strong correlations in our data between \(\Phi_\text{axi}/\Phi_\text{open}\) and \(\cos i_{B_r=0}\) (see Table~\ref{tab:correlations}).
This shows that the current sheet inclination is a good proxy for \(\Phi_\text{axi}/\Phi_\text{open}\) in our models, despite being calculated much closer to the stellar surface. It is also of note that when compared with the empirical relations of Table~\ref{tab:correlations} both the Solar models exhibit an axisymmetry excess of \( \mysim  0.1\); hence we expect that the relations will be less reliable for complicated Sun-like magnetic fields.

Although we have not carried out any multiple regressions it seems justified that \(\Phi_\text{axi}\propto \Phi_\text{open}  \cos i_{B_r=0}\); the axisymmetric flux is proportional to the open flux, and also modulated by the current sheet inclination. The amount of cosmic ray shielding would thus be affected by the surface field strength as \(\Phi_\text{axi}\propto B_r\) and be modulated by \(i_{B_r=0}\).

A full treatment of cosmic ray scattering would require a time-dependent wind model including CMEs in order to model the entrainment and formation of global merged interaction regions in the astrosphere, and a model of cosmic ray scattering from the outer reaches of the astrosphere and inwards. This is an area of active research. Using 3D wind models and a 2D model of cosmic ray propagation, \citet{2012ApJ...760...85C} found that the amount of cosmic rays would be reduced by a factor of 100 or more for rotation rates of 6--15 days and a 10 times stronger small-scale magnetic field. 
More recently 
\citet{2020MNRAS.499.2124R} found an attenuation of an order of magnitude for \(P_\text{rot}=\SI{9}{\day}\) and \(|\vec B| = \SI{5.5}{\gauss}\) using a one-dimensional model of cosmic ray propagation. 

While the magnetic field of a planet shields the surface from galactic cosmic rays, \citet{2009Icar..199..526G,2015A&A...581A..44G} found that the compression of the magnetosphere does not modulate the shielding effect in the steady state, instead the stronger curvature of the magnetic field compensates for the reduction in \(R_\text{m}\).  It has however been proposed that transient wind phenomena may open the Earth's magnetosphere to cosmic ray flux \citep[see the discussion in][]{PhysRevLett.117.171101,2017ICRC...35..133E,2018arXiv180310499M}. A non-steady wind would then contribute to shielding in the outer astrosphere, but could reduce shielding when disturbing the magnetosphere of an Earth-like planet.

A truly cataclysmic test of the power of the astrosphere would be a nearby supernova event. The study of \citet{2008ApJ...678..549F} found that passage through a supernova remnant could reduce the size of the heliosphere to \SI{1}{\astronomicalunit{}}. While this is an extreme example, it highlights how transient phenomena can easily dominate over steady state processes. This is should also be investigated further for young Solar-type stars. 
In the Solar case coronal mass ejection (CME) mass losses account for a few percent of the wind mass loss \citep{2010ApJ...722.1522V}.
Linking CME activity and flare activity, \citet{2013ApJ...764..170D} suggested that mass loss from active stars on the order of \(150 \dot M_\Sun\) may be dominated by transient CMEs rather than the steady state wind. 
The CME matter would be indistinguishable from the steady wind matter in the outer astrosphere where the Lyman-\(\alpha\) emissions of \citet{2014ApJ...781L..33W} originate; this could be proposed as a partial explanation of lower wind mass loss estimates from some wind models, but does not explain the presence of the wind dividing line. This phenomenon could however be caused by strong dipolar magnetic fields inhibiting the escape of CME material~\citep{2018ApJ...862...93A}.  For CME-dominated winds we would expect that the wind's influence on atmospheric erosion and cosmic ray shielding may also be dominated by the transient CMEs, challenging the finality of steady state wind models.

\section{Conclusions}\label{sec:Conclusions}
We have started out with a desire to study the space weather of the Sun at \SI{0.6}{\giga\year} using magnetogram data from the Hyades cluster. This is close to the age when life first arose on the Earth \citep{1996Natur.384...55M}. We have seen that even though the Solar system would have been a different, more active place when the Sun's age was \SI{0.6}{\giga\year} as the Hyades stars are today, the steady state wind pressure would not significantly hinder a young Earth from holding onto its atmosphere and shielding it from the Solar wind.

As the Solar-type stars in the Hyades cluster are of similar age and provenance, we have been able to directly study the effect of magnetic field strength and complexity on key parameters such as mass loss, angular momentum loss and wind pressure over a range of magnetic field strength values. We have studied the effect of the ZDI underreporting of absolute magnetic field strength on these parameters by considering two series of models in which the magnetic field differs by a scaling factor of five. 

In terms of power law trends with \(|B_r|\), we have found strong correlations between the open flux, mass loss, angular momentum loss, and the magnetogram scaling factors of 1 and 5 in the \(B_\ZDI\) and \(5B_\ZDI\) series of wind models. For the open magnetic flux and the mass loss values, there is also residual variation of \SI{50}{\percent} that we ascribe to differences in the shape and complexity of the magnetic field independent of the field strength. For the angular momentum loss this residual variation is \SI{140}{\percent}.

In order to validate our results we have compared our mass and angular momentum loss values to Solar values as well as a number of previously published studies (Fig.~\ref{fig:results-context}). We find general agreement with observed Solar values, with the caveat that our range of values may be greater than the observed differences between Solar minimum and Solar maximum.
We find good agreement between our results and other Alfvén wave driven models. Ideal MHD models with hot coronae have predicted higher values of both \(\dot M\) and \(\dot J\). 

The mass loss panel in Fig.~\ref{fig:results-context} give signs of saturation of the steady mass flux at a rate of \(10 \dot M_\Sun\) which is in agreement with the work of \citet{2007A&A...463...11H} but low compared to many other studies and the Lyman-\(\alpha\) observations of \citet{2014ApJ...781L..33W}. A possible resolution to this conflict could be that the wind mass loss of young Solar-type stars are dominated by transient phenomena such as CMEs rather than the steady state wind loss. 

\subsection{Future directions}

It would be interesting to extend this study by incorporating models of younger stars such as the ones in the Hercules-Lyra association and the Pleiades cluster. These stars will have shorter rotation periods as well as stronger magnetic fields, but will still be slow magnetic rotators \citep{1976ApJ...210..498B}. Going towards younger stars, we expect to see more dramatic differences to the Solar models and more energetic and dynamic space weather. Frequent coronal mass ejections may turn out to be the dominating wind loss mechanism, unlike the current-day Sun where CMEs only account for a few percent of the total wind loss. A CME dominated wind mass loss could explain the high mass loss values of \citet{2014ApJ...781L..33W}. The role photospheric currents 
\citep{2016MNRAS.455L..52V} past the wind dividing line play in powering flare and CME activity should be investigated.

The modelling of younger stars may lead to a greater need to modify \awsom{} parameters such as the Poynting flux-to-field ratio from the Solar default values. Comparisons with complementary measurements such as Zeeman broadening, X-ray and UV flux, and the Skumanich law could help inform these choices.

\section*{Acknowledgements}
    DE is funded by a University of Southern Queensland 
    (USQ) International Stipend Research Scholarship (ISRS) and a
    USQ International Fees Research Scholarship (IFRS).
    This research was undertaken using the University of Southern Queensland (USQ) Fawkes HPC which is co-sponsored by the Queensland Cyber Infrastructure Foundation (QCIF), see~\url{www.usq.edu.au/hpc}. 

    This work utilises data obtained by the \href{https://gong.nso.edu/}{Global Oscillation Network Group (GONG)} Program, managed by the National Solar Observatory, which is operated by AURA, Inc.\ under a cooperative agreement with the National Science Foundation. The data were acquired by instruments operated by the Big Bear Solar Observatory, High Altitude Observatory, Learmonth Solar Observatory, Udaipur Solar Observatory, 
    Instituto de Astrofísica de Canarias, 
    and Cerro Tololo Inter-American Observatory.
    This work has made use of the \href{http://vald.astro.uu.se/}{Vienna Atomic Line Database (VALD)}, operated at Uppsala University, the Institute of Astronomy RAS in Moscow, and the University of Vienna.
    We acknowledge the use of the \href{http://simbad.u-strasbg.fr/simbad/}{SIMBAD database}. 
    This research has made use of the VizieR catalogue access tool, CDS, Strasbourg, France (\href{https://www.doi.org/10.26093/cds/vizier}{DOI:\@ 10.26093/cds/vizier}). The original description of the VizieR service was published in~\defcitealias{2000A&AS..143...23O}{A\&AS~143,~23}\citetalias{2000A&AS..143...23O}.
    This research has made use of NASA's \href{https://ui.adsabs.harvard.edu/}{Astrophysics Data System}.
    We acknowledge use of NASA/GSFC's Space Physics Data Facility's OMNIWeb service, and OMNI data.

    This work was carried out using the \swmf{} tools developed at The University of Michigan \href{https://spaceweather.engin.umich.edu/the-center-for-space-environment-modelling-csem/}{Center for Space Environment Modelling (CSEM)} and made available through the NASA \href{https://ccmc.gsfc.nasa.gov/}{Community Coordinated Modelling Center (CCMC)}.
    This work has made use of the following additional numerical software and visualisation software: 
    NumPy     ~\citep{2011CSE....13b..22V},
    SciPy     ~\citep{2020SciPy-NMeth},
    Matplotlib~\citep{2007CSE.....9...90H},
    Tecplot and PyTecplot.

\section*{Data availability}
The data underlying this article will be shared on reasonable request
to the corresponding author.

\bibliographystyle{mnras}
\bibliography{bibliography} %

\appendix
\section{Statistical correlations}\label{sec:correlations}
In Section~\ref{sec:mag_scale_effects} we considered the effect of magnetic scaling by comparing the unscaled \(B_\ZDI\) wind models and the scaled \(5B_\ZDI\) models for each star. In this appendix we follow a complementary approach in which we look for power law relations in the \(B_\ZDI\) series of models and the \(5B_\ZDI\) series of models separately, followed by the analysis of a pooled series comprising both the \(B_\ZDI\) models and the \(5B_\ZDI\) models. 

As the Hyades stars in this work are all quite similar in terms of mass, radius, and period of rotation as given in Table~\ref{tab:observed_quantities}, this analysis can shed light on the isolated effect of magnetic field strength, and disentangle it from the effect of variations in magnetic field complexity in a sample of stars where all the fundamental parameters are similar. 

Unlike the analysis in Section~\ref{sec:mag_scale_effects}, this analysis requires the standard requirements of ordinary least square (OLS) regression to be met, see e.g. \citet{1998ara..book.....D}. 
With the caveat that the five data points in the \(B_\ZDI\) and \(5B_\ZDI\) series are on the low side, we observe that when log-transforming our data the assumptions of homoscedasticity (magnitude of residuals independent of the magnitude of the independent variable) and normality (normal distribution of residuals) are met in our sample. 

\begin{figure*}
    \centering
    \includegraphics{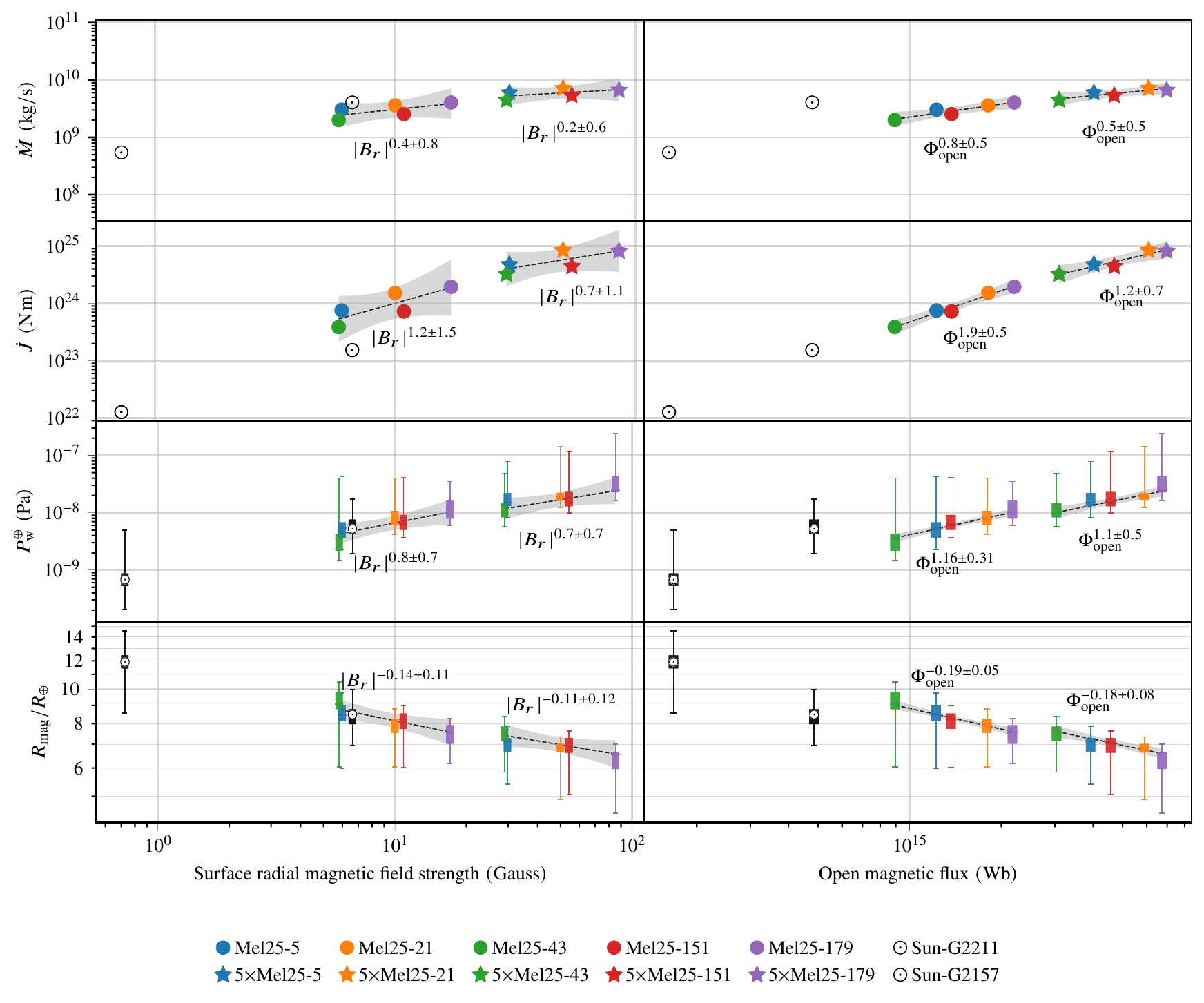}

    \caption{
    Mass loss, angular momentum loss, and total wind pressure at \SI{1}{\astronomicalunit} plotted the radial magnetic field values (left) and the open magnetic flux values (right).
    The dashed lines represent power law fits to the \(B_\ZDI\) and \(5B_\ZDI\) series with shaded \SI{95}{\percent} confidence bands, calculated using ordinary least squares (OLS) regression, the results of which are tabulated in Table~\ref{tab:correlations}. Subject to the usual assumptions of OLS, there is a \SI{95}{\percent} chance that the true power law lies inside the shaded region. The widening of the confidence bands away from the midpoint of the \(B_\ZDI\) values reflects the lack of data in that region, rather than an increase in the effect of magnetic geometry.  The Sun values are represented by the Sun symbol `$\Sun$'; the left symbol is the value at Solar minimum. 
    In the wind pressure panel, we use boxplots to indicate the variation with orbital phase and Solar rotation in each model. 
    When comparing the left column and the right column, it is noticeable that while Mel25-5 and Mel25-43 have very similar \(|B_r|\) values (differing by \SI{3}{\percent}), while the two \(\Phi_\text{open}\) values differ by about \SI{50}{\percent}. Similarly, Mel21-5 and Mel25-151 differ by \SI{10}{\percent} in \(|B_r|\), while the \(\Phi_\text{open}\) values differ by about \SI{50}{\percent} as well. By plotting against \(\Phi_\text{open}\) the effect of the magnetic field geometry is accounted for, leading to a more regular spread of values on the horizontal axis. 
    Compared to the confidence bands when fitting to \(|B_r|\) (left column), the confidence bands are significantly tighter for the when fitting to the open magnetic flux \(\Phi_\text{open}\) (right column), indicating that the open magnetic flux \(\Phi_\text{open}\) has superior predictive power than \(|B_r|\).
    The fitted power law exponents and the associated statistical information can also be found in Table~\ref{tab:correlations}.
    }\label{fig:dotmdotjpwind_broken_trend}
\end{figure*}
\begin{figure*}
    \centering
    \includegraphics{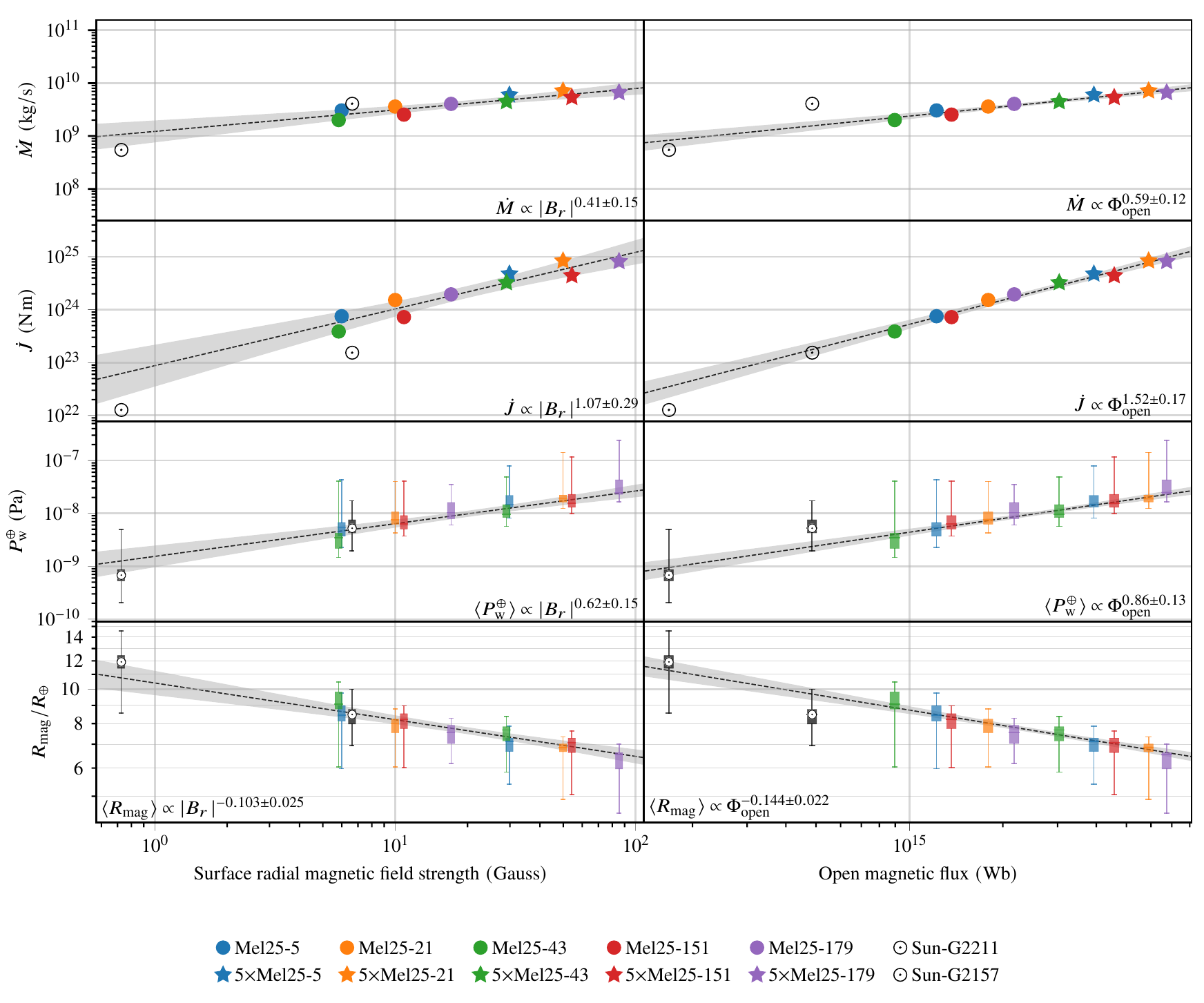}
    \caption{
    This plot is similar to Fig.~\ref{fig:dotmdotjpwind_broken_trend}, except that the \(B_\ZDI\) and \(5B_\ZDI\) series are treated as a single population. By pooling the data in this way, clear trends appear for each of the four parameters. 
    The symbols in this figure are the same as in Fig.~\ref{fig:dotmdotjpwind_broken_trend}.
    In each panel, the dashed line and shaded region represents a log-log fit and \SI{95}{\percent} confidence bands; the corresponding coefficients are displayed inside each panel. For the wind pressure and magnetospheric stand-off distance, the boxplot represents the variation of the wind in the region of the Earth's orbit as described in Section~\ref{sec:wind-pressure-planets}. 
    The Solar minimum and Solar maximum are both represented by the Sun symbol `\(\Sun\)'; the Solar minimum is left of the Solar maximum value as \(|B_r|\) is smaller at Solar minimum. 
    In the two bottom panes we use boxplots to indicate the variation with orbital phase and Solar rotation in each model.
    Comparing the left column and the right column of panels, it is evident that the confidence bands are tighter when fitting to \(\Phi_\text{open}\) than when fitting to \(|B_r|\); the effect is most pronounced for the angular momentum loss \(\dot J\). The fitted power law exponents and the associated statistical information can also be found in Table~\ref{tab:correlations}.
    }\label{fig:dotmdotjpwind_trend} 
\end{figure*}
\begin{table*}
    \centering
    \caption{
        Strong predictive power of open flux on angular momentum loss and wind pressure: 
        This table shows log-log correlation with \(|B_r|\) (top part) and \(\Phi_\text{open}\) (middle part) for the \(B_\ZDI\) series, \(5B_\ZDI\) series, and a pooled series. 
        In the \(a\) columns we give the exponent in the fitted power laws and a \SI{95}{\percent} confidence interval around the central value. The coefficients of determination (\(r^2\)values) are provided in the \(r^2\) columns, and the \(p\) values are provided in the \(p\) columns.
        It is notable that \(\dot J\), \(P_\Wind^\Earth\) and \(R_\text{mag}\), are all formally correlated with \(\Phi_\text{open}\) at \SI{5}{\percent} or \SI{1}{\percent} significance. The mass loss value is also weakly correlated with \(\Phi_\text{open}\). Meanwhile the magnetic quantities and Alfvén radius quantities are more strongly correlated with \(|B_r|\) than with \(\Phi_\text{open}\).
        The bottom part shows the correlation between the axisymmetric open flux fraction and the cosine of the current sheet inclination. 
    }\label{tab:correlations}
    \sisetup{
        table-figures-decimal=2,
        table-figures-integer=1,
        table-figures-uncertainty=2,
        table-number-alignment=center,
        add-decimal-zero=false,
        add-integer-zero = false,
        omit-uncertainty = false
        }

    \begin{tabular}{
        l
        S
        S[
            table-figures-decimal=3,
            table-figures-integer=1,
            table-figures-uncertainty=0,
            table-figures-exponent = 0
        ]
        S[
            table-figures-decimal=1,
            table-figures-integer=1,
            table-figures-uncertainty=0,
            table-figures-exponent = 3
        ]
        S
        S[
            table-figures-decimal=3,
            table-figures-integer=1,
            table-figures-uncertainty=0,
            table-figures-exponent = 0
        ]
        S[
            table-figures-decimal=1,
            table-figures-integer=1,
            table-figures-uncertainty=0,
            table-figures-exponent = 3
        ]
        S
        S[
            table-figures-decimal=3,
            table-figures-integer=1,
            table-figures-uncertainty=0,
            table-figures-exponent = 0
        ]
        S[
            table-figures-decimal=1,
            table-figures-integer=1,
            table-figures-uncertainty=0,
            table-figures-exponent = 3
        ]
        }
        \toprule
        {Quantity} & \multicolumn{9}{c}{Correlation with $a\log_{10}|B_r|+b$} \\
        \cmidrule(lr){2-10}
        {} & \multicolumn{3}{c}{$B_\ZDI$ series}& \multicolumn{3}{c}{$5B_\ZDI$ series}& \multicolumn{3}{c}{Pooled series}\\
        \cmidrule(lr){2-4}\cmidrule(lr){5-7}\cmidrule(lr){8-10}
        {} & 
        {$a$} & {$r^2$} & {$p$} &
        {$a$} & {$r^2$} & {$p$} &
        {$a$} & {$r^2$} & {$p$} \\
        \midrule
$\log_{10}\max|B_r|$                          &   1.09 \pm 0.74 &  0.881& 1.8e-02&   1.09 \pm 0.73 &  0.881& 1.8e-02&   1.02 \pm 0.15 &  0.969& 2.6e-07\\
$\log_{10}|\vec{B}|$                          &   0.96 \pm 0.10 &  0.997& 8.6e-05&   1.00 \pm 0.10 &  0.997& 6.2e-05&   0.98 \pm 0.02 &  0.999& 7.7e-14\\
$\log_{10}\Phi_0$                             &   0.98 \pm 0.50 &  0.929& 8.2e-03&   0.98 \pm 0.50 &  0.929& 8.2e-03&   1.00 \pm 0.10 &  0.985& 1.3e-08\\
$\log_{10}\Phi_\text{open}$                   &   0.68 \pm 0.63 &  0.795& 4.2e-02&   0.63 \pm 0.57 &  0.805& 3.9e-02&   0.72 \pm 0.12 &  0.958& 8.9e-07\\
$\log_{10} R_\Alfven$                         &   0.39 \pm 0.13 &  0.968& 2.5e-03&   0.26 \pm 0.19 &  0.870& 2.1e-02&   0.40 \pm 0.05 &  0.978& 6.6e-08\\
$\log_{10} |\vec{r_\Alfven} \times \uvec{z}|$ &   0.38 \pm 0.15 &  0.958& 3.8e-03&   0.25 \pm 0.20 &  0.842& 2.8e-02&   0.39 \pm 0.05 &  0.974& 1.3e-07\\
$\log_{10}\dot M$                             &   0.43 \pm 0.82 &  0.484& 1.9e-01&   0.24 \pm 0.61 &  0.333& 3.1e-01&   0.41 \pm 0.15 &  0.827& 2.7e-04\\
$\log_{10}\dot J$                             &   1.17 \pm 1.52 &  0.667& 9.2e-02&   0.67 \pm 1.14 &  0.539& 1.6e-01&   1.07 \pm 0.29 &  0.901& 2.7e-05\\
$\log_{10}P_\text{wind}^\Earth$               &   0.82 \pm 0.67 &  0.832& 3.1e-02&   0.67 \pm 0.71 &  0.748& 5.9e-02&   0.62 \pm 0.15 &  0.919& 1.2e-05\\
$\log_{10} R_\text{mag}$                      &  -0.14 \pm 0.11 &  0.832& 3.1e-02&  -0.11 \pm 0.12 &  0.748& 5.9e-02&  -0.10 \pm 0.02 &  0.919& 1.2e-05\\
        \midrule
        {} & \multicolumn{9}{c}{Correlation with $a\log_{10}\Phi_\text{open}+b$} \\
        \cmidrule(lr){2-10}
        {} & \multicolumn{3}{c}{$B_\ZDI$ series}& \multicolumn{3}{c}{$5B_\ZDI$ series}& \multicolumn{3}{c}{Pooled series}\\
        \cmidrule(lr){2-4}\cmidrule(lr){5-7}\cmidrule(lr){8-10}
        {} & 
        {$a$} & {$r^2$} & {$p$} &
        {$a$} & {$r^2$} & {$p$} &
        {$a$} & {$r^2$} & {$p$} \\
        \midrule
$\log_{10}|B_r|$                              &   1.17 \pm 1.09 &  0.795& 4.2e-02&   1.27 \pm 1.15 &  0.805& 3.9e-02&   1.33 \pm 0.23 &  0.958& 8.9e-07\\
$\log_{10}\max|B_r|$                          &   1.28 \pm 1.51 &  0.708& 7.4e-02&   1.41 \pm 1.55 &  0.737& 6.3e-02&   1.36 \pm 0.30 &  0.929& 7.0e-06\\
$\log_{10}|\vec{B}|$                          &   1.09 \pm 1.14 &  0.756& 5.6e-02&   1.26 \pm 1.21 &  0.786& 4.5e-02&   1.30 \pm 0.24 &  0.951& 1.6e-06\\
$\log_{10}\Phi_0$                             &   1.30 \pm 0.53 &  0.953& 4.4e-03&   1.41 \pm 0.55 &  0.957& 3.9e-03&   1.36 \pm 0.11 &  0.990& 2.4e-09\\
$\log_{10} R_\Alfven$                         &   0.46 \pm 0.45 &  0.779& 4.7e-02&   0.32 \pm 0.44 &  0.650& 9.9e-02&   0.53 \pm 0.11 &  0.943& 3.0e-06\\
$\log_{10} |\vec{r_\Alfven} \times \uvec{z}|$ &   0.44 \pm 0.46 &  0.759& 5.4e-02&   0.30 \pm 0.45 &  0.606& 1.2e-01&   0.53 \pm 0.11 &  0.937& 4.5e-06\\
$\log_{10}\dot M$                             &   0.76 \pm 0.52 &  0.879& 1.8e-02&   0.51 \pm 0.51 &  0.769& 5.1e-02&   0.59 \pm 0.12 &  0.946& 2.4e-06\\
$\log_{10}\dot J$                             &   1.86 \pm 0.53 &  0.976& 1.6e-03&   1.24 \pm 0.73 &  0.907& 1.2e-02&   1.52 \pm 0.17 &  0.982& 3.2e-08\\
$\log_{10}P_\text{wind}^\Earth$               &   1.16 \pm 0.31 &  0.979& 1.3e-03&   1.07 \pm 0.46 &  0.948& 5.1e-03&   0.86 \pm 0.13 &  0.967& 3.5e-07\\
$\log_{10} R_\text{mag}$                      &  -0.19 \pm 0.05 &  0.979& 1.3e-03&  -0.18 \pm 0.08 &  0.948& 5.1e-03&  -0.14 \pm 0.02 &  0.967& 3.5e-07\\
        \midrule
        {} & \multicolumn{9}{c}{Correlation with $a \cos i_{B_r=0} + b$} \\
        \cmidrule(lr){2-10}
        {} & \multicolumn{3}{c}{$B_\ZDI$ series}& \multicolumn{3}{c}{$5B_\ZDI$ series}& \multicolumn{3}{c}{Pooled series}\\
        \cmidrule(lr){2-4}\cmidrule(lr){5-7}\cmidrule(lr){8-10}
        {} & 
        {$a$} & {$r^2$} & {$p$} &
        {$a$} & {$r^2$} & {$p$} &
        {$a$} & {$r^2$} & {$p$} \\
        \midrule
        $\Phi_\text{axi} / \Phi_\text{open}$ & 
0.77 \pm   0.07 & 0.998 & 5.3e-05 &
0.71 \pm   0.09 & 0.996 & 1.2e-04 &
0.74 \pm   0.06 & 0.990 & 3.2e-09 
        \\
        \bottomrule
    \end{tabular}
\end{table*}

\subsection{\(B_\ZDI\) and \(5B_\ZDI\) series separately}

    In Fig.~\ref{fig:dotmdotjpwind_broken_trend} we provide log-log plots of the 
    wind mass loss \(\dot M\), 
    angular momentum loss \(\dot J\), 
    wind pressure for an Earth-like planet \(P_\Wind^\Earth\), 
    and 
    magnetospheric stand-off distance for an Earth-like planet \(R_\text{mag}/R_\Earth\)
    against 
    the average absolute radial surface magnetic field \(|B_r|\) 
    and 
    the open magnetic flux \(\Phi_\text{open}\). 

We fit power laws on the form \(y(x)\propto x^{a}\) to the results. This requires fitting a curve with equation
\begin{equation}\label{eq:loglog-fit}
    y(x) = b x^{a} \text{ so that } \log_{10} y(x) = \log_{10} b + a \log_{10} x
\end{equation}
to the set of model values. 

In the scaling laws in Fig.~\ref{fig:dotmdotjpwind_broken_trend}, such as \(\dot M \propto |B_r|^{0.4\pm0.8}\) for the \(B_\ZDI\) series in the top left panel, the \(\pm 0.8\) term represents \SI{95}{\percent} confidence intervals on the fitted parameter exponent \(a\), i.e, there is only a \SI{5}{\percent} chance that the true value of \(a\) lies outside the range \(0.4\pm0.8\), given that the magnitude of the residuals are fixed, the residuals are normally distributed, and the other requirements of OLS regression are met, see e.g.\ \citet{1998ara..book.....D}. The shaded regions in Fig.~\ref{fig:dotmdotjpwind_broken_trend} are \SI{95}{\percent} confidence bands on the whole fitted curve \(y(x) = b x^{a}\); they are a visual way of representing the cumulative effect of the uncertainty in both the \(a\) and \(b\) parameters in equation \eqref{eq:loglog-fit}. Subject to the same assumptions as the confidence intervals on \(a\), there is only a \SI{5}{\percent} chance that the true curve law lies outside the shaded regions. 

While all the fitted lines increase with increasing \(|B_r|\), it is clear from the fitted coefficients and confidence intervals that no trends in \(\dot M\) and \(\dot J\) may be inferred at \SI{95}{\percent} significance level from the data points in one individual series, as the confidence intervals on \(a\) include zero. Notably, Mel25-5 and Mel25-43 have very similar average field strengths (\SI{8.8}{\gauss} vs \SI{8.9}{\gauss}) but differ by a factor of \SI{50}{\percent} in \(\dot M\) (see the \(B_\ZDI\) series in Table~\ref{tab:main-table}). The exponent values in \(P_\Wind^\Earth\) and \(R_\text{mag}/R\) also include zero for the \(5B_\ZDI\) series.

Although the Sun differs from the Hyades stars in terms of age and period of rotation, the Solar values fall inside the \SI{95}{\percent} confidence band for the \(B_\ZDI\) series for \(\dot M\) and \(P_\Wind^\Earth\) when plotted against \(|B_r|\). 
The Solar maximum value of \(\dot J = \SI{1.5e23}{\newton\meter}\) falls outside the \SI{95}{\percent} confidence interval 
\SIrange{3.9e23}{1.6e24}{\newton\meter}; a lower value is however expected as angular momentum loss is roughly proportional to the stellar rate of rotation as in equation~\eqref{eq:angmom_loss} and one-dimensional models such as the one of \citet{1967ApJ...148..217W}. 

The lack of clear trends in the \(B_\ZDI\) and \(5B_\ZDI\) with \(B_r\) highlights that further variables are needed to explain differences we observe in our simulations. The Mel25-5 and Mel25-21 cases lie above the trend lines in Fig.~\ref{fig:dotmdotjpwind_broken_trend}, while Mel25-43 and Mel25-151 lie under the trend lines. The variation in radius between the stars is expected to explain some of this variance, as is variation in magnetic field geometry. 

The right column in Fig.~\ref{fig:dotmdotjpwind_broken_trend} show the same data as in the left column, but plotted against the open magnetic flux rather than the average radial field strength. The tighter correlations of \(\dot J\) and \(\dot M\) with \(\Phi_\text{open}\) in Fig.~\ref{fig:dotmdotjpwind_broken_trend} appear to support the argument of \citet{2015ApJ...798..116R} that rotational braking laws are better formulated in terms of the open flux than in terms of absolute field strength. When plotted against \(\Phi_\text{open}\), the models in either series spread out well on the \(x\) axis, in contrast to the similar \(|B_r|\) values of Mel25-5 and Mel25-43 seen in the left column.

\subsection{Pooled series of \(B_\ZDI\) and \(5B_\ZDI\) data}

In Fig.~\ref{fig:dotmdotjpwind_trend} we plot the pooled data of both our series \(B_\ZDI\) and \(5B_\ZDI\). In this way we are no longer directly representing the five stars in Table~\ref{tab:observed_quantities}, instead the pooled series should be thought of as a parameter study where the range of magnetic field strengths is extended to \SIrange{8.9}{120}{\gauss}, and where the sample size is 10 instead of 5. 
In each panel the dashed lines and shaded regions represent power law fits and \SI{95}{\percent} confidence intervals of the power law coefficients. The Sun symbols represent the Solar models at Solar maximum/minimum; we emphasise that these two models are excluded from the fit. 

For a plot with this range in \(|B_r|\) or \(\Phi_\text{open}\) values it is important to note that the models all have similar rates of rotation and ages. This is different from the stellar samples in the studies of \citet{2015MNRAS.449.4117V} and \citet{2018ApJ...856...53P} where rotation period, age, and magnetic field strength are themselves correlated due to the Skumanich law and the relation between rotation rates and magnetic field strengths \citep{2018haex.bookE..26V}. 

Unlike in Fig.~\ref{fig:dotmdotjpwind_broken_trend}, we do not see significantly tighter correlations with \(\Phi_\text{open}\) than with \(|B_r|\) in Fig~\ref{fig:dotmdotjpwind_trend} except for \(\dot J\). The larger number of data points in the pooled series thus appears to improve the correlations in the \(|B_r|\) series nearly to those of the \(\Phi_\text{open}\) series. More data points (i.e. more stellar wind models) would be required to investigate this further.

\subsection{
    Full correlations table
 }\label{sec:Correlations-with-br-and-phi}

Table~\ref{tab:correlations} shows the full set of calculated power law exponents, some of which have been given in Fig.~\ref{fig:dotmdotjpwind_broken_trend} and Fig.~\ref{fig:dotmdotjpwind_trend}. The table contains power law fits to all the quantities in Table~\ref{tab:magnetic_averages} and Table~\ref{tab:main-table} where fitting a power law is meaningful. 
In addition to 
\SI{95}{\percent} confidence intervals for the exponents \(a\) of the fitted power laws (see equation~\ref{eq:loglog-fit}) 
we provide
the \(p\) values associated with the \(a\) values
and 
the coefficients of determination (\(r^2\)-values) of the fitted power laws.
The \(p\) value represents the level of confidence in the observed trend;
the smaller \(p\), the greater the likelihood that the observed trend in the data will persist for a larger set of wind models. Note that the \SI{95}{\percent} confidence interval for \(a\) includes zero exactly when \(p \geq \SI{5}{\percent}\). The \(r^2\) values represent the proportion of variance explained by \(|B_r|\) or \(\Phi_\text{open}\). Both the \(p\) values and the \(r^2\) values of the fitted curve are calculated in the standard way \citep[see e.g.\@][]{1998ara..book.....D}. From the table one may draw the following tentative conclusions:
\begin{enumerate}

    \item From the upper part of the Table~\ref{tab:correlations}, it is clear that there are significant correlations between \(|B_r|\), \(|\vec B|\) and \(\max |B_r|\) in each of the series \(B_\ZDI\) and \(5B_\ZDI\), as well as between \(|B_r|\) and the average Alfvén distance \(R_\Alfven\), the torque-averaged Alfvén distance \(|\vec{r_\Alfven} \times \uvec{z}|\), and the open magnetic flux \(\Phi_\text{open}\).

    \item The \(p\) values are smaller in the \(B_\ZDI\) series than in the \(5B_\ZDI\) series, possibly indicating that the scaled magnetic field increases the importance of the magnetic geometry. This would be interesting to investigate by including stars with stronger magnetic fields so that the two series would overlap. We intend to return to this in a future paper.

    \item For the correlations with \(|B_r|\) we have \(r^2 \gtrsim \SI{80}{\percent}\) and  \(p < \SI{5}{\percent}\) \emph{except} for \(\Phi_\text{open}\) and the four last rows \(\dot M\), \(\dot J\), \(P_\Wind^\Earth\) and \(R_\text{mag}/R\). The mass loss, followed by the angular momentum loss, are the aggregate quantities that are the \emph{least well} explained by variations in \(|B_r|\), indicating that other factors need to be accounted for in a good model of wind mass loss.

    \item The open magnetic flux \(\Phi_\text{open}\) is well correlated with \(\dot M\), \(\dot J\), \(P_\Wind^\Earth\) and \(R_\text{mag}/R\). Here we have \(r^2 \gtrsim \SI{80}{\percent}\) and \(p < \SI{5}{\percent}\) (except for \(\dot M\) in the \(5B_\ZDI\) series where \(p=\SI{5.1}{\percent}\) and \(r^2=\SI{77}{\percent}\)). As was also noted in the previous section, the better correlation of \(\dot J\) and \(\dot M\) with \(\Phi_\text{open}\) appears to support the argument of \citet{2015ApJ...798..116R} that rotational braking laws are better formulated in terms of the open flux than in terms of absolute field strength.

    \item For the pooled series, all the parameters of Table~\ref{tab:correlations} are correlated with \(p\leq\SI{1}{\percent}\) and it is thus clear that all the quantities of our model are strongly correlated with \(B_r\) over this large range of magnetic field values.

    \item The fact that the exponent \(a\approx 1\) for the magnetic quantities, \(|\vec B|\), \(\max |B_r|\) and \(\Phi_0\) indicates that the magnetic geometry is largely independent from the average radial field strength \(|B_r|\). 

    \item While the Alfvén surface quantities \(R_\Alfven\) and \(|\vec{r_\Alfven} \times \uvec{z}|\) do not scale linearly with \(|B_r|\), the high \(r^2\) values show that \(|B_r|\) indicates that the magnetic geometry plays a small role in determining these Alfvén surface quantities.
    
    \item The strong correlation (\(p \leq \SI{1}{\percent}\)) between \(\Phi_\text{open}\) and the wind pressure at \SI{1}{\astronomicalunit}, \(P_\Wind^\Earth\) is particularly noteworthy as the calculation of \(\Phi_\text{open}\) requires a significantly smaller domain of simulation than the calculation of \(P_\Wind^\Earth\). Hence it may be justified to save computational resources modelling only the inner domain (the corona) where \(\Phi_\text{open}\) is calculated and not extending the model to \SI{1}{\astronomicalunit} and beyond. 
    \end{enumerate}

\section[Numerical comparison with See et al. (2019)]{Comparison with See et al. (2019)}\label{appendix:see-comparison}

Table~\ref{tab:See-comparison} shows the \(\dot M\) and \(\dot J\) values of this work and the \(\dot M\) and \(\dot J\) values of the \citet{2019ApJ...886..120S} mod-M15 and CS11 methods. The CS11 method has good agreement with the our series in mass loss, and excellent agreement with the \(B_\ZDI\) series for angular momentum loss. 
The mass loss values from the mod-M15 method values of mass loss fall between the \(B_\ZDI\) and \(5B_\ZDI\) for two stars and is otherwise higher than our results, the mod-M15 series for angular momentum loss is 5--10 times higher than our highest \(\dot J\) value series \(5B_\ZDI\). 

\begin{table} 
    \centering
    \caption{
        Comparison of stellar mass loss and angular momentum loss values calculated in this work and the mod-M15 and CS11 models in \citet{2019ApJ...886..120S} as described in Section~\ref{sect:see-comparison} and plotted in Figure~\ref{fig:results-context}. The \(B_\ZDI\) and \(5B_\ZDI\) columns refer to the \(B_\ZDI\) and \(5B_\ZDI\) in this work. 
    }\label{tab:See-comparison}
    \sisetup{
        range-phrase=--,
        range-units=single,
        scientific-notation=fixed,
        }
    \begin{tabular}{lcccc}
        \toprule
        {Star} & \multicolumn{4}{c}{Mass loss values  \(\dot M\)} \\
        {} & CS11 & mod-M15 & \(B_\ZDI\) & \(5B_\ZDI\)  \\ 
        \midrule
        Mel25-5       &  \num[fixed-exponent=9]{5.17E+09} & \num[fixed-exponent=9]{1.79E+10} & \num[fixed-exponent=9]{3.1e+09} & \num[fixed-exponent=9]{6.0e+09}{} \\
        Mel25-21      &  \num[fixed-exponent=9]{1.06E+10} & \num[fixed-exponent=9]{5.90E+09} & \num[fixed-exponent=9]{3.6e+09} & \num[fixed-exponent=9]{7.2e+09}{} \\
        Mel25-43      &  \num[fixed-exponent=9]{4.71E+09} & \num[fixed-exponent=9]{2.06E+10} & \num[fixed-exponent=9]{2.0e+09} & \num[fixed-exponent=9]{4.5e+09}{} \\
        Mel25-151     &  \num[fixed-exponent=9]{3.43E+09} & \num[fixed-exponent=9]{7.14E+09} & \num[fixed-exponent=9]{2.5e+09} & \num[fixed-exponent=9]{5.4e+09}{} \\
        Mel25-179     &  \num[fixed-exponent=9]{5.32E+09} & \num[fixed-exponent=9]{4.82E+09} & \num[fixed-exponent=9]{4.1e+09} & \num[fixed-exponent=9]{6.6e+09}{} \\ 
        \midrule
        {Star} & \multicolumn{4}{c}{Angular momentum loss values  \(\dot J\)} \\
        {} & CS11 & mod-M15 & \(B_\ZDI\) & \(5B_\ZDI\)  \\ 
        \midrule
        Mel25-5        & \num[fixed-exponent=23]{6.22E+23} &  \num[fixed-exponent=25]{3.10E+25} & \num[fixed-exponent=23]{7.5e+23} & \num[fixed-exponent=24]{4.8e+24}{}  \\
        Mel25-21       & \num[fixed-exponent=23]{1.48E+24} &  \num[fixed-exponent=25]{2.74E+25} & \num[fixed-exponent=23]{1.5e+24} & \num[fixed-exponent=24]{8.4e+24}{}  \\
        Mel25-43       & \num[fixed-exponent=23]{3.43E+23} &  \num[fixed-exponent=25]{1.93E+25} & \num[fixed-exponent=23]{3.9e+23} & \num[fixed-exponent=24]{3.3e+24}{}  \\
        Mel25-151      & \num[fixed-exponent=23]{6.18E+23} &  \num[fixed-exponent=25]{2.34E+25} & \num[fixed-exponent=23]{7.2e+23} & \num[fixed-exponent=24]{4.4e+24}{}  \\
        Mel25-179      & \num[fixed-exponent=23]{1.15E+24} &  \num[fixed-exponent=25]{2.76E+25} & \num[fixed-exponent=23]{1.9e+24} & \num[fixed-exponent=24]{8.2e+24}{}  \\
            \bottomrule
        \end{tabular}
    \end{table}

\section{Hipparcos catalog numbers}
Table~\ref{tab:hipparcos} gives \emph{Hipparcos} catalog numbers for the Hyades stars in this work. Elsewhere in this work we use abbreviated versions of the names used in~\citetalias{2018MNRAS.474.4956F}.
\begin{table} 
    \centering
    \caption{
        \emph{Hipparcos} catalog designations for the Hyades stars in this work. Further catalog designations for each star are given in~\citetalias{2018MNRAS.474.4956F}.
    }\label{tab:hipparcos}
    \begin{tabular}{ll}
        \toprule
        {Star} & Hipparcos \\
        \midrule
\href{http://simbad.u-strasbg.fr/simbad/sim-id?Ident=Cl+Melotte+25+5  }{Cl Melotte 25 5  }  & HIP 16908 \\
\href{http://simbad.u-strasbg.fr/simbad/sim-id?Ident=Cl+Melotte+25+21 }{Cl Melotte 25 21 }  & HIP 19934 \\
\href{http://simbad.u-strasbg.fr/simbad/sim-id?Ident=Cl+Melotte+25+43 }{Cl Melotte 25 43 }  & HIP 20482 \\
\href{http://simbad.u-strasbg.fr/simbad/sim-id?Ident=Cl+Melotte+25+151}{Cl Melotte 25 151}  & HIP 23701 \\
\href{http://simbad.u-strasbg.fr/simbad/sim-id?Ident=Cl+Melotte+25+179}{Cl Melotte 25 179}  & HIP 20827 \\
        \bottomrule
    \end{tabular}
\end{table}

\bsp	%
\label{lastpage} %
\end{document}